\documentclass[11pt,a4paper]{article}
\usepackage[utf8]{inputenc}
\usepackage[english]{babel}
\usepackage{amsmath,bm}
\usepackage{slashed}
\usepackage{amsfonts}
\usepackage[T1]{fontenc}
\usepackage{tasks}
\settasks{
    label = {}, 
    label-format = {\normalfont(\alph*)},
    label-align = left,
    label-width = 1.5em,
    column-sep = 1em, 
}
\usepackage{amssymb}
\usepackage{multirow}
\usepackage{cancel}
\usepackage[numbers,sort&compress]{natbib}
\usepackage[colorlinks = true,
            linkcolor = blue,
            urlcolor  = blue,
            citecolor = blue,
            anchorcolor = blue]{hyperref}
\usepackage{array}
\usepackage{booktabs}
\usepackage{makeidx}
\usepackage{authblk}
\usepackage{float}
\usepackage{mathtools}
\usepackage{graphicx}
\usepackage[dvipsnames,svgnames]{xcolor}
\usepackage{titlesec}
\usepackage{authblk}
\usepackage{subcaption}
\usepackage[normalem]{ulem}
\usepackage{booktabs} % for professional 
\usepackage{float}
\usepackage[table]{xcolor} % For row colors
\usepackage{graphicx}      % For resizebox
\usepackage[left=2cm,right=2cm,top=2cm,bottom=2cm]{geometry}
\usepackage{tikz}
\usetikzlibrary{arrows.meta}  % for better arrows
\usetikzlibrary{decorations.pathmorphing} % for snake lines (vector bosons)
\usetikzlibrary{decorations.markings}     % for arrows along lines
\usepackage{subcaption} % in preamble
\newcommand{\bea}{\begin{eqnarray}}
\newcommand{\bec}{\begin{columns}}
\newcommand{\eec}{\end{columns}}
\newcommand{\eea}{\end{eqnarray}}
\newcommand{\beq}{\begin{equation}}
\newcommand{\eeq}{\end{equation}}
\newcommand{\ec}{\end{center}}

\newcommand{\bc}{\begin{center}}

\title{\bf Implications of portal vector-like lepton on associated Higgs production 
at a multi-TeV muon collider}

\author[a,\thanks{\href{mailto:krishna.tewary.phy22@gm.rkmvu.ac.in}
{krishna.tewary.phy22@gm.rkmvu.ac.in}, \href{mailto:krishnatewary3@gmail.com}{krishnatewary3@gmail.com}}]{Krishna Tewary}

\author[a,\thanks{\href{mailto:sanjoy.phy@gm.rkmvu.ac.in}{sanjoy.phy@gm.rkmvu.ac.in}, \href{mailto:sanjoyhri@gmail.com}{sanjoyhri@gmail.com}}]{Sanjoy Biswas}

\affil[a]{{\it Department of Physics, Ramakrishna Mission Vivekananda Educational and Research Institute, Belur Math, Howrah 711202, India.}}

\author[b,\thanks{\href{mailto:shivamverma@prl.res.in}{shivamverma@prl.res.in}, \href{mailto:shivam.59910103@gmail.com}{shivam.59910103@gmail.com}}]{Shivam Verma}

\affil[b]{{\it Theoretical Physics Division, Physical Research Laboratory,
Shree Pannalal Patel Marg, Ahmedabad - 380009, Gujarat, India.} }

\date{}

\begin{document}

\maketitle

%%%%%%%%%%% Abstract %%%%%%%%%
\begin{abstract}

We {explore} a \textit{portal vector-like lepton} (pVLL) extension of the Standard Model (SM) and {study} its implications for Higgs and vector-boson associated production ($hV$, with $V = Z$-boson or dark photon) at a future muon collider facility. We show that while the $\mu^+ \mu^- \to hZ$ production rate remains close to its SM prediction in a wide range of parameter space, the rate for $\mu^+ \mu^- \to h\gamma_d$ can be substantially enhanced owing to the \textit{non-decoupling nature} of the interaction involving the heavy lepton, the muon and the dark photon. We demonstrate that the $h\gamma_d$ production rate can exceed the corresponding $hZ$ rate by a factor of $1$--$100$ within the {\it perturbative unitarity limit}, making it a promising channel for probing Higgs interactions and potential new physics effects. 

 {We also examine the role of  the pVLL state in the context of dark matter (DM) phenomenology and identify regions of parameter space consistent with the observed relic abundance by extending the simplified setup with a viable DM candidate.
The $h \, \gamma_d$ production} can also be used to constrain the dark photon mass ($m_{\gamma_d}$) and/or the dark gauge coupling ($g_d$) consistent with {various constraints including} the current muon $g-2$ measurements within the pVLL framework. We perform a detailed collider analysis of the $h\gamma_d$ process in the $b\bar{b}~+$ missing energy final state. A $2\sigma$ exclusion limit for $m_{\gamma_d}$ up to $80$ GeV is obtained assuming $g_d=0.05$, $\sin\theta_L=4\times10^{-5}$, $\sin\theta_s=0.05$, for a heavy lepton mass $~\sim 3$ TeV at a $3$ TeV muon collider with an integrated luminosity of $1$ ab$^{-1}$.

\end{abstract}

\clearpage

%%%%%%  Sec 1: Introduction %%%%%%%%%%%%%
\section{Introduction}
   
% Motivation

The discovery of the Higgs boson \cite{ATLAS:2012yve, CMS:2012qbp} marks a remarkable triumph of theoretical and experimental particle physics. To probe its properties with greater precision, however, we require an abundant production of Higgs bosons at high-energy (HE) and high-luminosity (HL) facilities such as the Large Hadron Collider (LHC) and future collider experiments. Higgs production in association with a gauge boson is a well-studied process at lepton colliders~\cite{Kilian:1995tr,Liu:2016zki,Akeroyd:1999gu, Kanemura:1999tg, Ouazghour:2024twx}, both in past experiments \cite{LEPWorkingGroupforHiggsbosonsearches:2003ing,OPAL:2002sbl} and in proposed future facilities \cite{FCC:2018evy, Abramowicz:2016zbo, Franceschini:2021aqd, Hamada:2024ojj}. The presence of new physics, particularly, scenarios predicting vector-like leptons (VLLs), can significantly alter the associated Higgs production cross section at lepton colliders -- either through modified Higgs couplings, the appearance of new diagrams, or both. Moreover, such scenarios may open up additional Higgs production channels with enhanced rates. In this work, we explore the implications of VLLs acting as portal matter on the {associated} Higgs production cross section at a future muon collider.

% Introduction of VLLs
 
VLLs constitute a simple yet well-motivated extension of the Standard Model (SM) of particle physics. VLL extension of the SM has been studied extensively in the literature in the context of various anomalies, including the muon $g-2$, the $W$-boson mass, the Cabibbo angle anomaly, and dark matter phenomenology~\cite{Chun:2020uzw,Asadi:2024jiy, Crivellin:2020ebi, Kirk:2020wdk, Wojcik:2023ggt, Poh:2017tfo, Dermisek:2021ajd, Lee:2021gnw, Kawamura:2019hxp, Endo:2020tkb, Manzari:2021prf, Capdevila:2020rrl, Abdallah:2023pbl, Lee:2022nqz}. They naturally arise in a variety of theoretical frameworks \cite{Lee:2016wiy,Gursey:1975ki,Kyae:2013hda,Cvetic:2001tj,Cleaver:1999mw,Marchesano:2013ega}, including grand unified theories, models addressing the Higgs mass hierarchy problem, and constructions inspired by string theory. The search for VLLs at collider experiments remains an active area of investigation, including both prompt and long-lived signatures, primarily in SM final states \cite{ATLAS:2024mrr,CMS:2022cpe, CMS:2019hsm}. VLL is also a potential candidate to search for at the future colliders such as future  $pp,~ e^+ e^-$, muon collider \cite{Yue:2024sds,Shang:2021mgn,Guo:2023jkz, Ghosh:2023xbj,Bhattiprolu:2019vdu}.

% VLL as portal matter

In recent years, the idea of VLLs as portal matter \cite{Rizzo:2022qan,Rueter:2019wdf,Rizzo:2018vlb,Wojcik:2023ggt} or portal VLLs (pVLLs) has gained significant interest owing to their rich phenomenology, their ability to evade existing collider constraints, and their potential connection to the dark sector, particularly dark matter phenomenology \cite{Belyaev:2025cgf,Lahiri:2024rxc}.

% Model description
In this work, we consider an extension of the Standard Model (SM) with an $SU(2)$-singlet VLL that transforms under both the SM hyper-charge group $U(1)_Y$ and an additional abelian dark gauge group $U(1)_D$ \cite{Rizzo:2022qan,Rueter:2019wdf,Rizzo:2018vlb,Wojcik:2023ggt}. The gauge boson associated with $U(1)_D$ is referred to as the dark photon ($\gamma_d$).
The model also contains an additional complex scalar field, singlet under the SM gauge group but carrying non-zero $U(1)_D$ charge. The full gauge symmetry, $SU(3)_C \times SU(2)_L \times U(1)_Y \times U(1)_D$, is spontaneously broken to $SU(3)_C \times U(1)_{\rm EM}$ through the non-vanishing vacuum expectation values (VEVs) of the SM Higgs doublet and the dark scalar. The VEV of the dark scalar not only generates the dark photon mass but also induces mixing between the pVLL and SM leptons through a Yukawa-type portal interaction.  
The $U(1)_Y$ hypercharge assignment of the pVLL considered here corresponds to an electric charge of $-1$, and it is allowed to mix only with one generation of SM charged leptons in order to evade stringent bounds from lepton-flavor violation.  In this work, we focus on pVLL mixing with the second-generation charged lepton. 

% Non-decoupling effect and description of muon collider 

An important feature of this model is that the right-handed second generation charged lepton ($\mu_R^\prime$) and the pVLL ($\mu_{p}^\prime$) carry identical $SU(2)_L \times U(1)_Y$ charges. Consequently, any rotation between $\mu_R^\prime$ and $\mu_{p_{_R}}^\prime$ preserves the diagonal structure of the $Z$-boson couplings, ensuring that no $\mu_R-\mu_{p_{_R}}-Z$ interaction arises after transforming to the mass basis. In contrast, since the muon is uncharged under $U(1)_D$, such a rotation induces an off-diagonal $\mu_R-\mu_{p_{_R}}-\gamma_d$ coupling, suppressed by $\sin\theta_R$, the sine of the corresponding mixing angle. One can have a sizable value of $\sin\theta_R \sim 0.1$ for heavy muon mass $\sim$ a few TeV even when the mixing angle ($\sin\theta_L$) involving the corresponding left-chiral fermions is kept at a tiny value $\sim 10^{-5}$. {The {\it non-decoupling} nature of the interaction is central to both the dark matter phenomenology and the collider signatures of the model.}

{ After examining the various constraints on the model parameter space arising from Higgs signal strength measurements, heavy Higgs and di-Higgs searches at the LHC, limits from lepton flavor universality and other electroweak precision observables, searches for vector-like leptons at the LHC, as well as consistency with current experimental results on the muon $g-2$, we proceed to explore the implications for dark matter relic abundance and associated Higgs production at a proposed future lepton collider, utilizing the {\it non-decoupling effect} within the {\it perturbative unitarity} limit.}

{We investigate the role of the pVLL in the context of DM relic density  calculations, assuming that the dark sector contains a Dirac DM particle ($\chi$) charged under $U(1)_D$ as a motivation behind the consideration of pVLL extension of the SM. In the absence of the pVLL state, the annihilation rate in the $\chi \bar\chi \to f \bar f$ channel is typically suppressed due to the small gauge kinetic mixing (KM), often resulting in an overabundant relic density. The indirect effect of the pVLL is to {\it facilitate} DM annihilation into a pair of muons by modifying the $\mu^+ \mu^- \gamma_d$ vertex, without introducing any additional Feynman diagrams. Furthermore, its presence introduces a non-trivial interplay between the gauge kinetic mixing and fermion mixing parameters.}

{The non-decoupling effect also show up in the Higgs production in association with a vector boson at the proposed future muon collider} ($\mu^+ \mu^- \to h V$ with $V= Z, \gamma_d$).
The muon collider \cite{Delahaye:2019omf,Schulte:2022brl,Black:2022cth} is envisioned as a high-energy circular lepton collider that combines the clean experimental environment of lepton collisions with access to multi-TeV center-of-mass energies ($\sqrt{s}$), in contrast to circular $e^+e^-$ machines. It is designed to operate at energy stages from an initial 3 TeV to 10 TeV (or higher), corresponding to circumferences of about 4.5 km and 10 km, respectively, with expected peak integrated luminosities of 1 ab$^{-1}$ at $\sqrt{s} = 3$ TeV and 10 ab$^{-1}$ at $\sqrt{s} = 10$ TeV \cite{InternationalMuonCollider:2024jyv}.  

 We have shown that while the $hZ$ production cross section remains essentially unchanged from the SM prediction, the $h\gamma_d$ production rate can be significant even for very small $\sin\theta_L$, owing to the non-decoupling effect. In this scenario, the dark photon is treated as an invisible particle\footnote{This will be the case {for light DM with $m_\chi < m_{\gamma_d}/2$, where the dark photon} predominantly decays to {an} invisible final state.}. We analyze this process in the $b \bar{b}$ {\it plus missing energy} final state at $\sqrt{s}=3$ and $10$~TeV muon colliders, where the $b \bar{b}$ system is highly boosted due to the $2\to 2$ kinematics and large center-of-mass energy. The dominant SM backgrounds to this channel are processes leading to $b\bar{b}\nu\bar{\nu}$ final state, with $h\nu\bar{\nu}$ and $Z\nu\bar{\nu}$ accounting for more than $90\%$ of the total cross section. We demonstrate that the requirement of jet substructure, together with additional kinematic selections such as missing energy and the $b\bar{b}$ invariant mass, can efficiently suppress these backgrounds.

% 
 
% Exploring dark photon even without KM

The main highlight of this work is that the $h\gamma_d$ process not only yields a higher Higgs production rate than $hZ$ production at a muon collider, but also offers a noble way to explore dark photons in the mass range $10$--$100$~GeV, independent of the gauge kinetic mixing parameter ($\varepsilon$)~\cite{Holdom:1986eq,Fayet:1990wx,Fabbrichesi:2020wbt,Alexander:2016aln}.
This is because of the fact that the muon--heavy muon--Higgs vertex is sensitive to the dark photon mass ($m_{\gamma_d}$) and the dark gauge coupling ($g_d$).  Consequently, the absence of any observable enhancement in Higgs production along with missing energy at a muon collider would imply a stringent limit on $m_{\gamma_d}$ or $g_d$, for fixed values of other relevant model parameters.
We illustrate this using several representative benchmark points and show that, even for fermion mixing angle as small as $\sim 10^{-6}$ and heavy muon masses in the range $1$--$3$~TeV, a wide range of dark photon masses can be excluded.
We illustrarte that a substantial region of the parameter space consistent with current measurement of muon $g-2$ {and DM relic abundance} can be probed at the future muon collider using our analysis.
% Organization of rest of the article

The remainder of the manuscript is organized as follows. In Section~\ref{sec:model}, we present the theoretical framework considered in this work to illustrate the impact of portal VLLs on associated Higgs production at muon collider, along with a discussion of relevant constraints on the model parameters.
{Section~\ref{sec:DM} contains the parameter space consistent with observed DM relic abundance.} In
Section~\ref{sec:collider_analysis}, we describe the collider analysis of Higgs production in association with a vector boson. The results of our analysis are presented in Section~\ref{sec:results}, and finally we conclude in Section~\ref{sec:conclusion}.
% 
% 
%%%%%%  Sec 2:Theoretical Framework %%%%%%%%%%
% 
\section{Theoretical Framework}
\label{sec:model}
We consider a framework in which the Standard Mode{l} is extended by a vector-like lepton, singlet under $SU(3)_C \times SU(2)_L$ with  $U(1)_Y$ hypercharge  -$1$. The VLL plays the role of a portal matter, linking the visible SM sector with a dark sector.  A simplified realization of such a scenario is to have a VLL transforming non-trivially under an additional local $U(1)_D$ symmetry of the dark sector with corresponding  gauge boson referred to as the dark photon ($\gamma_d$). The dark photon acquires mass through spontaneous breaking of $U(1)_D$, achieved by introducing a complex scalar field $\Phi_d$ which is a singlet under the SM gauge group but charged under $U(1)_D$. Furthermore, if $\Phi_d$ carries the same $U(1)_D$ charge as the pVLL, a Yukawa-type portal interaction $\omega_f \, \Phi_d \, \overline{\mu}^{\prime}_{p_L} \mu^{\prime}_R $ is permitted.
The Yukawa interaction reduces to a bilinear term between the pVLL and the SM lepton once $\Phi_d$ gets a non-zero VEV. In this work, we focus on the case where the pVLL mixes with the second-generation charged lepton only to avoid constraints coming from flavor violating observables. The assignment of charge of the relevant field content is presented in Table~\ref{tab:charge_assignment}. 
% 
% Tab-1: Charge assignment of the relevant fields %%
% 
\begin{table}[H]
\centering
\resizebox{0.6\columnwidth}{!}{
\begin{tabular}{lcccc}
\toprule
Fields & $SU(3)_C$ & $SU(2)_L$ & $Y$ & $Y_d$\\
\midrule
\midrule
$\mu^{'}_{R}$ & \textit{1} & \textit{1} & -1 & \textcolor{red}{0} \\
$L_{L}=\begin{pmatrix}
\nu_{\mu_L}\\
\mu^{'}_{L}
\end{pmatrix}$ & \textit{1} & \textit{2} & -1/2 & \textcolor{red}{0} \\
$\Phi$ & \textit{1} & \textit{2} & 1/2 & \textcolor{red}{0} \\
$\mu^{\prime}_p$ & \textit{1} & \textit{1} & -1 & \textcolor{red}{1} \\
% $\mu^{'}_{p_R}$ & \textit{1} & \textit{1} & -1 & \textcolor{red}{1} \\
$\Phi_d$ & \textit{1} & \textit{1} & 0 & \textcolor{red}{1} \\
\bottomrule
\bottomrule
\end{tabular}}
\caption{Representation of relevant fields and their $U(1)$ charge assignment{s} under the extended gauge group.}
\label{tab:charge_assignment}
\end{table}

In Table~\ref{tab:charge_assignment},  $\mu_R^\prime$ and $L_{L}$ represent the second-generation SM right-handed charged lepton and left-handed lepton doublet, respectively, and $\Phi$ corresponds to the SM complex scalar field doublet. The additional VLL singlet under $SU(2)_L$ is represented by $\mu_p^\prime$. $\Phi_d$ is the complex scalar field that does not carry charges under the SM gauge group. All fields are expressed in their interaction basis. 
% 
% Lagrangians
% 
The complete Lagrangian of the model invariant under $SU(3)_C \times SU(2)_L \times U(1)_Y \times U(1)_D$ can be expressed as 
% Total Lagrangian
\begin{eqnarray}
\label{eq:L_total}
      \mathcal{L} &=& \mathcal{L}_{\rm Gauge} +  \mathcal{L}_{\rm Scalar} + \mathcal{L}_{\rm Fermions}  + \mathcal{L}_{\rm Yukawa} 
\end{eqnarray}
where $\mathcal{L}_{\rm Gauge},~ \mathcal{L}_{\rm Scalar},~\mathcal{L}_{\rm Fermions}$ are part of the full Lagrangian containing gauge invariant kinetic terms involving the gauge fields, scalar fields including corresponding scalar potential, and fermion fields including the mass term of the pVLL, respectively. Finally, $\mathcal{L}_{\rm Yukawa}$ represents the trilinear Yukawa interactions involving scalar and  fermions.

In particular, the gauge sector of the theory is given by 
% Gauge sector Lagrangian
\begin{eqnarray}
\label{eq:L_gauge}
         \mathcal{L}_{\rm Gauge} &=&
         -\frac{1}{4}G_{\mu\nu}^a G^{a,\mu\nu}
         -\frac{1}{4}W_{\mu\nu}^i W^{i,\mu\nu}
         -\frac{1}{4}B_{\mu\nu} B^{\mu\nu} 
         -\frac{1}{4}B_{d,\mu\nu} B_d^{\mu\nu}
        { - \frac{\varepsilon}{2}}B_{d,\mu\nu}B^{\mu\nu} ,
\end{eqnarray}
where $G_{\mu\nu}^a$ denotes the QCD field strength tensor with $a=1,\dots,8$ labeling the adjoint of $SU(3)_C$, 
$W_{\mu\nu}^i$ ($i=1,~2,~3$) are the field strength tensors of $SU(2)_L$, 
$B_{\mu\nu}$ corresponds to the $U(1)_Y$ hypercharge, 
and $B_{d,\mu\nu}$ is the field strength tensor of the dark $U(1)_D$. 
{The non-canonical form of the gauge boson kinetic terms is due to the presence of kinetic mixing involving the field strength tensors corresponding to two abelian gauge groups, $U(1)_Y$ and $U(1)_D$, respectively. The strength of this gauge kinetic mixing is parametrized by the parameter $\varepsilon$. In order to bring the gauge kinetic terms into canonical form, we perfom the following non-orthogonal transformation between the hypercharge gauge boson $B_\mu$ and corresponding $U(1)_D$ gauge boson: }
\begin{eqnarray}
   { B_\mu^\prime} &{=}& {B_\mu + \varepsilon \, B_{d, \mu}} \nonumber\\
    {B_{d,\mu}^\prime} &{=}& {\sqrt{1-\varepsilon^2} \, B_{d,\mu}}
\end{eqnarray}
{In the primed basis, the gauge kinetic terms have canonical form. However, after the breaking of $SU(2) \times U(1)_Y \times U(1)_D \to U(1)_{\rm EM}$, the mass matrix involving the neutral gauge boson in the basis ($B_\mu^\prime~,~ W_\mu^3~,~ B_{d,\mu}^\prime$) has off-diagonal terms. The physical neutral gauge boson mass eigen states ($A_\mu$, $Z_\mu$, and $A_{d,\mu}$) are related to this by the following orthogonal transformation that involves two mixing angles, the weak mixing angle (${\theta}_{_W}$) and $\theta_d$ \cite{Babu:1997st}. }

\begin{eqnarray}
  {  \begin{pmatrix} 
A_\mu \\ 
Z_{\mu} \\ 
A_{d, \mu} 
\end{pmatrix}} {=} 
{
\begin{pmatrix} 
\cos{\theta}_{_W} & \sin{\theta}_{_W} & 0 \\ 
- \cos\theta_d \sin{\theta}_{_W} & \cos\theta_d \cos{\theta}_{_W} & \sin\theta_d  \\ 
\sin\theta_d \sin{\theta}_{_W} & - \sin\theta_d \cos{\theta}_{_W} & \cos\theta_d 
\end{pmatrix} }
{
\begin{pmatrix} 
B^\prime_\mu \\ 
W_\mu^3 \\ 
B^\prime_{d,\mu} 
\end{pmatrix}}
\end{eqnarray}

{{The angle of rotation $\theta_d$ is given by}
\begin{eqnarray*}
  {  \tan{2\theta_d}} &{=}& {\left( 
\frac{-2 \, \varepsilon \, \sqrt{1-\varepsilon^2} \, \sin{\theta}_{_W} \, M_{Z_{0}}^2}
{M_{A_{d_{0}}}^2 - M_{Z_{0}}^2 \left(1 - \varepsilon^2 \, (1 + \sin^2{\theta}_{_W} ) \right)}
\right)}
\end{eqnarray*}
}
{{where, $M_{Z_{0}}=v_{_{\text{EW}}}(g^2 +g'^2)/2 $ and $M_{A_{d_{0}}}=v_d \,g_d$ are the $Z$ boson mass and dark photon mass, respectively, without the kinetic mixing.}}
% 

% Scalar sector Lagrangian
The scalar-sector Lagrangian is 
\begin{eqnarray}
\label{eq:L_scalar}
        \mathcal{L}_{\rm Scalar} 
        &=& |D_{\mu}\Phi|^2 + |D_{\mu}\Phi_d|^2 - V(\Phi, \Phi_d) \, ,
\end{eqnarray} 
% Scalar potential
with the scalar potential
\begin{eqnarray}
\label{eq:V_potential}
          V(\Phi, \Phi_d) &=& - \mu_h^2|\Phi|^2 - \mu_{h_d}^2 |\Phi_d|^2 
          + \lambda_h|\Phi|^4 + \lambda_{h_d}|\Phi_d|^4 
          + \lambda_{hh_d}|\Phi|^2 |\Phi_d|^2 \, .
\end{eqnarray} 

% Symmatry breaking condition
  The symmetry breaking condition and boundedness of the potential require $\mu_h^2, \mu_{h_d}^2 >0$ and $(\lambda_h \lambda_{h_d} - \lambda_{h h_d}^2), \lambda_h, \lambda_{h_d} >0$. The field configurations of $\Phi$ and $\Phi_d$ at the minima of the scalar potential in unitary gauge is given by 
\begin{eqnarray}
\label{eq:vevs}
    \Phi=
        \begin{pmatrix}
             0 \\
            \dfrac{v_{_{\rm EW}}+h_1}{\sqrt{2}}
        \end{pmatrix},
\quad 
    \Phi_d=\dfrac{1}{\sqrt{2}}(v_d+h_2).
\end{eqnarray}
where $h_1$ and $h_2$ denote the fluctuation around the minimum field configurations and $v_{_{\rm EW}}$ and $v_d$ represent the VEV of the respective scalar fields. We denote the fields in the mass basis by $h$ and $h_d$ which correspond to the observed Higgs  and the hypothetical dark Higgs fields, respectively. The physical scalar fields ($h,~h_d$) are related to  the ($h_1,~h_2$) by the following unitary transformation 
%% Scalar mixing
%
\begin{eqnarray}
\label{eq:scalar_mixing}
    \begin{pmatrix}
        h \\
         h_d
    \end{pmatrix}
&=&
    \begin{pmatrix}
        \cos\theta_{s} & -\sin\theta_{s} \\
        \sin\theta_{s} & \cos\theta_{s}
    \end{pmatrix} 
    \begin{pmatrix}
        h_1 \\
        h_2
    \end{pmatrix}
\end{eqnarray}
Here, $\theta_s$ is the scalar mixing angle.

% Fermion sector Lagrangian
The expression for $\mathcal{L}_{\rm Fermion}$ and $\mathcal{L}_{\text{Yukawa}}$ appearing in Eq.~\eqref{eq:L_total} involving second generation leptons and pVLL consistent with gauge symmetry are given by
\begin{eqnarray} \label{eq:L_fermion}
\mathcal{L}_{\text{Fermions}} &\supset& 
i [
\bar{L}_L \slashed{D} L_L  +
\bar{\mu}_R' \slashed{D} \mu_R' +\overline{\mu}^{\prime}_p \, \slashed{D} \, \mu_p^{\prime} 
]
- m_{\mu^{\prime}_p}\, \overline{\mu}^{\prime}_p\, \mu_p^{\prime}.
\end{eqnarray} 
%
% Yukawa terms
%
\begin{eqnarray}
\label{eq:L_yukawa}
\mathcal{L}_{\text{Yukawa}} &\supset& 
 - {y_m \, \bar{L}_\mu \Phi \, \mu_R'}
- \omega_f\, \Phi_d\, \overline{\mu}^{\prime}_{p_L} \mu^{\prime}_{R}
+ \mathrm{H.c.}
\end{eqnarray} 

% Mixing between muon and VLL
The mixing between $\mu^\prime$ and $\mu^\prime_p$ is generated once the dark Higgs assumes non-zero VEV.
%
% Mass matrix in interaction basis
After symmetry breaking, the mass matrix in the interaction basis $\mu^\prime$ and $\mu^\prime_p$ takes the form
\begin{eqnarray} \label{eq:fermion_mass_matrix}
        \mathcal{M}
        =
            \begin{pmatrix}
                \dfrac{y_m\, v_{_{\rm EW}}}{\sqrt{2}} & 0 \\[0.3cm]
                \dfrac{\omega_f \, v_d}{\sqrt{2}} & m_{\mu^{\prime}_p}
            \end{pmatrix}
\end{eqnarray} 
 The mass eigenbasis $\mu$ and $\mu_p$ in which the mass matrix is diagonal, represent the physical muon and the corresponding heavy muon partner, respectively. The mass eigenbasis and the interaction basis are connected by the following bi-unitary transformation,
\begin{equation}
\label{eq:fermion_mixing}
        \begin{pmatrix}
            \mu \\ 
            \mu_p
        \end{pmatrix}_{{L,R}}
        = 
        \begin{pmatrix}
            \cos{\theta_{_{L,R}}} & -\sin{\theta_{_{L,R}}}\\ 
            \sin{\theta_{_{L,R}}} & \cos{\theta_{_{L,R}}}
        \end{pmatrix}
        \begin{pmatrix}
            \mu^\prime\\ 
            \mu_p^\prime
        \end{pmatrix}_{{L,R}}
\end{equation} 
    
where, ${\theta_{_{L,R}}}$ is the {mixing angle} between the left (right) chiral muon and the corresponding {heavy muon counter part with the same chirality}.

% Model parameters
The model parameters $y_m,~\omega_f,~ m_{\mu^{\prime}_p}$ appearing in the Lagrangian of Eq.~\eqref{eq:L_fermion} and (\ref{eq:L_yukawa}) can be expressed in terms of the physical parameters $m_{\mu}, ~m_{{\mu_{_p}}}~\text{and}~\sin\theta_L$ as
\begin{subequations}
    \begin{eqnarray}
        y_m &=& \dfrac{\sqrt{2}}{v_{_{\text{EW}}}} m_{_{\mu_p}} D \\ [6pt]
        \omega_f &=& \dfrac{m_{_{\mu_p}} (1- \dfrac{m_\mu^2}{m_{_{\mu_p}}^2})}{\sqrt{2} v_d D}  \sin2\theta_L \\ [6pt]
        m_{_{\mu^{\prime}_p}} &=& \dfrac{m_{\mu}}{D}  \\ [6pt] \nonumber 
    \end{eqnarray}
\end{subequations}
where $D$ is a dimensionless quantity given by $ D=\sqrt{\dfrac{m_{{\mu}}^2}{m_{_{\mu_p}}^2}\cos^2{\theta_L}+ \sin^2{\theta_L}}$.

The absolute value of the sine of the mixing angle between the left chiral second generation charged lepton and the pVLL is expressed as 
\begin{eqnarray}
\label{eq:stL}
    \left| \sin \theta_{L}\right| =\frac{1}{2}\sqrt{\frac{2 m_{_{\mu_p}}^{2}-2m_{\mu}^{2}-\omega_{f}^{2}v_{d}^{2}}{m_{_{\mu_p}}^{2}-m_{\mu}^{2}}\left(1-\sqrt{1-\frac{8\omega_{f}^{2}v_{d}^{2}m_{\mu}^{2}}{\left(2 m_{_{\mu_p}}^{2}-2m_{\mu}^{2}- \omega_{f}^{2}v_{d}^{2}\right)^{2}}}\right)}
\end{eqnarray}
The reality of $\sin\theta_L$ requires 
\begin{eqnarray}
    \left|\omega_f \right| < \sqrt{2}~ \dfrac{(m_{_{\mu_p}}-m_{\mu})}{v_{d}}
\end{eqnarray}
On the other hand, the perturbative unitarity \cite{Kim:2019oyh} demands
\begin{eqnarray}
    \left| \omega_f \right| < 4~\sqrt{2\pi} 
\end{eqnarray}
These two conditions can be integrated as
\begin{eqnarray} \label{eq:perturbative_unitarity_bound}
    \left|\omega_f \right| < \sqrt{2}~\min \left(\frac{m_{_{\mu_p}}-m_{\mu}}{v_{d}},  4  \sqrt{\pi} \right)
\end{eqnarray}
{The corresponding mixing angle ($\sin\theta_R$) between the right chiral muon and heavy muon is given by}
{
\begin{equation}
    \sin\theta_R = \dfrac{\sin\theta_L} {\sqrt{\dfrac{m_\mu^2}{m_{\mu_p}^2}\cos^2\theta_L+\sin^2\theta_L}}
\end{equation}
}
{It turns out that $\sin\theta_R \sim \mathcal{O}(0.1)$ for $m_{\mu_{_p}}$ in the TeV range even for a tiny $\sin\theta_L$ ($\sim 10^{-5}$) and this {\it non-decoupling} effect plays a crucial role both in the context of dark matter and collider analysis}.

This framework naturally gives rise to three portal interactions connecting the visible and dark sectors: the \emph{gauge kinetic mixing}, \emph{Yukawa}, and \emph{scalar} portals. In particular, gauge kinetic mixing between the SM $U(1)_Y$ and dark $U(1)_D$ arises at one loop and is strongly constrained by low-energy precision measurements \cite{Fabbrichesi:2020wbt,Bauer:2018onh,APEX:2024jxw}, with allowed values sensitive to the dark photon mass. The dark photon couples to the SM fermions exclusively via gauge kinetic mixing, except for muons, which also couple through the muon--heavy muon mixing. While the kinetic-mixing portal doesn't play any crucial role, the scalar and Yukawa portals significantly affect Higgs-associated production at a muon collider in this scenario.

The list of free parameters relevant for the discussions in the later sections are  mass of the heavy muon ($m_{\mu_{_p}}$), mixing angle between the muon and its heavier counter part ($\sin\theta_L$), mass of the dark photon ($m_{\gamma_d}$), the dark gauge coupling ($g_d$), and the scalar mixing angle ($\sin\theta_s$) along with the SM parameters. Mass of the additional scalar ($m_{h_d}$) is an independent parameter of the model, however, its impact on our analysis is less significant unlike the scalar mixing angle ($\sin\theta_s$). {The kinetic mixing parameter $\varepsilon$ which will play a crucial role in the DM phenomenology along with the fermion mixing does not have significant impact on the collider analysis as long as its value is small enough to evade the existing constraints coming from low energy observations in the relevant dark photon mass range considered in our work.}

%%%%% subsection-1: Constraints %%%%%%
\subsection{Constraints}
\label{subsec:constraints}
In this subsection we discuss various observational constraints on the allowed values of the relevant model parameters discussed above. 
%
%
 % LHC constraints
 
\subsubsection{{Collider constraints}}
\label{subsubsec:Collider_constraints}
\begin{itemize}
\item[] \textbf{{Scalar search:}}
% \label{subsubsec:scalar_mixing_angle}
% 
The scalar mixing angle which simply plays a role of overall coupling rescaling factor in the analysis presented in this paper is constrained by various observations across the mass of the heavy scalar~\cite{Robens:2015gla, Robens:2019ynf,Robens:2022oue,Adhikari:2022yaa,Lane:2024vur,Robens:2022cun, Robens:2025nev}. For scalar mass below $100$ GeV there are constraints from the LEP experiment and invisible decay of the Higgs boson. For example, if $~m_{h_d} < \frac{m_h}{2}~$ the corresponding bound on the scalar mixing angle coming from the measurement of the Higgs to invisible final state at LHC is of the order of $0.01$ for $v_d \sim 100$ GeV \cite{ATLAS:2023tkt,CMS:2023sdw}. The LEP experiment suggests $\sin{\theta_s}$  in the range ($0.9-0.998$) is ruled out for $m_{h_d}$ in the range $10-100$ GeV and $v_d\sim v_{_{\rm EW}} $ \cite{Robens:2015gla,Robens:2019ynf,Robens:2022oue,Adhikari:2022yaa,Robens:2022cun}. For masses beyond $100$ GeV the constraint coming from blueLHC experiment is more relevant.
The constraint  on $\sin\theta_s$ in the range $~100~\text{GeV}<m_{h_d} <200~\text{GeV}~$ is primarily due to the Higgs signal strength measurement and $\lvert \sin\theta_s \rvert>0.1$ is ruled out in this mass range \cite{Robens:2022cun} with $v_d \sim 0.1 \times v_{_{\rm EW}}$.  For $~m_{h_d}>200~$ GeV, $~\sin{\theta_s} > 0.2~$ is disfavored by the LHC data \cite{Adhikari:2022yaa,Papaefstathiou:2022oyi}.

{The search for resonant and non-resonant di-Higgs production at the LHC can give indirect constraint on the scalar sector parameter as well. The new physics effect can alter the di-Higgs cross section at the LHC either via modification of the trilinear Higgs coupling or via new resonances or both. The CMS observation of non-resonant di-Higgs process at 13 TeV LHC center of mass energy at an integrated luminosity of 138 fb$^{-1}$ sets 95\%  Confidence Level (CL) limit on modified trilinear Higgs coupling ($\lambda_{hhh}$). The modification to this coupling w.r.t. its SM value is parametrized by $\kappa_{_\lambda}$ and defined as
\begin{eqnarray}
    \kappa_{_\lambda} = \frac{\lambda_{hhh}}{\lambda_{hhh}^{\rm SM}}
\end{eqnarray}
where $\lambda_{hhh}$ has the following form 
\begin{eqnarray}
\label{eq:lamda_hhh}
    \lambda_{hhh} = \frac{3 \, m^2_h}{v_{_{\rm EW}}} \left(
\cos^3\theta_s - \frac{v_{_{\rm EW}}}{v_d} \sin^3\theta_s
\right).
\end{eqnarray}
The current 95\% CL limit on $\kappa_{_\lambda}$ is $-1.39 < \kappa_{_\lambda} < 7.02$ \cite{CMS:2025ngq}, where  $\kappa_{_\lambda}=1$ corresponds to the SM prediction. This constraint on $\kappa_{_\lambda}$ implies a restriction on the choice of $\sin\theta_s$ as indicated by Eq.~(\ref{eq:lamda_hhh}).
In Fig.~\ref{fig:kappa_vs_sts} we plot $\kappa_{_\lambda}$ vs. $\sin\theta_s$ which suggests 
 For $v_d > 100$ GeV, $-1<\sin\theta_s <0.85$ is consistent with the current LHC limit on $\kappa_{_\lambda}$.
\begin{figure}[H]
    \centering
    \includegraphics[width=0.6\linewidth]{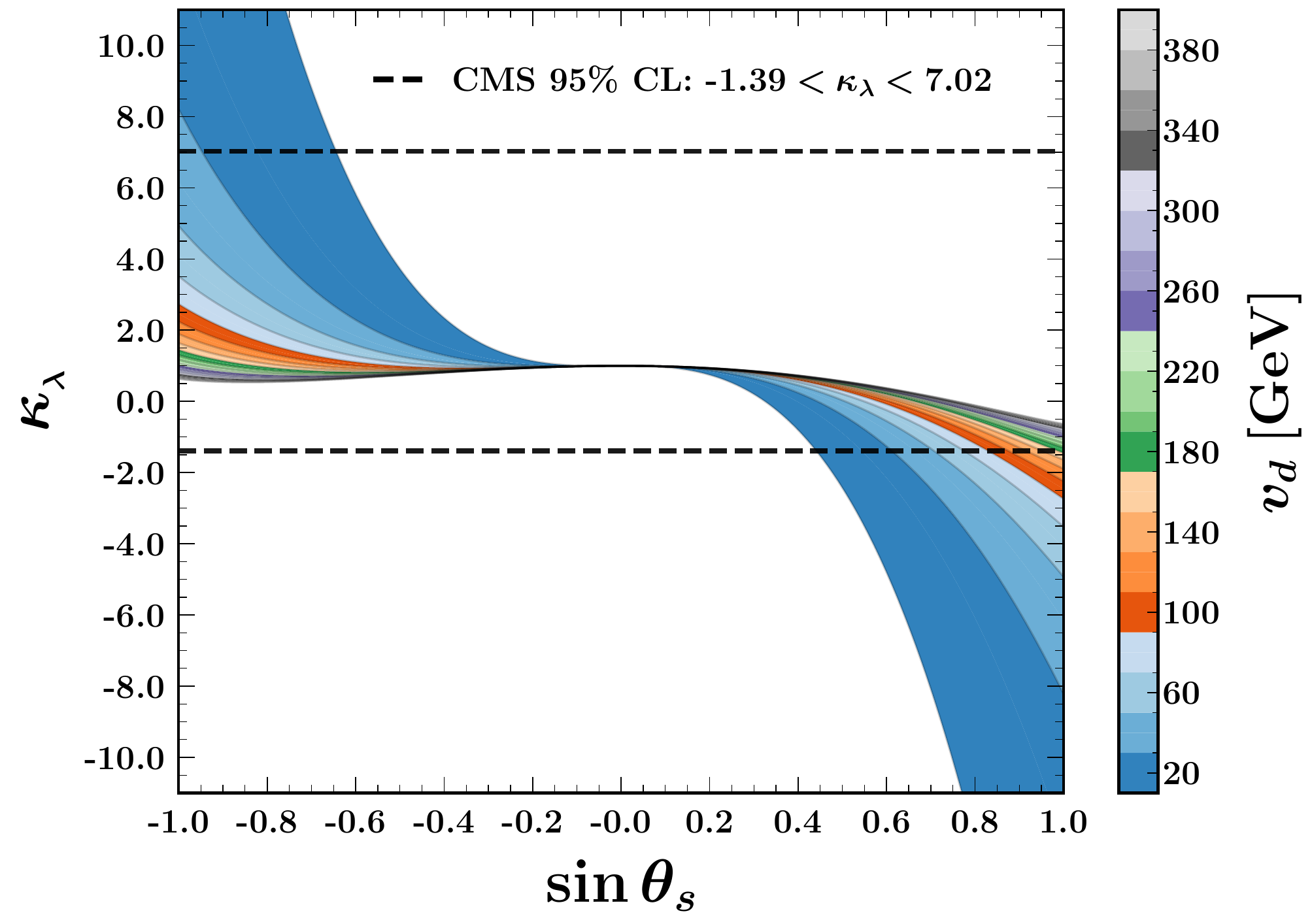}
    \caption{Deviation of the triple Higgs coupling compared to its SM value as a function of $\sin\theta_s$ in the present framework. The colorbar represents the range of the dark Higgs VEV considered above. }   
    \label{fig:kappa_vs_sts}
\end{figure}
%
%
% \textcolor{red}{Resonant di-Higgs:}
{
The search for resonant di-Higgs production at the LHC at $\sqrt{s}=13$ TeV by ATLAS collaboration \cite{Cheng:2025aev} sets upper limit on the resonant di-Higgs cross section as a function of the mass of the heavy resonance in the range $250$ GeV--$600$ GeV at 95\% CL limit. In  Fig.~\ref{fig:resonant_di_Higgs} we plot both the observed 95\% CL limit on the resonant  di-Higgs cross section obtained by ATLAS collaboration \cite{Cheng:2025aev} and that predicted by the present model as a function of the mass of the dark Higgs ($m_{h_d}$) for different choices of dark Higgs VEV.
\\
For $v_d=100$ GeV and $\sin\theta_s=0.2$, the allowed upper limit on the dark Higgs mass is around $325$ GeV.
 This constraint will be relaxed further for higher $v_d$ values and/or smaller $\sin\theta_s$. 
\begin{figure}[H]
    \centering
    \includegraphics[width=0.5\linewidth]{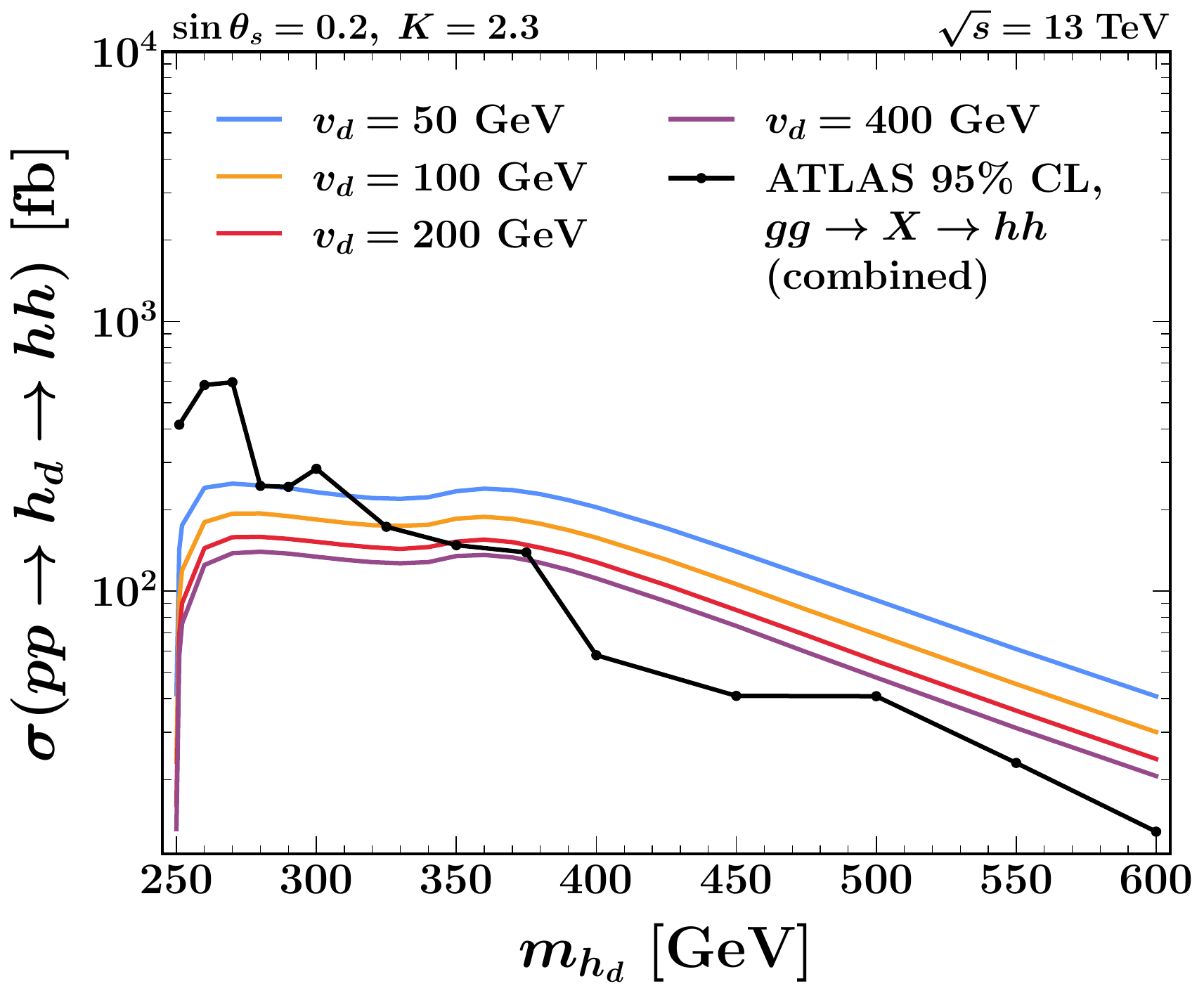}
    \caption{Resonant di-Higgs cross section (in fb) as a function of the mass of the dark Higgs for various choices of its vacuum expectation value at $\sqrt{s} = 13$ TeV and $\sin\theta_s$ $= 0.2$. The estimated cross section also assumes a $K$-factor of $2.3$. The solid black line represents observed 95\% CL limit on the resonant di-Higgs cross section by ATLAS experiment \cite{ATLAS:2023vdy}.}   \label{fig:resonant_di_Higgs}
\end{figure}
Additionally, the scalar mixing angle is also sensitive to studies at the future collider experiments such as HL-LHC, ILC, CLIC, CEPC etc. The stringent bound on $\sin\theta_s$ comes from FCC-ee which suggests $\sin\theta_s< 0.032$ can be ruled out. The projected sensitivity on the scalar mixing angle in the context of rest of the future collider experiments listed above are all consistent with $\sin\theta_s<0.045$ \cite{Buttazzo:2015bka}. 
However, these projections assume $1\sigma$ deviations from SM predictions.}}
\item[] \textbf{{VLL search:}} 
Various extensions of the SM containing the VLLs have been probed at the LHC as well as at the LEP experiment. Due to the limited center-of-mass energy available at the LEP collider, VLLs with mass above $102.6~\text{GeV}$ \cite{Sopczak:1999ua} are not excluded by the LEP experiment. VLL decaying in the SM mode i.e, $\nu W,~ \ell Z$ has been probed at the LHC in the multi-lepton \cite{ATLAS:2024mrr,CMS:2022nty} and hadronically decaying tau lepton with additional jets and/or missing transverse energy channels \cite{ ATLAS:2023sbu,  CMS:2025urb}.
The ATLAS experiment has searched for VLL mixing with first two generations of charged leptons and has obtained an exclusion limit up to 320 GeV (1220 GeV) on the allowed values of the singlet (doublet) -like electron leptons \cite{ATLAS:2024mrr}. The corresponding bound for VLL mixing with muons is 400 GeV (1270 GeV) obtained by same experiment if it is an $SU(2)_L$ singlet (doublet). The CMS experiment at the LHC sets a lower limit on the allowed value of the doublet (singlet) vector-like tau lepton mass to be $\gtrsim 1045~\text{GeV}~(150~\text{GeV})$ with 95 \% {CL} \cite{CMS:2022nty} using $138~\mathrm{fb}^{-1}$
 data collected during 2016--2018. The corresponding lower limit obtained by ATLAS experiment on the allowed value of mass of the tau VLL transforming as $SU(2)_L$ doublet  is $900$ GeV~\cite{ATLAS:2023sbu}. Moreover, these limits can be relaxed further if the VLLs are allowed to decay in the non-standard modes such as pVLL  decaying to a lepton and a dark photon as considered in the present model.
%
% Scalar mixing angle
\end{itemize}

%
% LFU
\subsubsection{Lepton flavor universality constraint}
\label{subsubsec:LFU}
In the SM, the universality of the gauge coupling ensures that the $Z$ boson decays in the leptonic final states, in particular, in the final state with electron or muon with equal rate and the ratio of these two rates is predicted to be close to unity in the SM \cite{Poh:2017tfo}. This has been verified in the LEP experiment at CERN and the observed value of the ratio of these two decay rates ($R$) is found to be \cite{ALEPH:2005ab}
\begin{eqnarray}
\label{eq:LFU}
    R &=&  {\Gamma(Z \to \mu^+ \mu^-)}/{\Gamma(Z \to e^+ e^-)} =  1.0009 \pm 0.0028 
\end{eqnarray} 
The model considered in this work introduces a mixing between the second-generation charged lepton and the pVLL. As a consequence, the $Z-\mu-\mu$ vertex, in particular, the $Z-\mu^-_{_L}-\mu^-_{_L}$ interaction vertex is modified to $\dfrac{g_z}{2} (2 \sin^2\theta_W - \cos^2\theta_L)$ (See Eq.~\eqref{eq:mur_mul_z_ver} in Appendix) compared to that {in the SM}\footnote{The corresponding factor for the SM $Z-\mu^-_{_L}-\mu^-_{_L}$ vertex is  $\dfrac{g_z}{2} (2 \sin^2\theta_W - 1)$ } while corresponding right-handed vertex factor remains the same due to the choice of our charge assignment. Therefore, for large mixing one expects deviation in the predicted value of $R$ inconsistent with the above observation. It turns out that, the lepton flavor universality constraint strongly disfavor ($\sin{\theta_L} > 0.030 ~(0.047)$) at $1\sigma$ ($2\sigma$) level \cite{Lee:2022nqz}.   

% EWPO
\subsubsection{Electroweak precision constraints} 
\label{subsubsec:EWPO}
The presence of VLL has various observable consequences and can potentially modify the predictions of the SM. It can contribute to the modification of muon life-time, the branching ratio of $Z \to \mu^+ \mu^-$, total width of the Z and W bosons and correction to the oblique parameters etc. Since the VLL carries electromagnetic charge it can modify the Higgs to di-photon decay rate as well. In our model, the pVLL can couple to the Higgs via muon-heavy muon mixing  and/or Higgs-dark Higgs mixing. Therefore, the corresponding interaction vertices will be proportional to $\sim (y_m \cos\theta_R \cos\theta_L \cos\theta_s~+~\omega_f \cos\theta_R \sin\theta_L \allowbreak \sin\theta_s$). Hence, for small $\sin\theta_L$ and $\sin\theta_s$ the deviation from the SM $h-\mu-\mu$ vertex factor is less significant. The correction to the oblique parameters, in particular, S and T due to the singlet VLL is expressed as \cite{Chen:2017hak}
% delta_T and delta_S expression
%
\begin{eqnarray}
    \Delta T &=& \frac{ m_{\mu}^{2}}{16\pi \sin^{2}_{W} m_{W}^{2}}\sin^{2}\theta_{L}\left[-(1+\cos^{2}\theta_{L})+2\cos^{2}\theta_{L}\frac{r_{\mu}}{r_{\mu}-1}\log(r_{\mu})+r_{\mu} \sin^{2}\theta_{L}\right]
\label{eq:delta_t}\\
\nonumber
\Delta S &=& \frac{1}{18\pi}\sin^{2}\theta_{L}\left[\log(r_{\mu})+\cos^{2}\theta_{L}\left(\frac{5(r^{2}_{\mu}+1)-22 r_{\mu}}{\left(1-r_{\mu}\right)^{2}}\right. \right.\\
&& \left. \left. \hspace{2cm} +\frac{3(r_{\mu}+1)\left(r^{2}_{\mu}-4r_{\mu}+1\right)}{\left(1-r_{\mu}\right)^{3}}\log(r_{\mu})\right)\right]
\label{eq:delta_s}
\end{eqnarray}
where, $r_{\mu} = m_{_{\mu_{p}}}/m_{\mu}$.

The updated global fit to the the electroweak precision observables tells that these corrections are constrained as follows \cite{ParticleDataGroup:2024cfk}: 
\begin{eqnarray}
    S &=& -0.05 \pm 0.07 \\ 
    T &=&  0.00 \pm 0.06
\end{eqnarray}
%

% muon g-2
\subsubsection{\texorpdfstring{Muon $g-2$}{Muon g-2}}
\label{subsec:muon_g-2}

The presence of muon--heavy muon--dark photon vertex can give leading contribution to the anomalous muon $g-2$ at one-loop in this pVLL framework. In the limit $m_{\mu_p} \approx m_{\mu^\prime_p}$ and $m_\mu \approx m_{\mu^\prime}= \dfrac{1}{\sqrt{2}}y_m v_{_{\rm EW}}$, this contribution is given by \cite{Lee:2021gnw, Lee:2022nqz}

\bea\label{eq:amu}
\Delta a_\mu & \approx &
	\begin{cases}
		~~ \dfrac{g^2_d\, m_{\mu^\prime_p}\, m_\mu}{16 \pi^2 \, m^2_{\gamma_d}} \sin2\theta_R \, \sin2\theta_L, & ~ m_{\mu^\prime_p} \gg m_{\gamma_d}  \\
		~~\dfrac{g^2_d \, m_{\mu^\prime_p} \, m_\mu}{4 \pi^2 \,m^2_{\gamma_d}} \sin2\theta_R \, \sin2\theta_L, & ~ m_\mu \ll m_{\mu^\prime_p} \ll m_{\gamma_d}.
	\end{cases}
\eea

The current world average value of $\Delta a_\mu = a^{\rm exp}_\mu - a^{\rm SM}_\mu$, combining Fermilab Run 1--6 data~\cite{Muong-2:2025xyk} with the results of  Brookhaven experiment~\cite{Muong-2:2006rrc} is 
\bea \Delta a_\mu = 38(63) \times 10^{-11}~~ \text{\cite{Aliberti:2025beg}} \eea

which corresponds to $0.6\sigma$ deviation compared to the SM prediction ($a^{\rm SM}_\mu = 0.00116592033(62)$ \cite{Aliberti:2025beg}).

\begin{figure}[H]
    \centering
    \subfloat[\label{subfig:muon_gm2_mzp_vs_gd}]{\includegraphics[width=0.49\columnwidth]{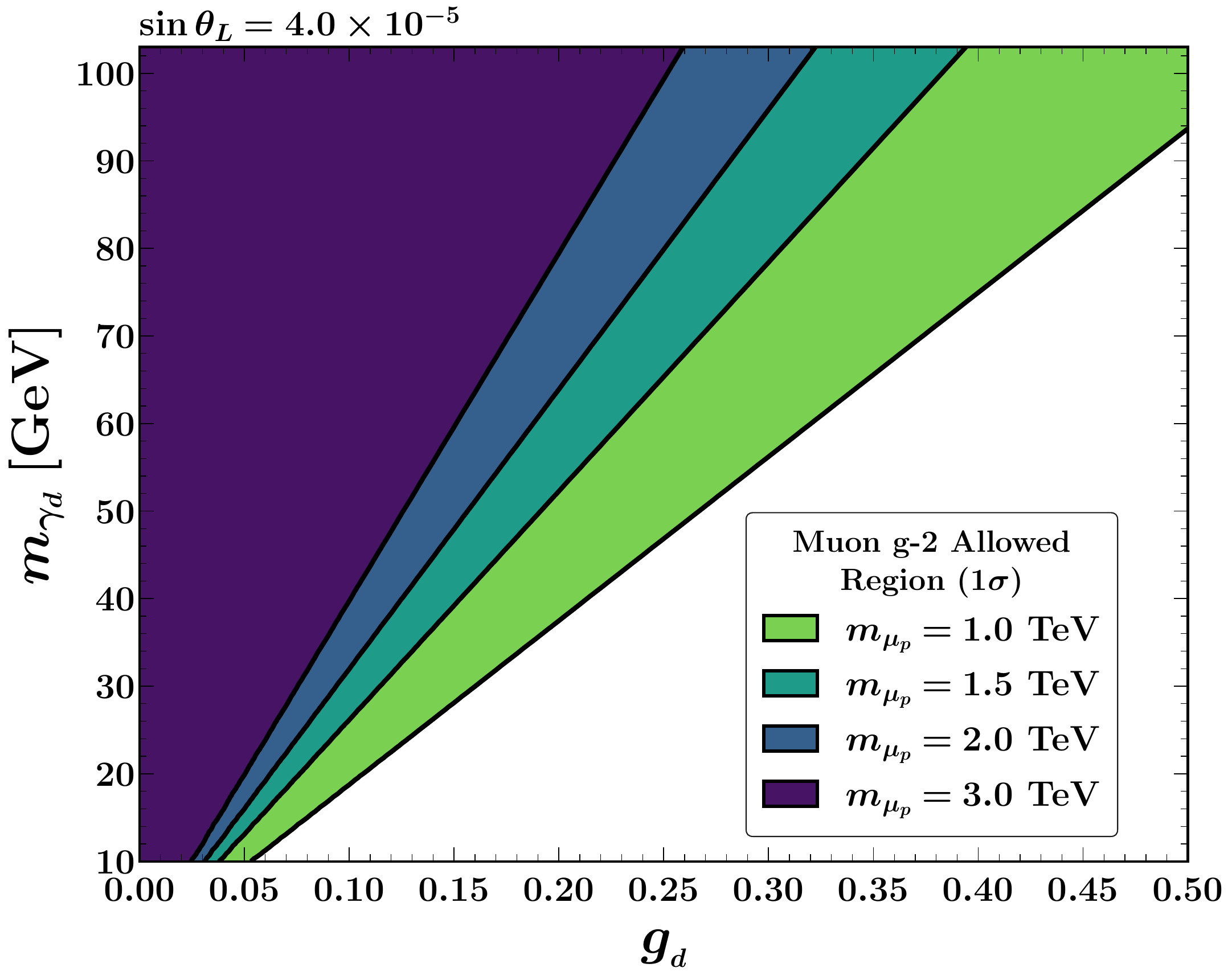}}~
    \subfloat[\label{subfig:muon_gm2_constraint_mzp_vs_mmup}]{\includegraphics[width=0.49\columnwidth]{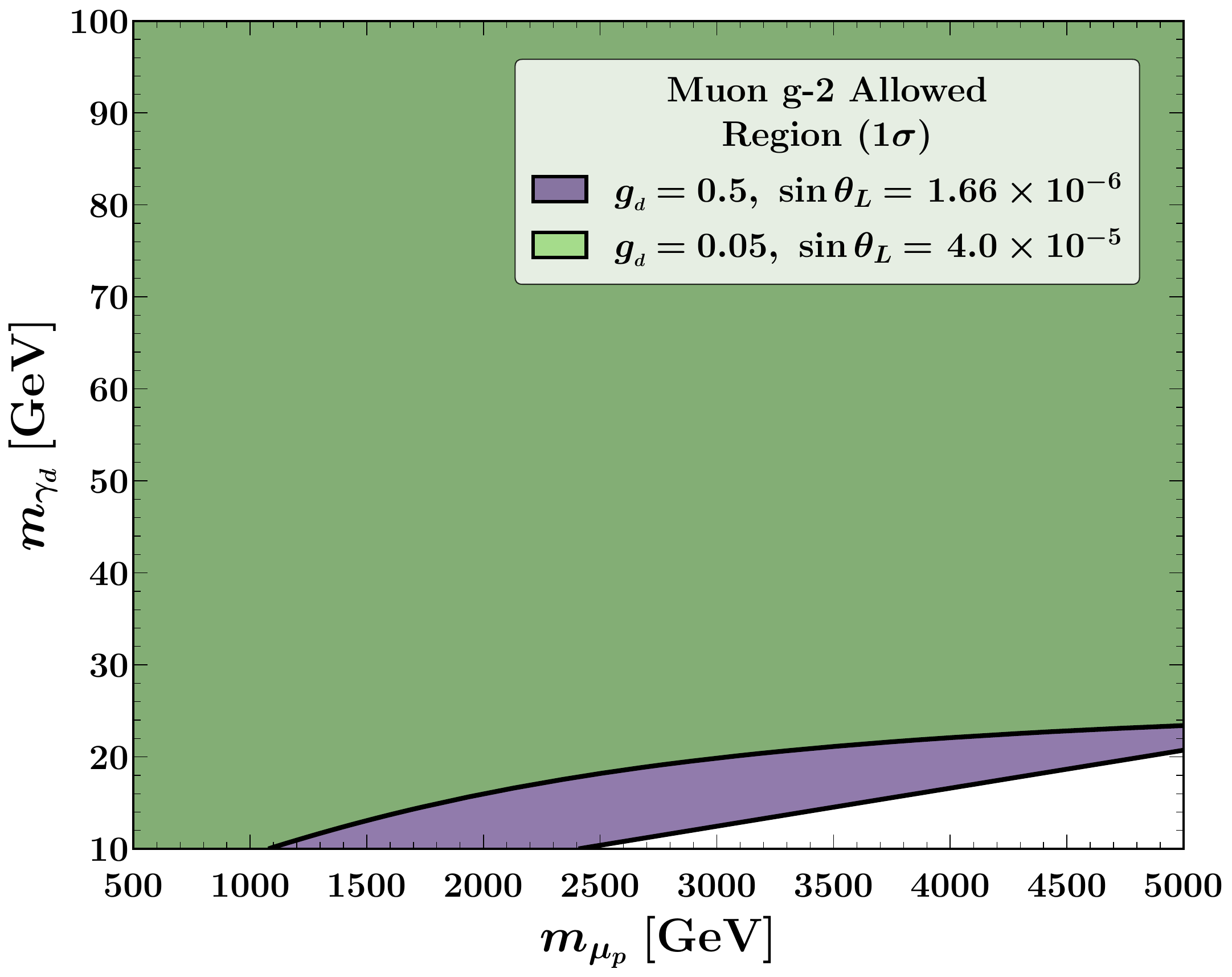}}~
    \caption{Region (coloured) of parameter space allowed by muon $g-2$ measurement at $1\sigma$ level~\cite{Aliberti:2025beg} in the (a) $m_{\gamma_d}-g_d$ plane for
    $\sin\theta_L = 4 \times 10^{-5}$ and four different $m_{\mu_p}$ values ($1$ TeV, $1.5$ TeV, $2$ TeV and $3$ TeV) and (b) $m_{\mu_p}-m_{\gamma_d}$ plane for two sets of ($g_d, ~\sin\theta_L$).
        }
    \label{fig:muon_gm2_constraint}
\end{figure}

In Fig.~\ref{fig:muon_gm2_constraint}, we show the muon $g-2$ allowed region of parameter space both in the $m_{\gamma_d}-g_d$ and $m_{\mu_p}-m_{\gamma_d}$ planes consistent with $\Delta a_\mu$ mentioned in Eq.~\eqref{eq:amu} at $1 \sigma$ level. Fig.~\ref{subfig:muon_gm2_mzp_vs_gd} corresponds to  $\sin\theta_L = 4 \times 10^{-5}$ and four different heavy muon masses (1 TeV, 1.5 TeV, 2 TeV and 3 TeV), whereas Fig.~\ref{subfig:muon_gm2_constraint_mzp_vs_mmup} corresponds to two distinct sets of $\left(g_d,~\sin\theta_L\right)$ with values $\left(0.05,~4 \times 10^{-5} \right)$ and $\left(0.5,~1.66 \times 10^{-6}\right)$.

\noindent
{In the remainder of this paper, we work with benchmark points consistent with all the constraints mentioned above.

%%%%%%%% DM relic abundance
\section{DM relic density}
\label{sec:DM}
In this section, we present the estimation of DM relic density assuming a viable DM candidate to demonstrate that the simplified pVLL scenario can naturally accommodate the observed DM relic density even when the kinetic mixing parameter ($\epsilon$) assumes tiny values or essentially absent due to strong constraints coming from low-energy observations and DM direct-detection experiments.

The observed dark matter relic density reported by the Planck 2018 collaboration suggests
\begin{eqnarray*}
    \Omega h^2 &=& 0.120 \pm 0.001 \, \text{\cite{Planck:2018vyg}}
\end{eqnarray*}

We augment the simplified model introduced earlier with a Dirac fermion ($\chi$) charged under the local $U(1)_D$ and neutral under SM gauge interaction as a potential DM candidate to account for the observed DM relic abundance. The Lagrangian presented in Eq.~(\ref{eq:L_total}) is extended by a DM sector ($ \mathcal{L}_{\rm DM}$), and is expressed as 
\begin{eqnarray}
    \label{eq:L_DM}
        \mathcal{L}_{\rm DM} &=& i \bar{\chi}\gamma^{\mu} D_{\mu}\chi - m_\chi \bar{\chi}\chi
        % V(\Phi, \Phi_d) \, ,
\end{eqnarray}

\begin{figure}[H]
    \centering
        \resizebox{\columnwidth}{!}
            {\subfloat[\label{subfig:2chi_to_2f}]{\includegraphics[width=0.35\columnwidth]{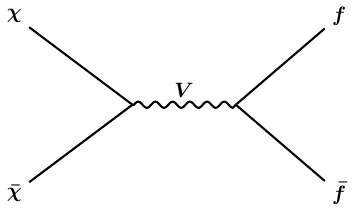}}~
            \subfloat[\label{subfig:2chi_to_2V}]{\includegraphics[width=0.3\columnwidth]{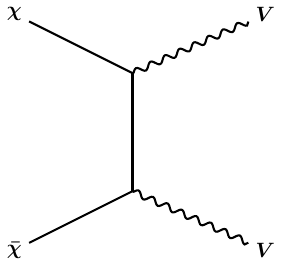}}~
            \subfloat[\label{subfig:2chi_to_VS}]{\includegraphics[width=0.35\columnwidth]{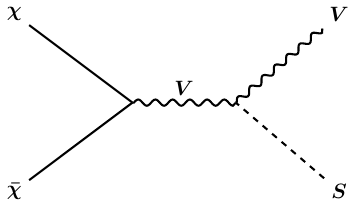}}
            }
    \caption{Feynman diagrams for DM-DM annihilation processes. Here, $V= Z, \gamma_d$  and $S= h, h_d$.} 
    \label{fig:DM_annihilation}
\end{figure}

In absence of the pVLL state, the DM annihilation mainly proceeds via the following annihilation channels: 
 \begin{equation*}
     \chi \bar{\chi} \to f \bar{f} \, , \qquad \chi \bar{\chi} \to V \, V \, , \qquad \chi \bar{\chi} \to V \, S
 \end{equation*}
where $f$ represent the SM fermions, $V= Z, \gamma_d$ and $S= h, h_d$. The corresponding Feynman diagrams are depicted in Fig.~\ref{fig:DM_annihilation}. For light DM in the mass regime $m_f < m_\chi < m_{\gamma_d}$, the dominant annihilation channel is $\chi \bar{\chi} \to f \bar{f}$ while for $m_\chi > m_{\gamma_d}$ or $2 m_\chi > m_{\gamma_d} + m_{_S}$ either or both of the remaining two channels dominate. The annihilation rate for $\chi \bar{\chi} \to f \bar{f}$ via s-channel $Z/\gamma_d$ mediation scales as $g_d^2 \varepsilon^2 {m^2_\chi}/{m^4_{_{V}}}$ away from resonance, and are typically suppressed due to the small gauge kinetic mixing, resulting in an overabundance of DM. The presence of a strong first order phase transition (SFOPT) in the dark sector may overcome this issue, as suggested in recent studies \cite{Mahapatra:2026fyv, Huang:2026qdc}.

In the present model, the issue of an overabundant DM relic density for light DM scenario, can be addressed through the portal vector-like lepton state without invoking the requirement of a SFOPT. Moreover, this framework can evade the direct detection bound, as long as the gauge kinetic mixing is small.
The role of pVLL state exclusively mixing with the second generation SM charged lepton is to {\it enhance} the annihilation rate $\chi \bar{\chi} \to \mu^+ \mu^-$ for light DM {\it away from  resonance}, through the modification of the $\mu^+ - \mu^- - \gamma_d$ vertex without introducing additional Feynman diagrams. 
 The other two channels ($\chi \bar{\chi} \to \gamma_d \gamma_d, \, \gamma_d h_d$) which are dominant in the mass region $m_{\chi} > m_{\gamma_d}$ or $2 m_{\chi} > m_{\gamma_d}+m_{h_d}$, on the other hand are controlled by the dark gauge coupling and remain unaffected due to the presence of the pVLL.

In presence of the portal matter state, the dark photon coupling to the muon receives an additional contribution proportional to $g_d\sin^2\theta_{R/L}$ beyond the usual vertex factor arising from the KM alone, while the couplings to all other SM fermions remain unaltered. The modified $\mu^+ - \mu^- - \gamma_d$ vertex has the following structure
\begin{eqnarray*}
    \mu^+_R - \mu^-_L - \gamma_d &\sim&
\frac{i\,\cos\theta_d}{\sqrt{1-\varepsilon^2}}
\left[
\frac{e\varepsilon}{\cos\theta_w}
- \frac{e\varepsilon\,\cos^2\theta_L}{2\cos\theta_w}
+ g_d\,\sin^2\theta_L
\right]
+ i e \sin\theta_d
\left[
\frac{\cos^2\theta_L}{2\,\sin\theta_w \cos\theta_w}
- \frac{\sin\theta_w}{\cos\theta_w}
\right] \\
\mu^+_L - \mu^-_R - \gamma_d &\sim&
\frac{i\,\cos\theta_d}{\sqrt{1-\varepsilon^2}}
\left[
\frac{e\varepsilon}{\cos\theta_w}
+ g_d \sin^2\theta_R
\right]
- i e \sin\theta_d \frac{\sin\theta_w}{\cos\theta_w}
\end{eqnarray*}
\noindent
As a result, in this scenario $\chi \bar{\chi} \to \mu^+\mu^-$ can be a dominant annihilation channel for light DM ( $m_{\mu} m_{\chi} < m_{\gamma_d}$) even for a small KM ($\varepsilon < 10^{-3}$) away from resonance. Importantly, this channel does not suffer from strong suppression factors, since $\sin\theta_R$ can be sizable ($\sim$ 0.1 to 0.3) even when $\sin\theta_L$ is small as already pointed out -- thanks to the same {\it non-decoupling effect}.

We have implemented the full Lagrangian ($\mathcal{L}+\mathcal{L}_{\rm DM}$) in the \textsc{Feynrules} 2.0 \cite{Alloul_2014} incorporating gauge kinetic, scalar and fermion mixings and interfaced the resulting model files with the \textsc{micrOMEGAs} \cite{Alguero:2023zol} to estimate the DM relic density.
In Fig.~\ref{fig:omega_vs_mx}, we plot the relic density
as a function of the DM mass ($m_\chi$) in the range $10-130$ GeV for two different choices of dark photon masses ($m_{\gamma_d}$ = 50 GeV and 90 GeV) and three different sets of $\varepsilon$ and $\sin\theta_L$: (a) $\varepsilon = 10^{-3}, ~10^{-5}~ ; ~\sin\theta_L=0$, (b) $\varepsilon=0;~ \sin\theta_L = 4 \times 10^{-5}$,  and (c) $\varepsilon = 10^{-3}, ~10^{-5};~ \sin\theta_L=4 \times 10^{-5}$.
\begin{figure}[H]
    \centering
        \subfloat[\label{subfig:relic_density_vs_mchi_stl_0pt0}]{\includegraphics[width=0.45\columnwidth]{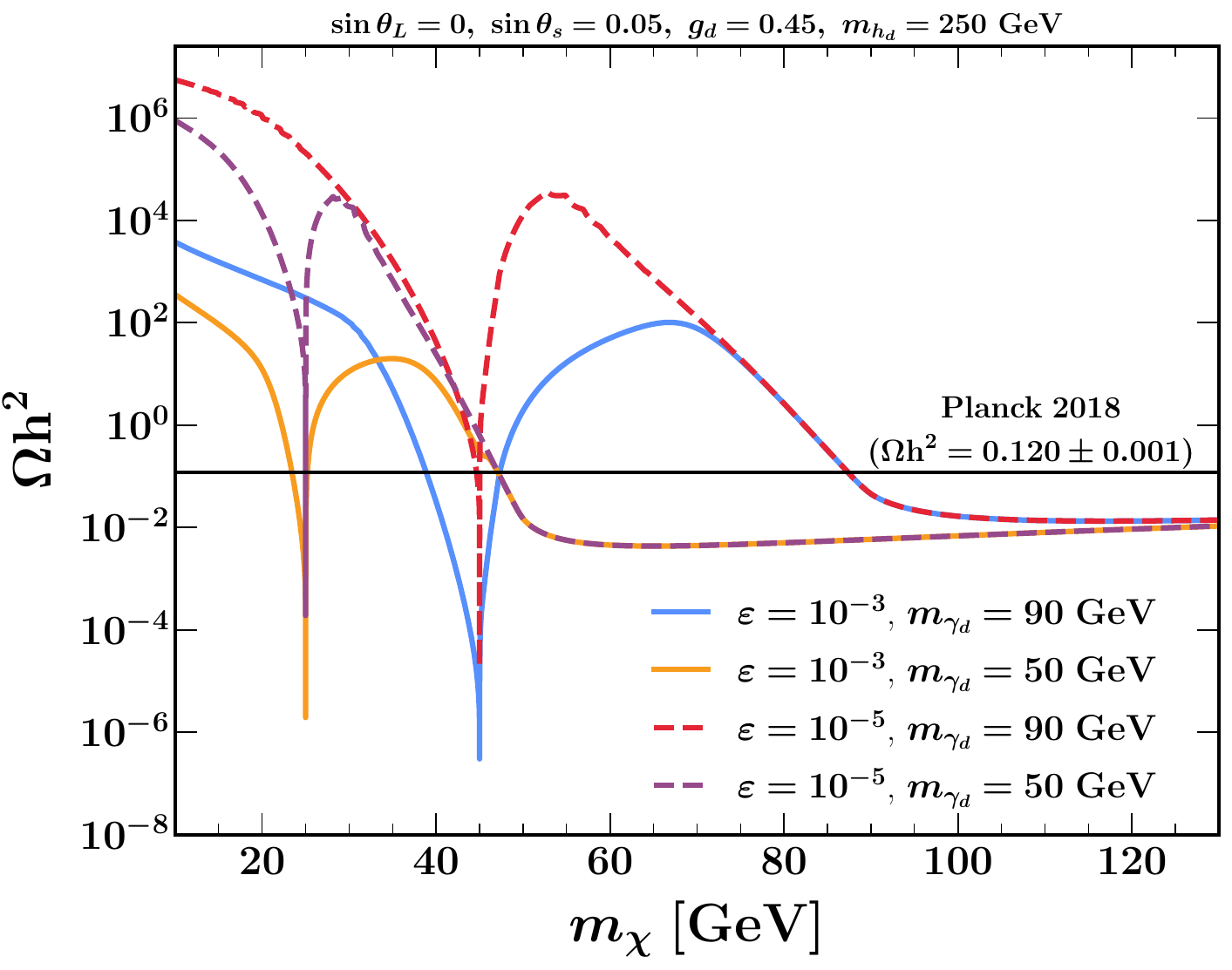}}
        ~\subfloat[\label{subfig:relic_density_vs_mchi_stl_4em05_skmix_0}]{\includegraphics[width=0.45\columnwidth]{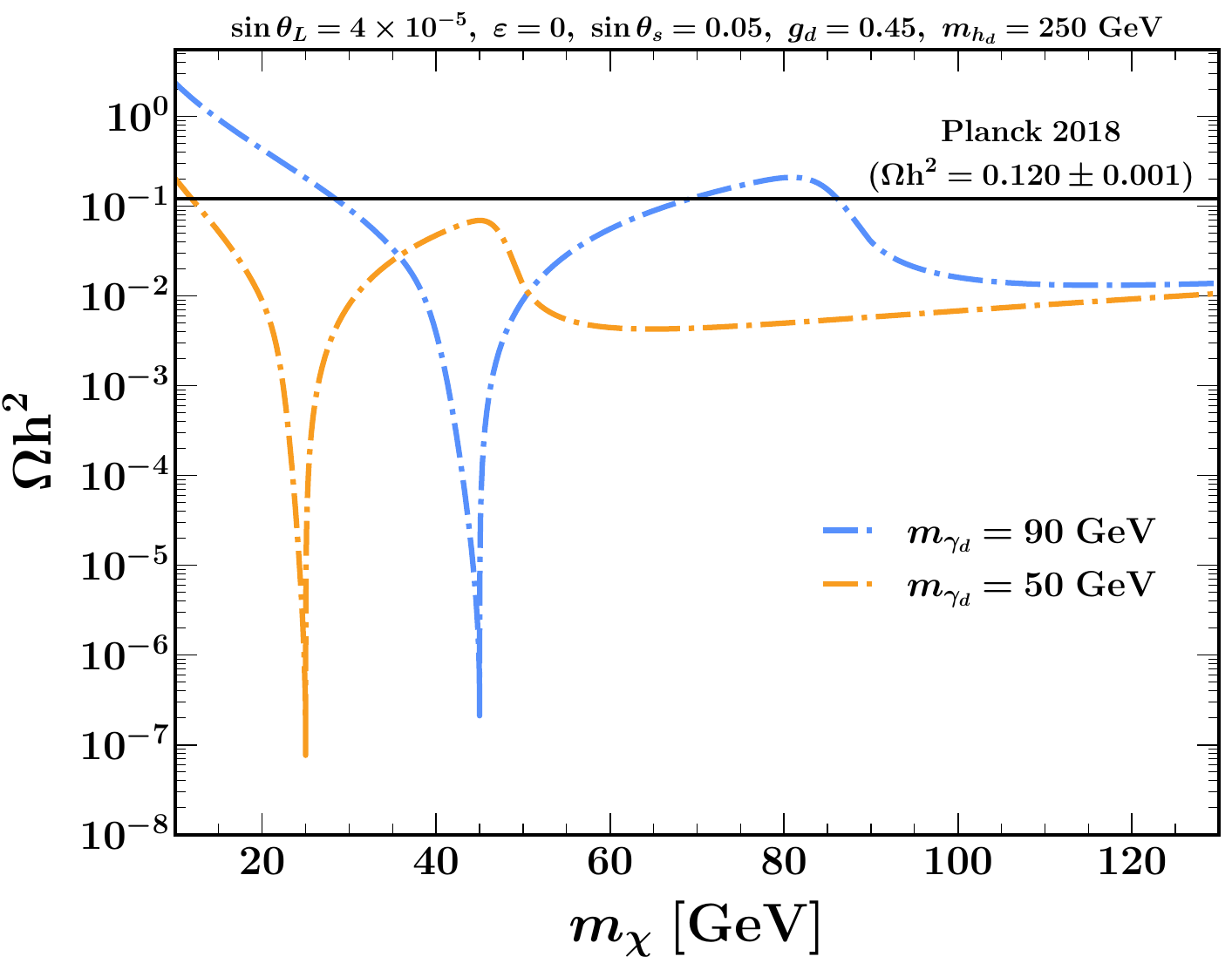}}
        
        \subfloat[\label{subfig:relic_density_vs_mchi_stl_4em05_nonzero}]{\includegraphics[width=0.45\columnwidth]{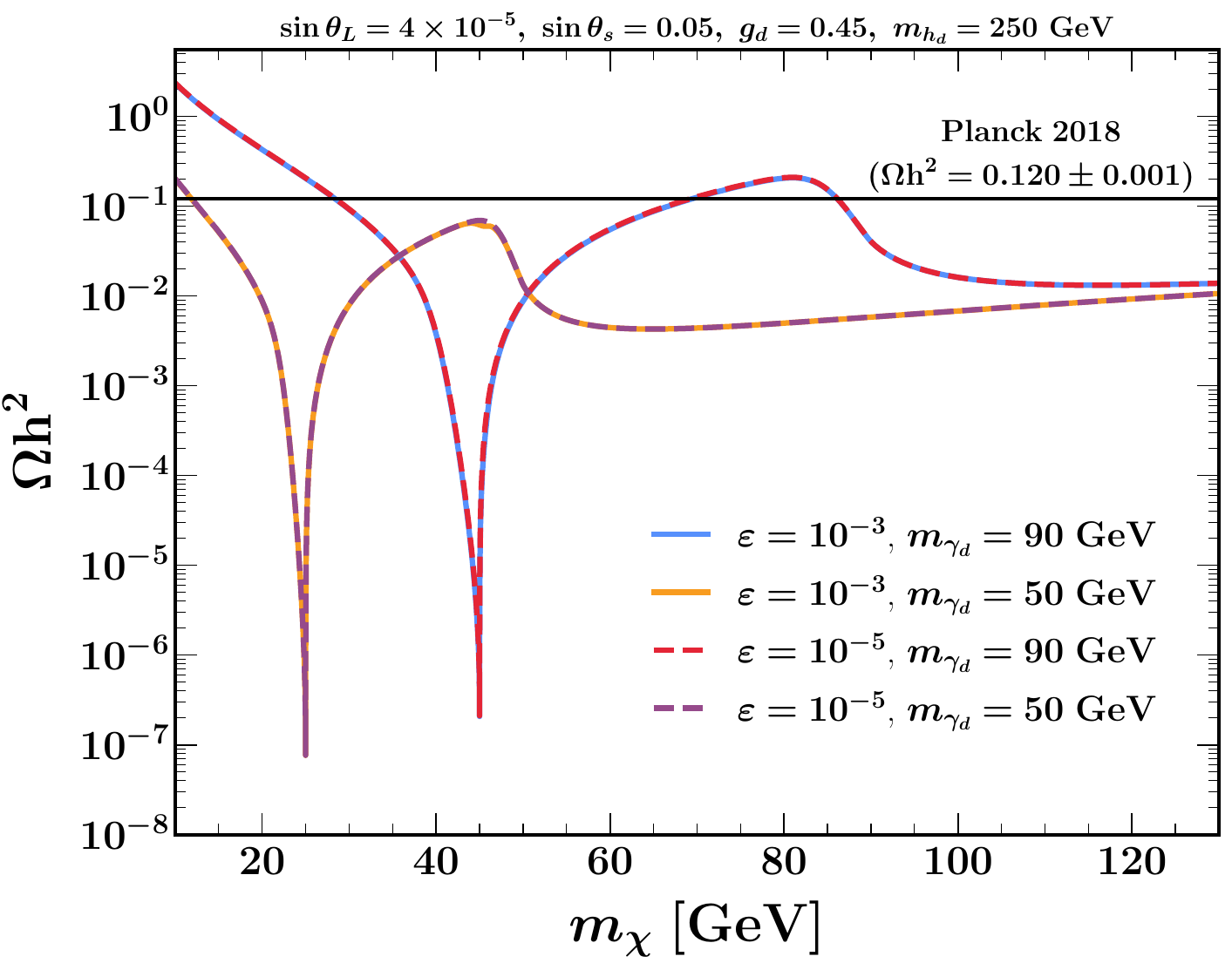}}
            
    \caption{Dark matter relic abundance as a function of dark matter mass for (a) $\varepsilon$ = $10^{-3}$, $10^{-5}$, $\sin\theta_L=0$, (b) $\varepsilon=0,~ \sin\theta_L = 4 \times 10^{-5}$, and (c) $\varepsilon=10^{-3}, ~10^{-5},~ \sin\theta_L=4 \times 10^{-5}$ for two different dark photon masses ($50$ GeV and $90$ GeV) assuming $g_d=0.45,~\sin\theta_s = 0.05$ and $m_{\mu_p}=1$ TeV. } 
    \label{fig:omega_vs_mx}
\end{figure}

In absence of the pVLL state, for light dark matter ( $ m_{\chi} < m_{\gamma_d}$ or $ 2m_{\chi} < m_{\gamma_d} + m_{h/h_d}$), the relative annihilation rate to individual SM fermion modes ($f=e,\mu , \tau, \nu, d,u,s,c, b$) are predicted by their respective SM gauge charge and color factor. One needs $\varepsilon \sim 10^{-2}$ or higher for $g_d \sim 0.45$ to achieve underabundant DM relic density away from resonance. The value of the KM parameter in this range is not only strongly constrained by the various low energy experiments in the context of dark photon searches depending on the mass of the dark photon but is also extremely sensitive to the DM direct detection experiments, as we will discuss later.
For $m_\chi > m_{\gamma_d}$ the predicted DM relic abundance is largely independent of the KM and a dark gauge coupling of the order of $0.45$ typically results in an underabundant DM relic density (Fig.~\ref{subfig:relic_density_vs_mchi_stl_0pt0}). In presence of the pVLL, assuming no (Fig.~\ref{subfig:relic_density_vs_mchi_stl_4em05_skmix_0}) or small (Fig.~\ref{subfig:relic_density_vs_mchi_stl_4em05_nonzero}) KM, the relic abundance in the light DM regime away from resonance are primarily controlled by the fermion mixing and dark gauge coupling. At the resonance point the relic abundance depends on the interplay among KM, fermion mixing and dark gauge coupling. In the mass regime where $\chi \bar{\chi} \to \gamma_d \gamma_d, ~\text{or}~\gamma_d h_d$ is dominant DM relic abundance remains essentially unaltered in the presence of pVLL.

To identify the DM relic abundance compatible parameter space in the DM mass range 10 -- 130 GeV, we perform a parameter space scan in the $g_d$ {\it{vs}} $m_\chi$ plane for various combinations of fermion and gauge kinetic mixings. This is depicted in Fig.~\ref{fig:gd_vs_mx} for two representative values of dark photon mass ($m_{\gamma_d}=50$ GeV and $90$ GeV), kinetic mixing parameter ($\varepsilon=0,~ 10^{-3}$) and fermion mixing angle ($\sin\theta_L=0,~ 4 \times 10^{-5}$).
% The dip observed in Fig.~\ref{5a} around $m_\chi \sim 45$ GeV and $\varepsilon \sim 0.01$ corresponds to on-shell $Z$ boson contribution in the $\chi \bar{\chi } \to f \bar{f}$ channel.
% 
We keep the scalar mixing angle ($\sin\theta_s$), masses of the heavy scalar ($m_{h_d}$) and the heavy muon ($m_{\mu_p}$) fixed at $0.05$, $250$ GeV and $1$ TeV, respectively, throughout the DM analysis.

 \begin{figure} [H]
    \centering
            % \subfloat[\label{subfig:relic_density_vs_mchi_stl_4em05_skmix_0}]{\includegraphics[width=0.6\columnwidth]{Manuscript/relic_density_vs_mchi_stl_4em05_skmix_0.pdf}}
            \subfloat[\label{subfig:gd_vs_mchi_50}]{\includegraphics[width=0.45\columnwidth]{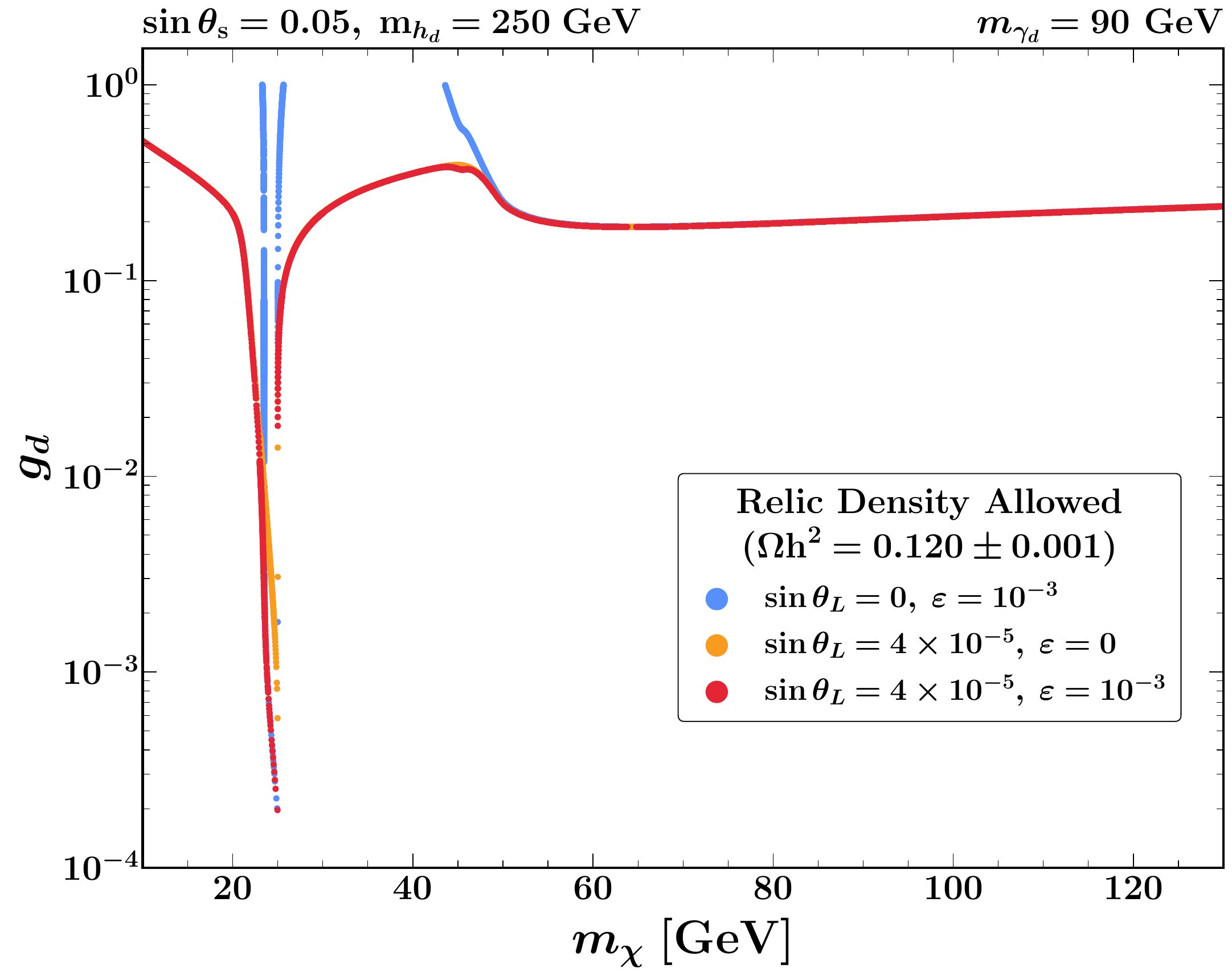}}~
            \subfloat[\label{subfig:gd_vs_mchi_90l}]{\includegraphics[width=0.45\columnwidth]{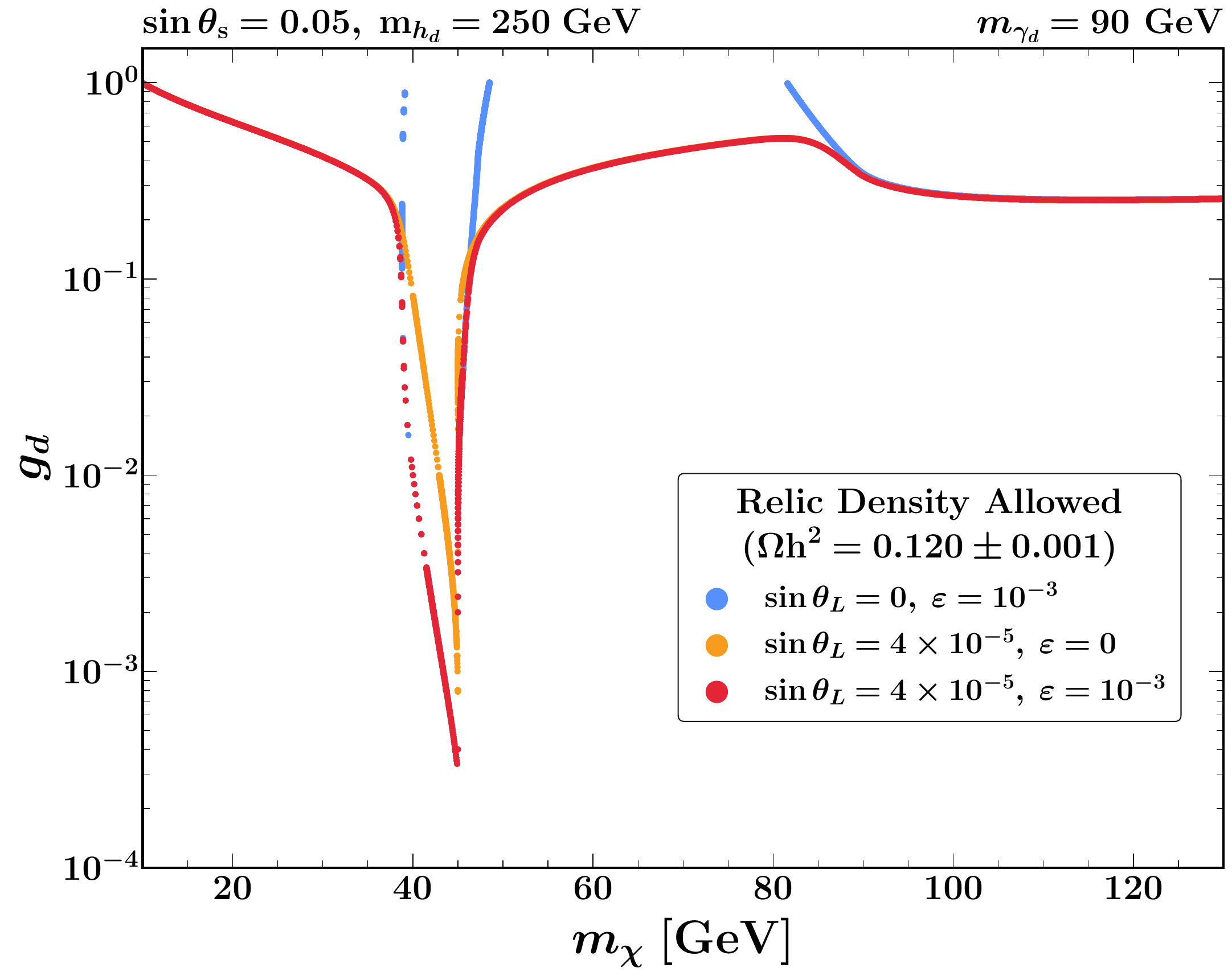}}
            
    \caption{Dark matter relic abundance allowed parameter space in the $g_d$ vs $m_{\chi}$ plane for various combinations of $\sin\theta_L$ and $\varepsilon$ assuming $\sin\theta_s = 0.05$ and $m_{\mu_p}=1$ TeV. The figure on the left-panel corresponds to $m_{\gamma_d}=50$ GeV and that on the right panel considers $m_{\gamma_d}=90$ GeV. }
    \label{fig:gd_vs_mx}
\end{figure}

{\bf Direct detection:} The value of the KM parameter is strongly constrained by several low-energy observations related to dark-photon searches~\cite{Fabbrichesi:2020wbt,Bauer:2018onh}. These observations typically set an upper bound, $\varepsilon \lesssim 10^{-3}$, on the allowed values of the KM parameter for dark-photon masses in the range $10$--$100$ GeV.
However, if the dark photon couples to a DM particle, the most stringent constraints on the KM parameter arise from direct-detection experiments. In Fig.~\ref{fig:DM_nucleon_scattering}, we display the observed upper limits on the DM--nucleon scattering cross section as a function of the DM mass in the range $10~{\rm GeV}$--$10~{\rm TeV}$, obtained from several direct-detection experiments, such as XENON1T (2018) \cite{XENON:2018voc}, XENONnT (2025) \cite{XENON:2025vwd}, and LZ (2025) \cite{LZ:2024zvo}. We also plot the theoretical prediction for the DM--nucleon scattering cross section in the same mass range for different choices of the KM parameter and the dark-photon mass, in the absence of any pVLF state.

The strongest bound on the KM parameter comes from the LZ (2025) experiment which suggests for $g_d=0.45$ and dark photon mass, $m_{\gamma_d}> 90$ GeV, $\varepsilon \lesssim 10^{-5}$ can remain allowed over a wide range of DM masses. This provides a strong motivation to consider DM models with portal vector-like leptons. Their effect in the context of DM relic abundance can enter at tree-level if mixing with one or more SM fermion is assumed, whereas in the context of direct detection their contribution arises via vacuum polarization diagrams at one loop. Consequently, the direct-detection cross section becomes highly model dependent and sensitive to the presence of pVLF states, especially their charge assignments and the number of such portal VLF states \cite{Rizzo:2018vlb}, and therefore on the detailed structure of the dark-sector (DS) physics.

%

%
%%
%
% 
%}
%
 \begin{figure}[H]
     \centering
     % \label{fig:DM_nucleon_scattering}
     \includegraphics[width=0.6\linewidth]{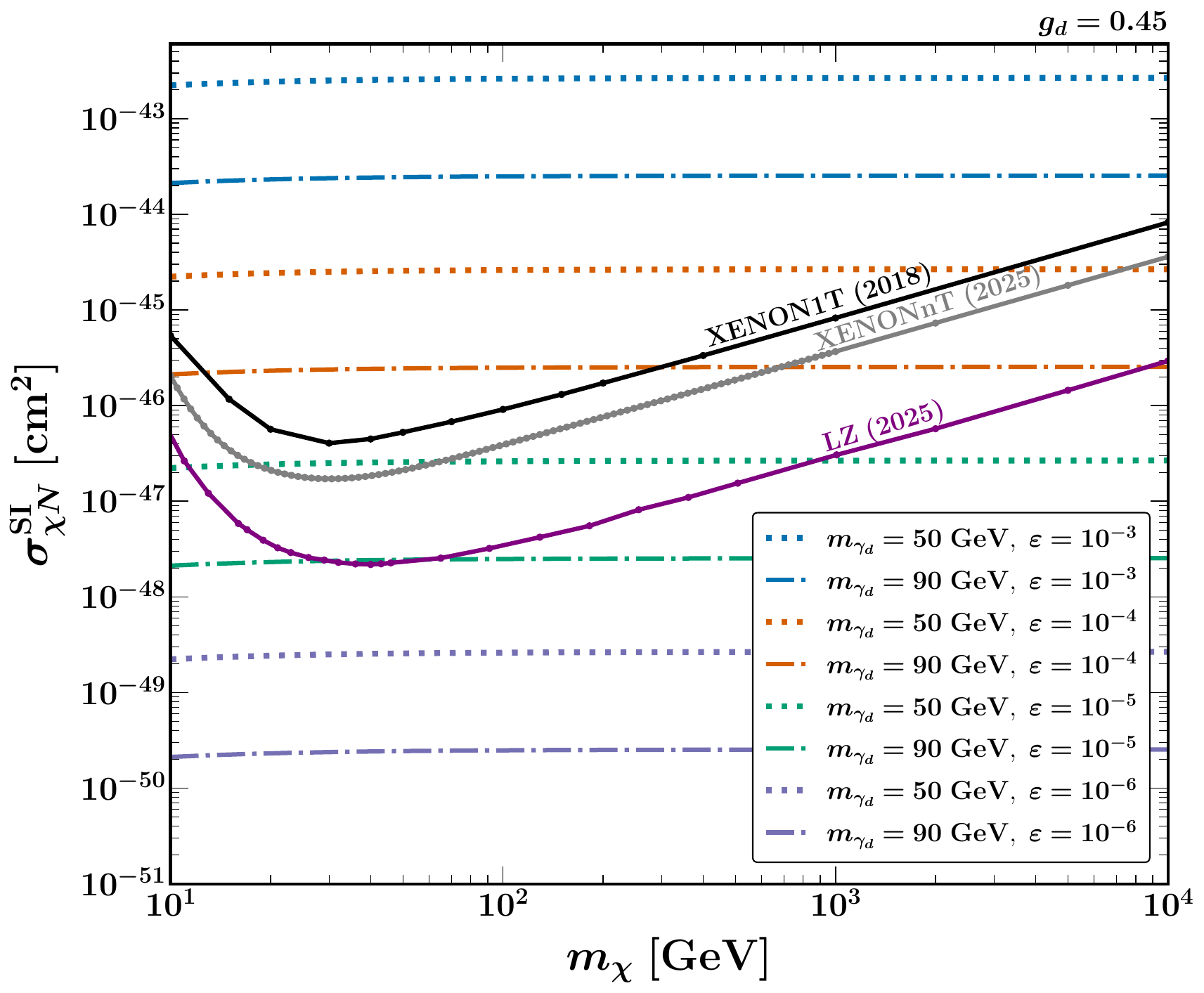}
     \caption{Observed or $-1\sigma$ power constrained upper limits (solid lines), as applicable, on the spin-independent direct detection cross-sections as a function of the dark matter mass from XENON1T 2018 (black) \cite{XENON:2018voc}, XENONnT 2025 (gray) \cite{XENON:2025vwd} and LZ 2025 (purple) \cite{LZ:2024zvo} experiments. The dotted and dot-dashed lines represent the theoretical prediction of spin-independent DM-nucleon scattering cross sections as a function of the dark matter mass in presence of the kinetic mixing alone ($\varepsilon= 10^{-3},~10^{-4},~ 10^{-5}~{\rm and}~ 10^{-6}$) for two different choices of the dark photon masses $m_{\gamma_d} = 50$ GeV and $90$ GeV assuming $g_d=0.45$. }
     \label{fig:DM_nucleon_scattering}
 \end{figure}}

{However, in this work we restrict ourselves to a simplified framework, focusing primarily on the implications of a single pVLF state mixing with the SM muon for associated Higgs production at a muon collider. As long as only one of the pVLF states is allowed to mix with the SM second generation charged lepton, the collider consequences presented in the latter section remain unaltered. In the remaining part of this work we also ignore the KM as its effect in the  collider analysis is essentially insignificant.}

% 
%% Sec 3: Collider analysis %%%%%
\section{Collider Analysis}
\label{sec:collider_analysis}
\begin{figure}[H]
    \centering
            \subfloat[\label{subfig:2mu_to_Vh_s_channel}]{\includegraphics[width=0.38\columnwidth]{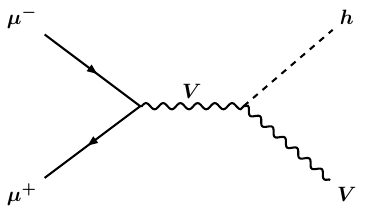}}~
            ~\subfloat[\label{subfig:2mu_to_Vh_t_channel}]{\includegraphics[width=0.3\columnwidth]{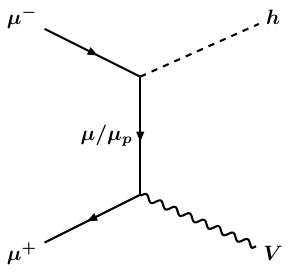}}
            ~\subfloat[\label{subfig:2mu_to_Vh_u_channel}]{\includegraphics[width=0.3\columnwidth]{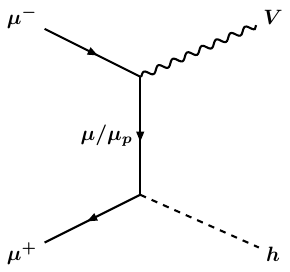}}
    \caption{Feynman diagrams for associated production of Higgs and a vector boson ($V=Z,\gamma_d$) at muon collider ($\mu^+ \mu^- \to  h V  $).} 
    \label{fig:VLL_PM}
\end{figure}

In this section, we demonstrate the effect of the portal vector-like lepton on the Higgs production in association with a vector boson ($V= Z, \gamma_d$) at future muon collider. We first estimate the effect of pVLL on the production rate of $\mu^+ \mu^- \to  h V$ and compare this w.r.t. the SM prediction of $\mu^+ \mu^- \to h Z$. The Feynman diagrams for $\mu^+ \mu^- \to hV$ process are depicted in Fig.~\ref{fig:VLL_PM}.
%%%% Fig. : Feynman diagrams: 2mu_to_hV %%%
%
In order to generate the events and estimate the cross section{, we use the Universal \textsc{Feynrules} Output \cite{Degrande:2011ua}, obtained from the implementation mentioned earlier in Section~\ref{sec:DM}, and interface it with the event generator  \textsc{MadGraph5\_aMC@NLO} (version $3.5.5$) \cite{Alwall_2014}. We then compute the leading order (LO) cross section in the context of future muon collider assuming center-of-mass energies $3$ TeV and $10$ TeV.}

%%%%%%%%%%% BPs %%%%%%%%%%%%%%%

We have chosen six benchmark points with parameter values\footnote{For the collider analysis, the kinetic mixing parameter ($\varepsilon$) is set to zero without loss of generality, as it does not affect the estimated cross sections within its allowed range of values for the range of dark photon masses considered in this work.} consistent with the constraints mentioned in section \ref{subsec:constraints}, as shown in Table~\ref{tab:BPs_parameters}.

% Table-2 : Benchmark parameters
\begin{table}[H]
\centering
\resizebox{0.8\columnwidth}{!}{
\begin{tabular}{c c c c c c }
  \toprule
  Benchmark Points & $m_{{\mu_{_p}}}$ [TeV] & $\sin{\theta_L}$ & $g_d$ & $\sin{\theta_s}$ & $m_{{\gamma_{_d}}}$ [GeV] \\
  \midrule
  \midrule
  BP1 & $1$ & $4.0 \times 10^{-5}$ & $0.45$ & $0.05$ & $90$  \\
  BP2 & $2$ & $2.0 \times 10^{-5}$ & $0.36$ & $0.05$ & $90$  \\
  BP3 & $3$ & $1.7 \times 10^{-6}$ & {$0.50$} & $0.05$ & $90$   \\
  BP4 & $1$ & $4.0 \times 10^{-5}$ & {$0.36$} & $0.05$ & $50$  \\
  BP5 & $2$ & $2.0 \times 10^{-5}$ & $0.30$ & $0.05$ & $50$  \\
  BP6 & $3$ & $1.7 \times 10^{-6}$ & {$0.45$} & $0.05$ & $50$   \\
  \bottomrule
\end{tabular}
}
\caption{Representative benchmark points {for collider analysis}}
\label{tab:BPs_parameters} 
\end{table}
% 
%%%% Subsection: hZ vs Zph %%%%%
\subsection{\texorpdfstring{$hZ$}{hZ} vs \texorpdfstring{$h \gamma_d$}{h gamma\_d} at muon collider}
\label{subsec:hZ_vs_hzp}
\begin{figure}[H]
    \centering
        \subfloat[\label{subfig:ration_bar_plot_Ecm_3tev}]
        {\includegraphics[width=0.45\columnwidth]{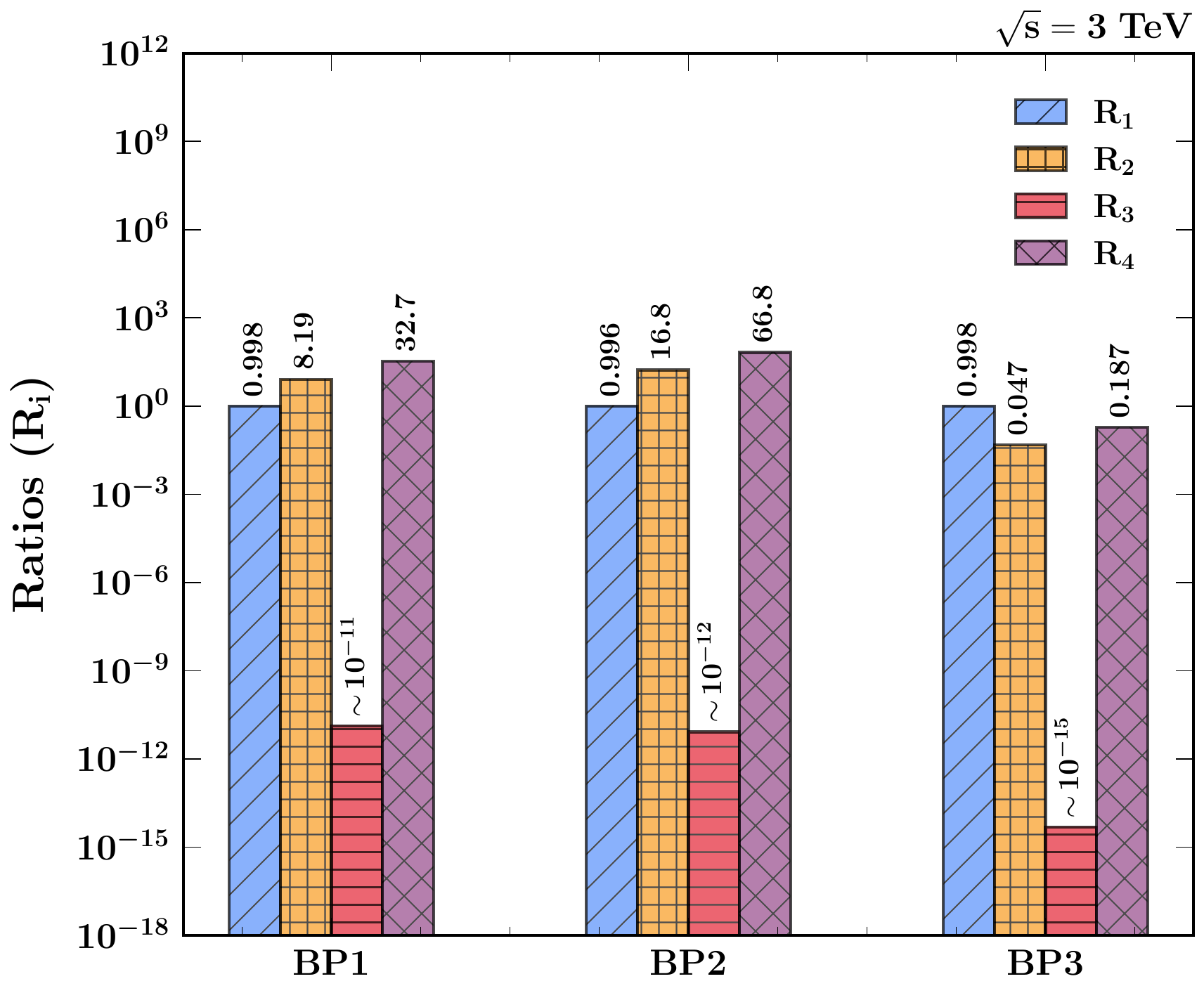}}~
        ~
        \subfloat[\label{subfig:ratio_bar_plot_Ecm_10tev}]
        {\includegraphics[width=0.45\columnwidth]{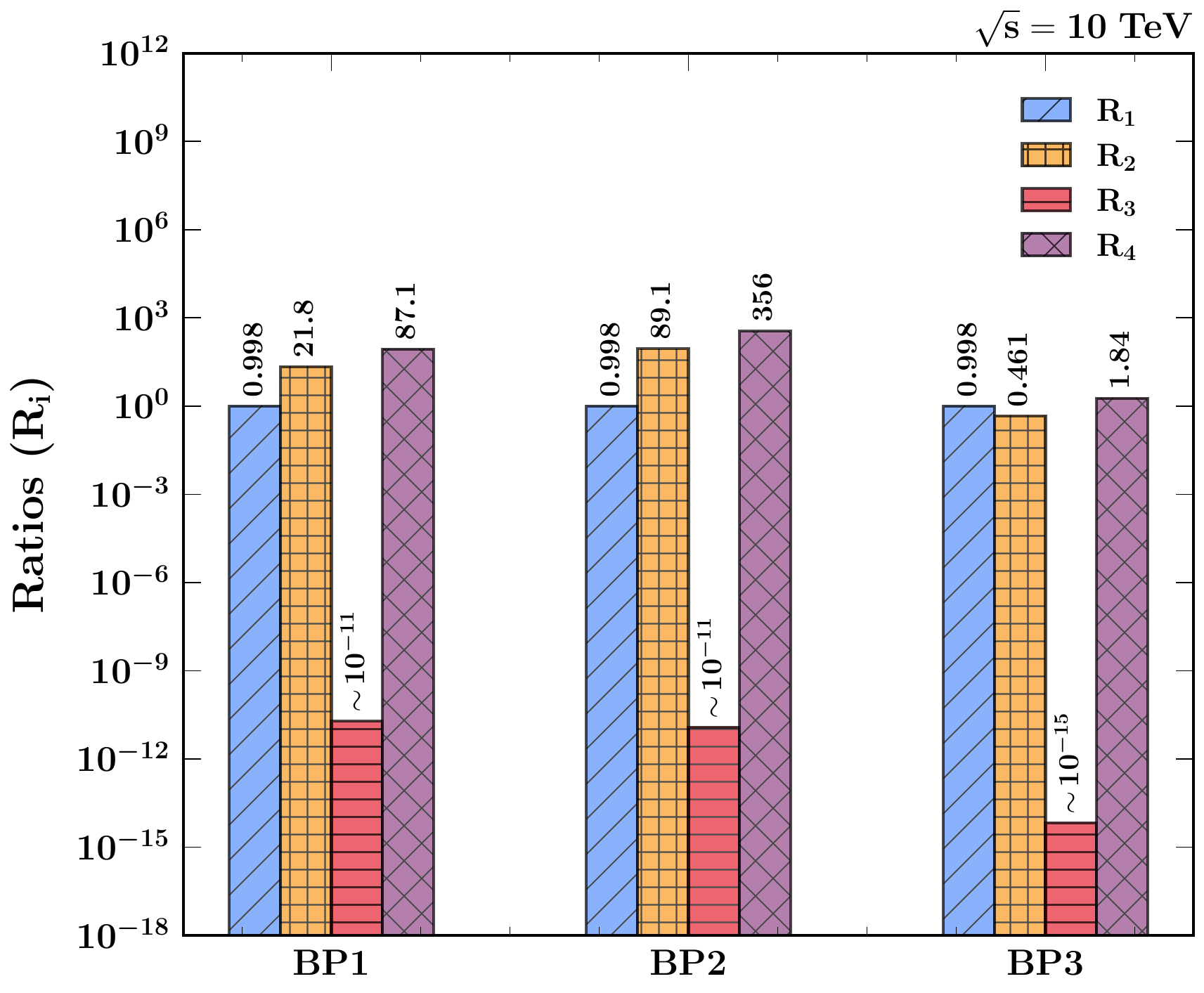}}~
    \caption{Ratios of cross sections ($R_i$'s defined in Eq.~\eqref{eq:ratio_equations}) for different benchmark points {(BP1-BP3) corresponding to dark photon mass 90 GeV at} (a) $\sqrt{s}=3$ TeV and (b) ${\sqrt{s}=}10$ TeV.}
    \label{fig:ratio_bar_plots}
\end{figure}

Here we present a comparison between $\mu^+ \mu^- \to h \gamma_d$ and $\mu^+ \mu^- \to hZ$ production rates in the context of the model described in Section \ref{sec:model} w.r.t. the SM $hZ$ production rate at future muon collider. In Fig.~\ref{fig:ratio_bar_plots},  we depict various ratios of cross sections defined in Eq.~\eqref{eq:ratio_equations} for three different benchmark points.
%
%% Ratios

\begin{subequations}
\begin{eqnarray}
    R_1 &=& \frac{\sigma_{_{\rm BSM}}(\mu^+ \mu^- \to h Z)}{\sigma_{_{\rm SM}}(\mu^+ \mu^- \to h Z)} \\[6pt]
    R_2 &=& \frac{\sigma_{_{{\rm BSM}}}(\mu^+ \mu^- \to h \gamma_d)}{\sigma_{_{\rm SM}}(\mu^+ \mu^- \to h Z)} \\[6pt]
    {R_3} &=& {\frac{\sigma_{_{\rm BSM}} (\mu^{+}_{_R} \mu^{-}_{_L} \to h \gamma_d)}{\sigma_{_{\rm SM}} ( \mu^+ \mu^- \to h Z)}} \\[6pt]  
    {R_4} &=&{\frac{ \sigma_{_{\rm BSM}} (\mu^{+}_{_L} \mu^{-}_{_R} \to h \gamma_d)}{\sigma_{_{\rm SM}} ( \mu^+ \mu^- \to  h Z)}}
\end{eqnarray}
\label{eq:ratio_equations}
\end{subequations}
where,
\bea\label{eq:SM_zh_xsec}
\sigma_{_{\rm SM}}(\mu^+ \mu^- \to hZ)&= &
	\begin{cases}
		~1.36~{\rm fb}, & {\rm for }~ \sqrt{s} = 3 ~{\rm TeV} \\
		~0.12~{\rm fb}, & {\rm for }~ \sqrt{s} = 10 ~{\rm TeV} 
	\end{cases}
\eea

One can see that while the cross section for $\mu^+ \mu^- \to hZ$ production in this scenario remains the same as that in the SM, a higher Higgs production rate is predicted in the $\mu^+ \mu^- \to h \gamma_d$ channel. This is true even for a $\sin\theta_L$ as small\footnote{Small $\sin{\theta_L}$ ($\sim 10^{-4}$ or less) is favoured by perturbative unitarity requirements for heavy muon mass beyond TeV scale} as $\sim 10^{-5}$ with reasonable choices of other relevant parameters (see Table~\ref{tab:BPs_parameters} for specification of benchmark points). The underlying reasons behind this is that $\mu_R^\prime$ and $\mu_{p_{_R}}^\prime$ share identical $SU(2)_L \times U(1)_Y$ charges. Hence, after transforming them into mass eigen-basis using Eq.~\eqref{eq:fermion_mixing} no off-diagonal $\mu^-_R-\mu^-_{p_{_R}}-Z$ interaction arises\footnote{This is actually true in the limit of zero kinetic mixing.}. However, the absence of $U(1)_D$ charge for the muon induces an off-diagonal $\mu_R^--\mu^-_{p_{_R}}-\gamma_d$ interaction in the mass basis. It turns out that the corresponding interaction vertex
% 
% $\mu^+_{L} -\mu_{p_{_R}}^- -Z$ is not present in this model while the
% 
$\mu^+_{L} -\mu_{p_{_R}}^- -\gamma_d$ (or $\mu^-_{R} -\mu_{p_{_L}}^+ -\gamma_d$ )  is proportional to $\sim g_d \sin\theta_R \cos\theta_R$ ( See Eq.~\eqref{Appn_eq:vf2} ) {which therefore controls the $h \, \gamma_d$ production rate due to sizable values of $\sin\theta_R$ as mentioned earlier.}
% The mixing angle ($\sin\theta_R$) between the right chiral muon and heavy muon is given by 
% %
% 
% %
%
% 
% 
{This can also be independently understood by the result that the
polarized cross section $\mu^+_L \mu^-_R \to h \gamma_d$ constitutes the dominant
part of the average unpolarized cross section for $\mu^+ \mu^- \to h \gamma_d$.
Note that $\mu^+_L \mu^-_R \to h \gamma_d$ gets contributions from $t$- and $u$- channel heavy muon mediated diagrams as well as that from $s$-channel dark photon mediation.  The latter contribution is {\it suppressed} due to the fact that $\mu^+_L-\mu^-_R-\gamma_d$ vertex is $\sim g_d\sin^2\theta_R$ in addition to the usual $s$-channel suppression at high center-of-mass energy.}

\begin{figure}[H]
    \centering

        \subfloat[\label{subfig:Mzp_vs_mmup_set1_3tev}]{\includegraphics[width=0.47\columnwidth]{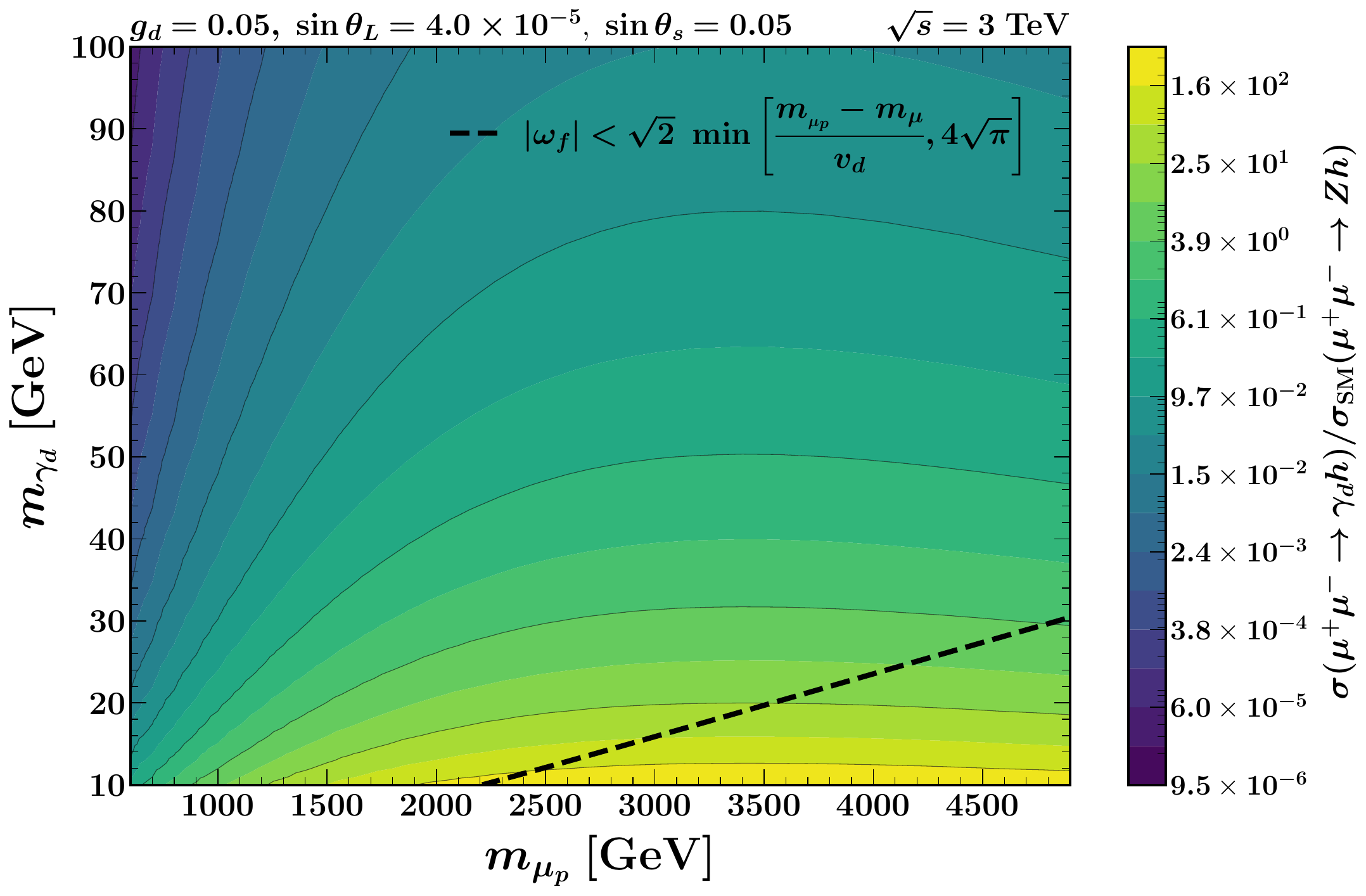}}~
        \subfloat[\label{subfig:Mzp_vs_mmup_set2_3tev}]{\includegraphics[width=0.47\columnwidth]{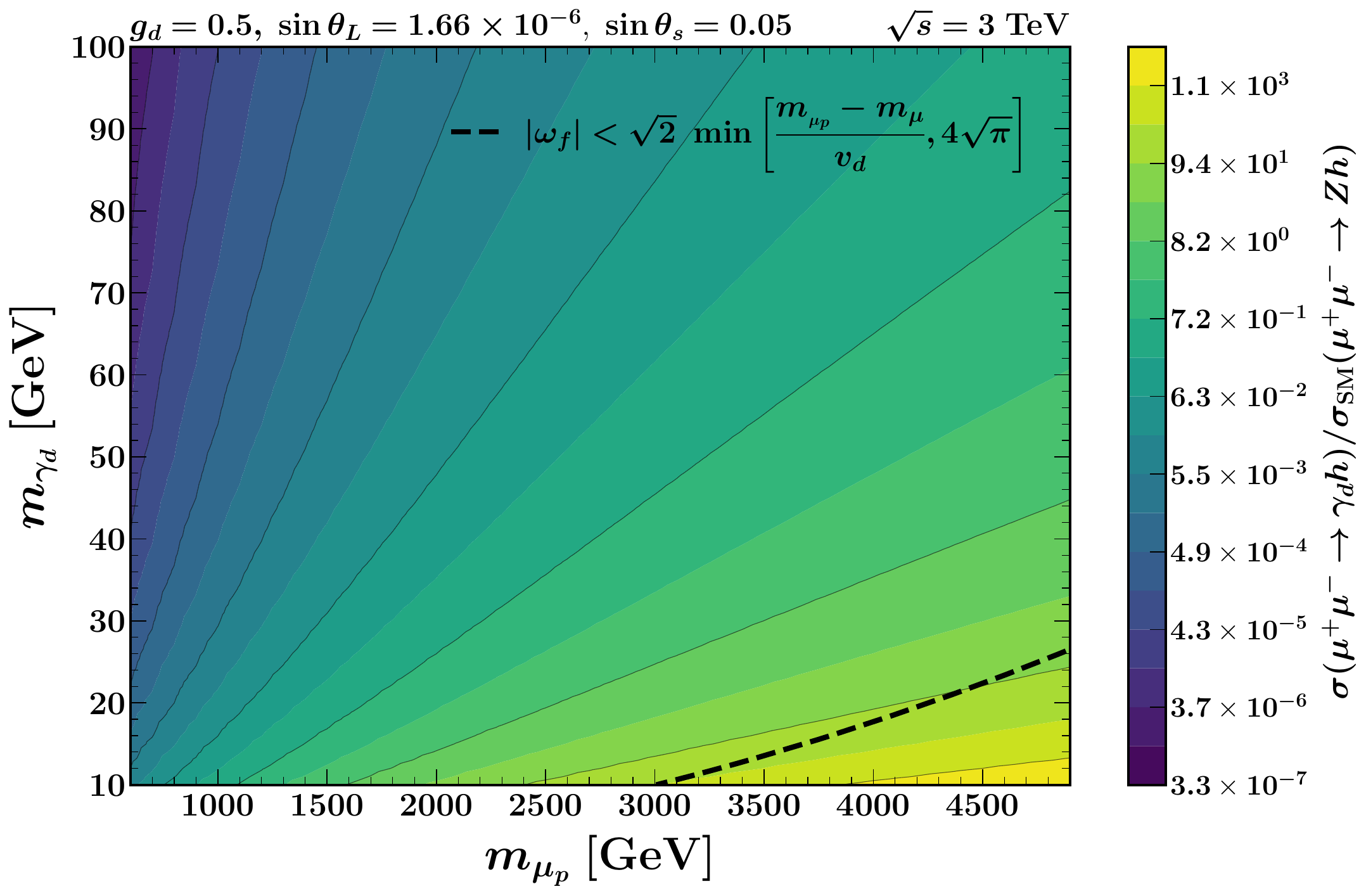}} 
    \caption{Variation of ratio of cross sections ($R_2$) for the process $\mu^+ \mu^- \to h \gamma_d$  with respect to the SM $hZ$ production rate in $m_{\gamma_d}$--$m_{\mu_{_p}}$  plane assuming $\sqrt{s}=3$ TeV for two different set of ($g_d,~ \sin\theta_L$) and a fixed value of $\sin\theta_s$. (a) The left plot corresponds to $g_d=0.05,~\sin\theta_s = 0.05,~\text{and}~ \sin\theta_L=4 \times 10^{-5}$ and (b) the right plot corresponds to $g_d=0.5,~\sin\theta_s = 0.05,~\text{and}~\sin\theta_L=1.66 \times 10^{-6}$. The region above the dashed (black) line is allowed by the constraint on $\omega_f$ described in Eq.~\eqref{eq:perturbative_unitarity_bound}. }
    \label{fig:mzp_vs_mmup_3tev}
\end{figure}

\begin{figure}[H]
    \centering
        \subfloat[\label{subfig:Mzp_vs_mmup_set1_10tev}]{\includegraphics[width=0.49\columnwidth]{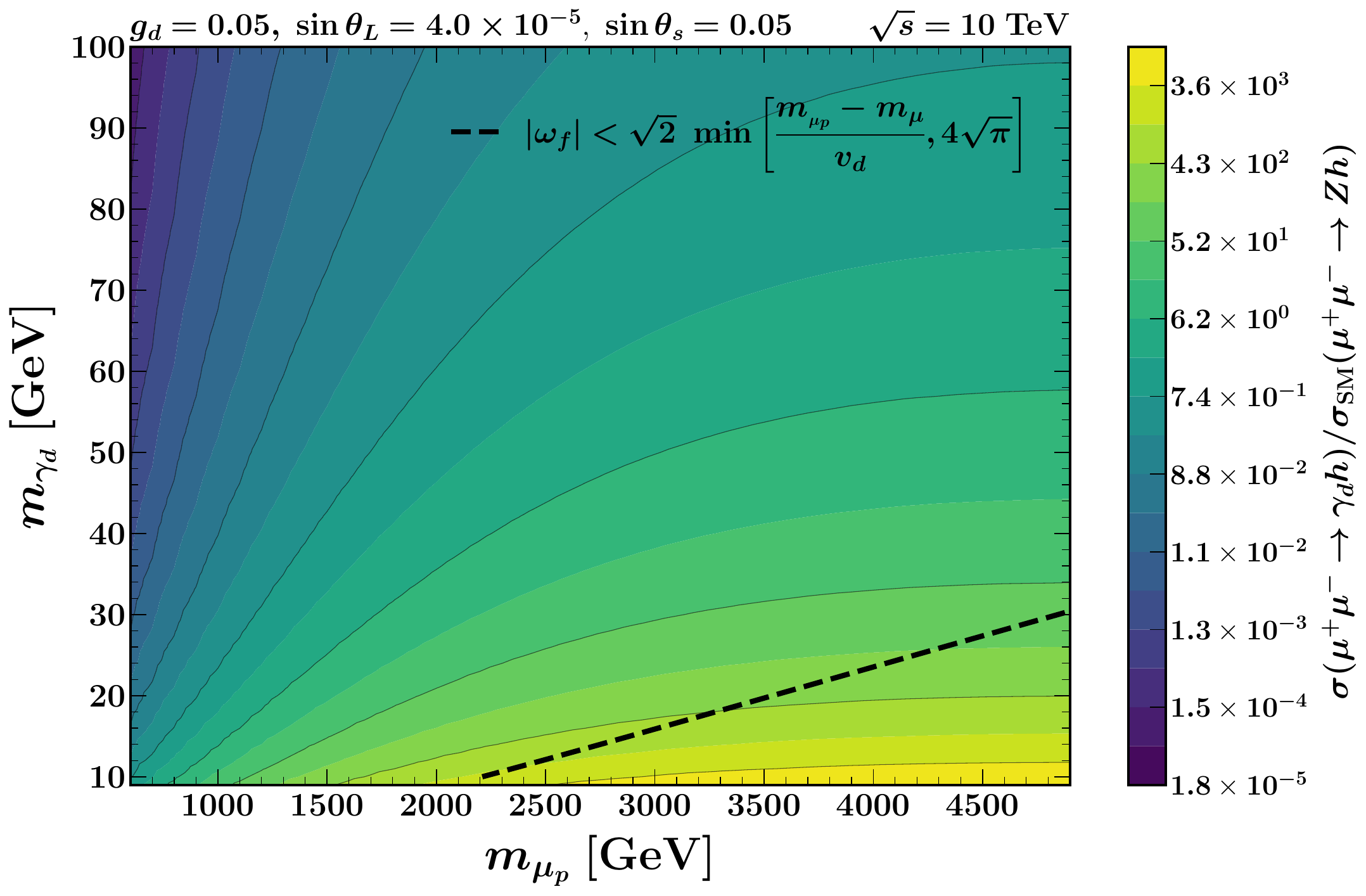}}~
        \subfloat[\label{subfig:Mzp_vs_mmup_set2_10tev}]{\includegraphics[width=0.49\columnwidth]{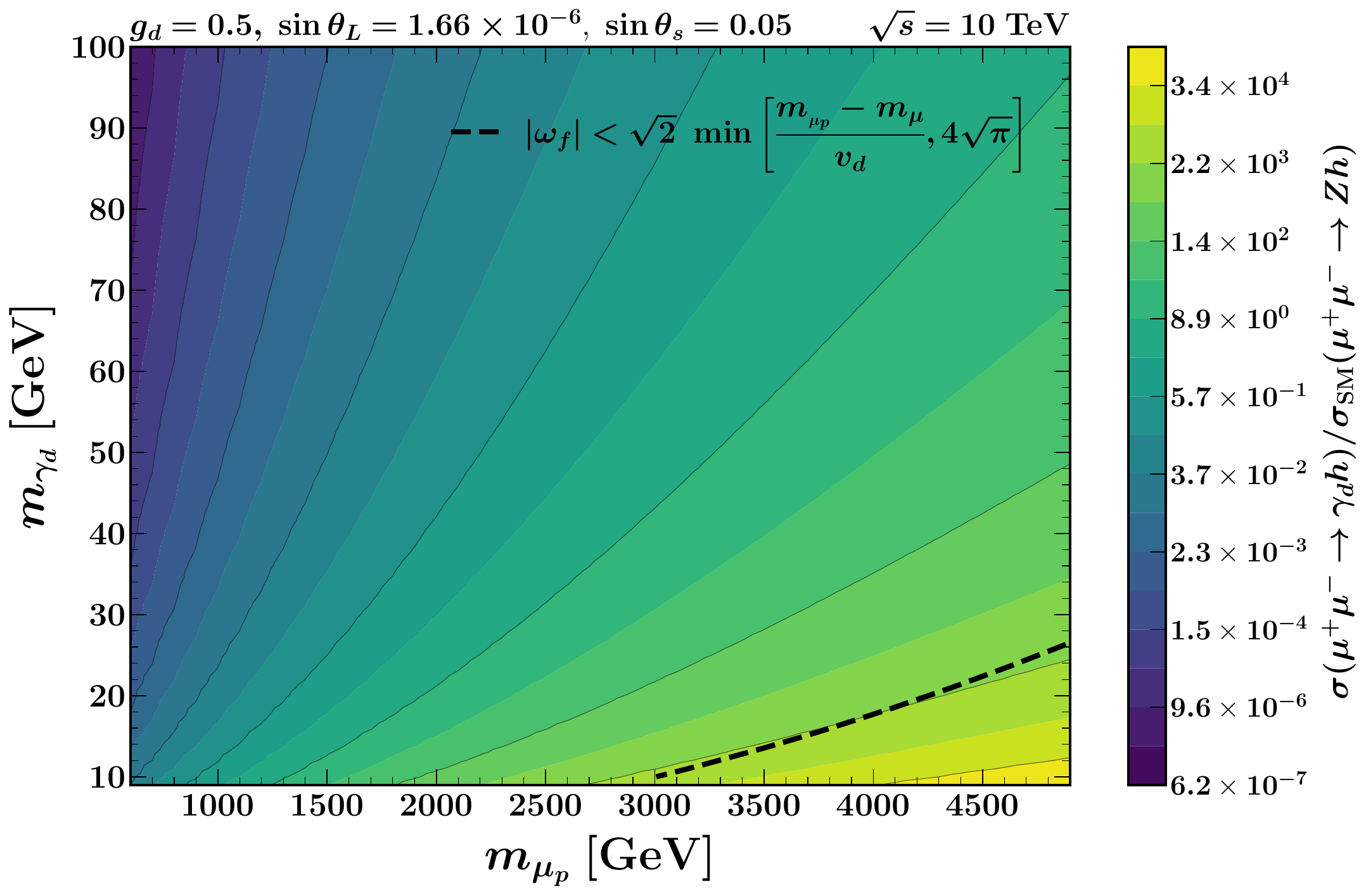}}
    \caption{Variation of ratio of cross sections ($R_2$) for the process $\mu^+ \mu^- \to h \gamma_d$  with respect to the SM $hZ$ production rate in $m_{\gamma_d}$--$m_{\mu_{_p}}$  plane assuming $\sqrt{s}=10$ TeV for two different set of ($g_d,~ \sin\theta_L$) and a fixed value of $\sin\theta_s$. (a) The left plot corresponds to $g_d=0.05,~\sin\theta_s = 0.05,~\text{and}~ \sin\theta_L=4 \times 10^{-5}$ and (b) the right plot corresponds to $g_d=0.5,~\sin\theta_s = 0.05,~\text{and}~\sin\theta_L=1.66 \times 10^{-6}$. The region above the dashed (black) line is allowed by the constraint on $\omega_f$ described in Eq.~\eqref{eq:perturbative_unitarity_bound}.}
    \label{fig:mzp_vs_mmup_10tev}
\end{figure}

In Fig.~\ref{fig:mzp_vs_mmup_3tev} and Fig.~\ref{fig:mzp_vs_mmup_10tev}, we plot $R_2$ (\,{defined in Eq.~\eqref{eq:ratio_equations}}\,) which represents the ratio of average unpolarized cross section for $\mu^+ \mu^- \to h \gamma_d$ to that of {the} SM $\mu^+ \mu^- \to hZ$ in the $m_{\gamma_d}$-- $m_{\mu_p}$ plane at $\sqrt{s}=3$ TeV and $\sqrt{s}=10$ TeV, respectively, to illustrate the effect of these two parameters on $R_2$ and hence the $h \gamma_d$ cross section for two different $g_d$ and $\sin\theta_L$ and fixed $\sin\theta_s$. The perturbative unitarity limit obtained using Eq.~\eqref{eq:perturbative_unitarity_bound} is also displayed on the same plot. 
Since the requirement of perturbative unitarity disfavors low values of $v_d$, one cannot achieve an arbitrary high values of $R_2$ by going to lower values of dark photon mass or higher values of $g_d$  and a trade off is maintained in the choice of the two independent parameters $m_{\gamma_d}$ and ${g_d}$. 

% R2 plot in mzp-mmup plane with perturbative unitarity at Ecm = 3 TeV
Therefore, depending on the parameter space, Higgs production in association with a dark photon at muon collider can provide us a Higgs production rate higher than $hZ$ production by a factor ranging from $1$ to $100$. 

%% Subsection-2 : Zph process analysis %%%
\subsection{Analysis of \texorpdfstring{$h \gamma_d$}{h gamma\_d} process}
\label{subsec:hzp_process_analysis}
In the remaining part of this article{,} we focus on the $h\gamma_d$ production at the future muon collider. As discussed above, the rate of this process can be significantly large compared to the $hZ$ production in a wide range of dark photon mass. Hence, it can contribute as an additional channel to the Higgs production at muon collider. 

\begin{figure}[H]
    \centering
        \resizebox{\columnwidth}{!}
        {
        \subfloat[\label{subfig:Ecm_variation_set1}]{\includegraphics[width=0.5\columnwidth]{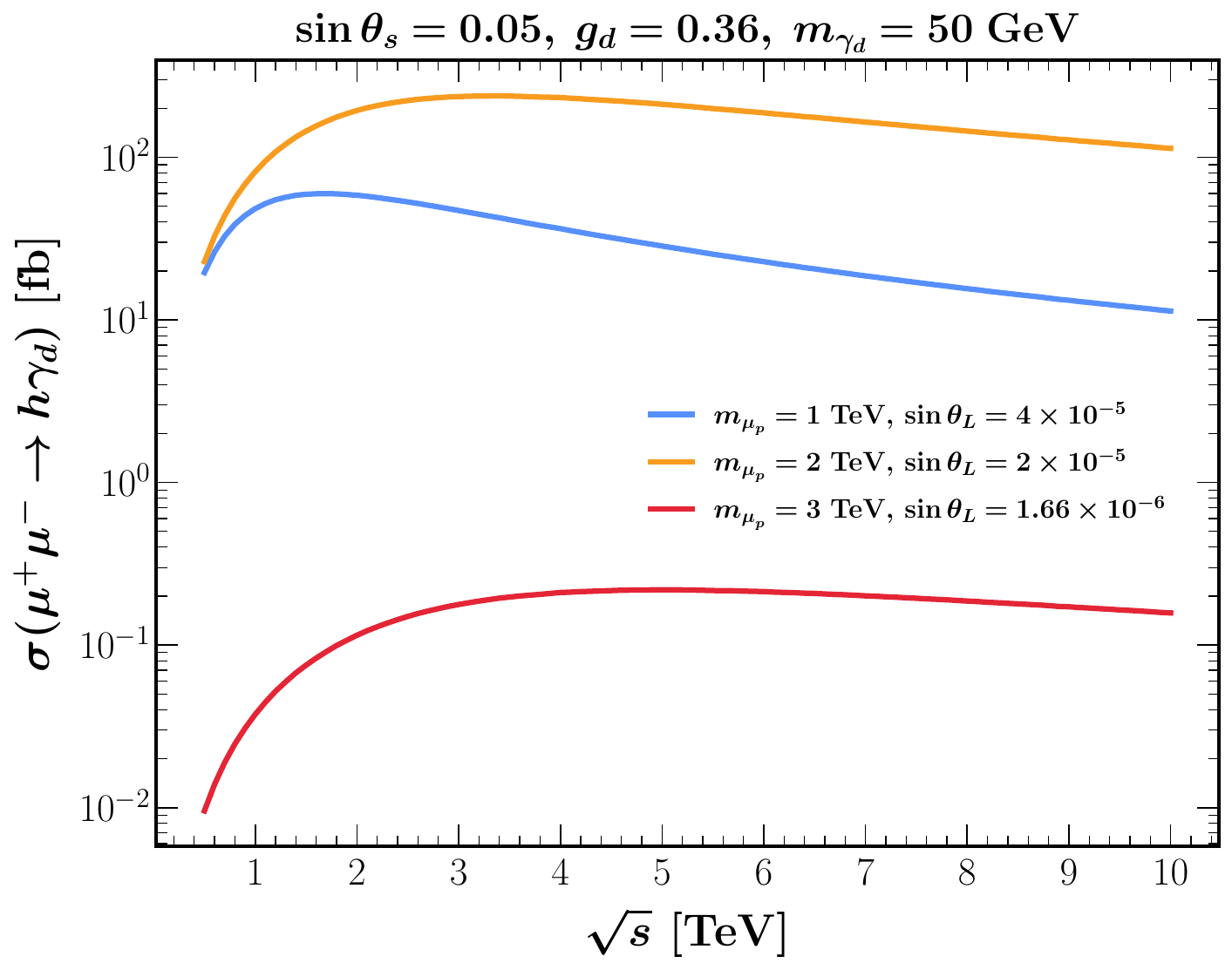}}~
        \hspace{0.4cm}
        \subfloat[\label{subfig:Ecm_variation_set2}]{\includegraphics[width=0.5\columnwidth]{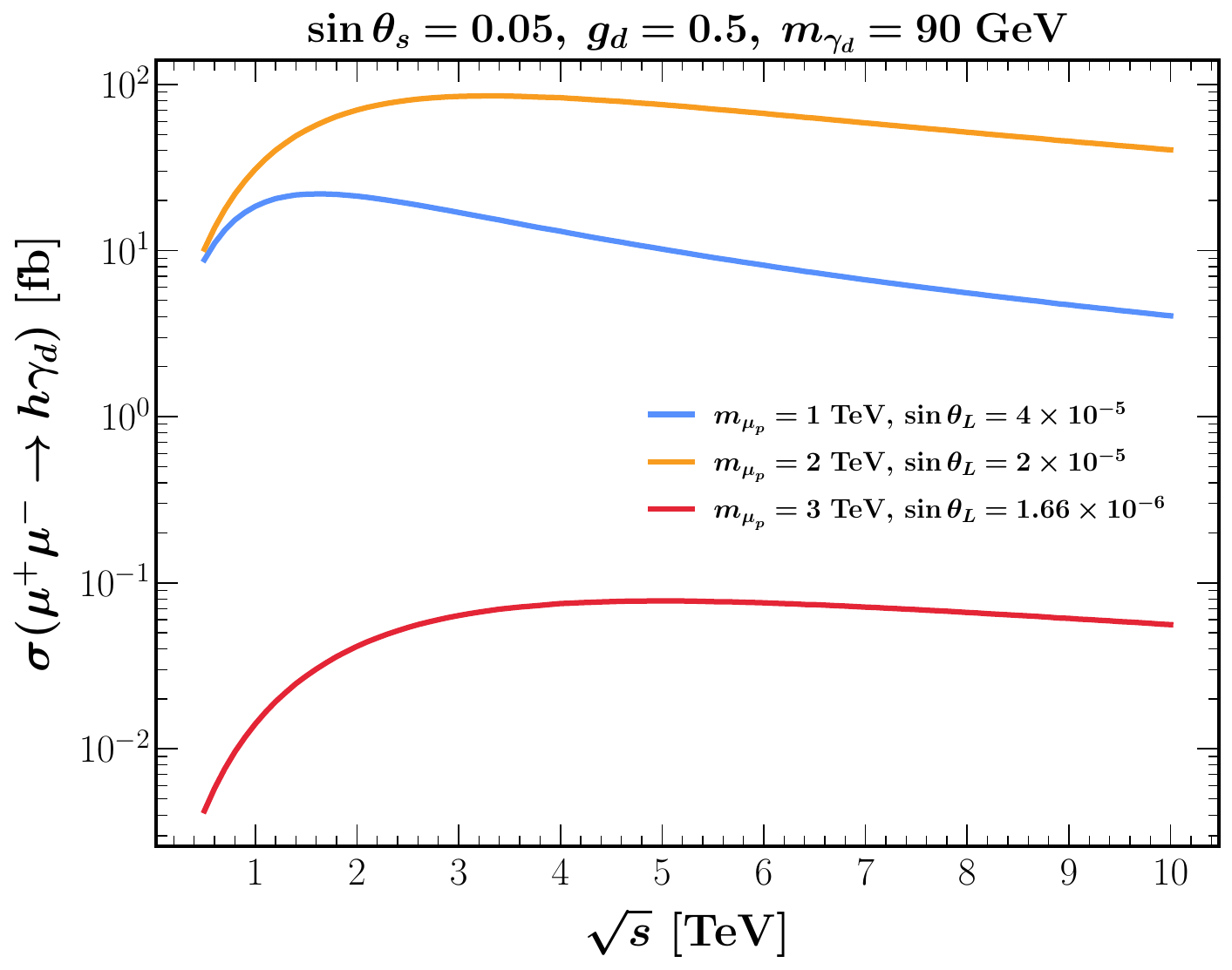}}
        }
    \caption{ LO cross section{s (in fb)} for $\mu^+ \mu^- \to h \gamma_d$ as a function of the muon collider center-of-mass energy for various choices of the $m_{\mu_p}$ and $\sin\theta_L$ assuming (a) $\sin\theta_s=0.05,~ g_d= 0.36,~ m_{\gamma_d}=50~\text{GeV}$ and (b) $\sin\theta_s=0.05,~ g_d= 0.50,~ m_{\gamma_d}=90~\text{GeV}$.
     }
    \label{fig:Ecm_variation}
  \end{figure}

 In Fig.~\ref{fig:Ecm_variation}{,} we present the LO cross sections for $\mu^+ \mu^- \to h \gamma_d$ as a function of the center-of-mass energy.
%
%
%  
%%%%%%%%%%% x-sec: BPs %%%%%%%%%%%
%
The LO cross sections for the process $\mu^+ \mu^- \to h \gamma_d$  {at the muon collider center-of-mass energies $\sqrt{s} = 3 $ TeV  and $10$ TeV for  various benchmark points mentioned in Table~\ref{tab:BPs_parameters}} are presented in Table~\ref{tab:xsec_BPs_standard}.
%
% Table-3: Cross section of BPs

\begin{table}[H]
    \centering
        \resizebox{0.7\columnwidth}{!}{
        \begin{tabular}{ c c c c c c c }
            \toprule
            \toprule
            $\sqrt{s}$& \multicolumn{6}{c}{$\sigma (\mu^+ \mu^- \to h \gamma_d)$~[fb]} \\
            ~[TeV]& BP1 & BP2 & BP3 & BP4 & BP5 & BP6 \\
            \midrule
            3  & $11.1$ & $22.8$ & $0.064$ & $47.2$ & $114$ & $0.436$  \\
            10 & $2.66$ & $10.9$ & $0.056$ & $11.4$ & $55.0$ & $0.388$  \\
            \bottomrule
            \bottomrule
        \end{tabular}}
    \caption{{LO cross section{s (in fb)} for $\mu^+ \mu^- \to h \gamma_d$ at $\sqrt{s}=3$ TeV and $10$ TeV for various choices of our benchmark points}}
    \label{tab:xsec_BPs_standard}
\end{table}
 The cross section for $h\gamma_d$ production strongly depends on dark photon mass and dark gauge coupling as well as on other model parameters (see Fig.~\ref{fig:mzp_vs_mmup_3tev}, Fig.~\ref{fig:mzp_vs_mmup_10tev}, as well as the corresponding vertex factors in Appendix~\ref{app:vert_fact} and the squared matrix elements in Appendix~\ref{app:matrix_element}).
 
%  
% %
% 
% %

% 
% %

% 

\begin{figure}[H]
    \centering
    \resizebox{\columnwidth}{!}{
    \subfloat[\label{subfig:2mu_to_zph1_s_channel_with_decay}]{\includegraphics[width=0.38\columnwidth]{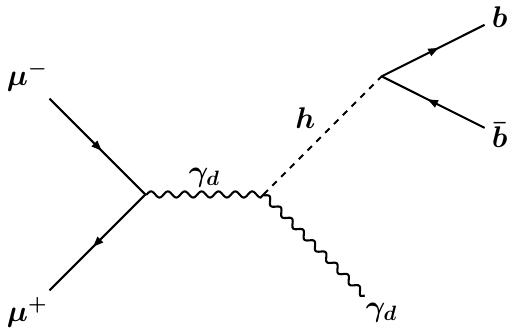}}~
    \subfloat[\label{subfig:2mu_to_zph1_t_channel_with_decay}]{\includegraphics[width=0.3\columnwidth]{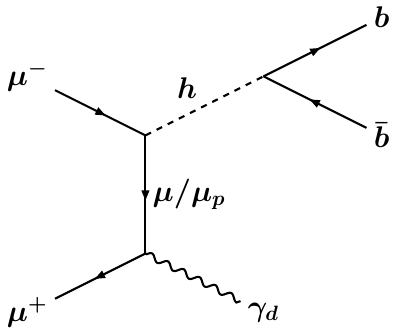}}~
    \subfloat[\label{subfig:2mu_to_zph1_u_channel_with_decay}]{\includegraphics[width=0.3\columnwidth]{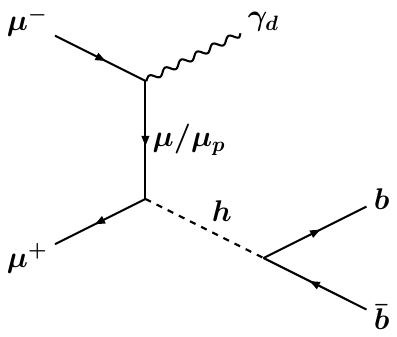}}
    }
    \caption{ {Feynman diagrams for various sub-processes representing $\mu^+ \mu^- \to h(b \bar{b}) + \gamma_d $ leading to {\it $b\bar{b}$ plus missing energy} final state, in absence of gauge kinetic mixing}.
    }
    \label{fig:final_state_signal}
  \end{figure}

{The dark photon in this scenario is assumed to be coupled to a light dark matter particle and therefore gives rise to invisible signature in terms of missing energy at the collider detector.} The Higgs on the other hand predominantly decays to a pair of bottom quarks. Therefore, the final state consists of {\it $b \bar{b}$ along with missing energy} at muon collider (see Fig.~\ref{fig:final_state_signal}).
%
% 2mu_to_zph_with_decay

\begin{figure}[H]
    \centering
        \subfloat[\label{subfig:2mu_to_hZ_with_decay}]{\includegraphics[width=0.3\columnwidth]{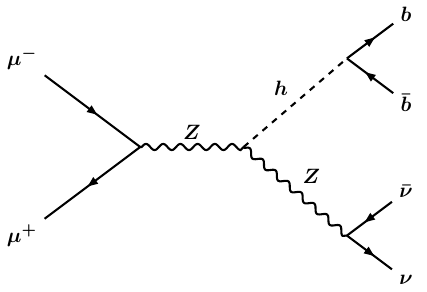}}~
        \subfloat[\label{subfig:2mu_to_zz_with_decay}]{\includegraphics[width=0.25\columnwidth]{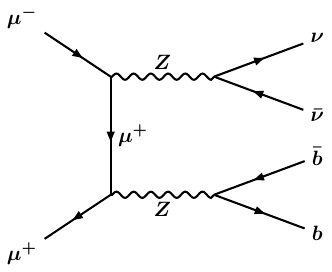}} 
        
        \vspace{0.5cm}
       \subfloat[\label{subfig:2mu_to_hvv_with_decay}]{\includegraphics[width=0.21\columnwidth]{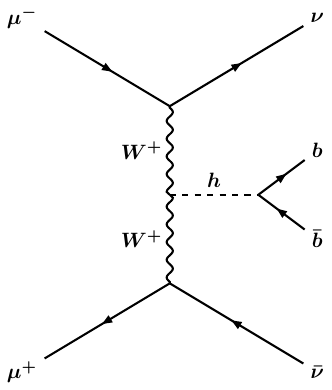}}~
        \subfloat[\label{subfig:2mu_to_zvv_with_decay}]{\includegraphics[width=0.21\columnwidth]{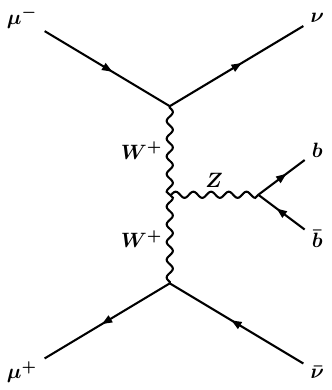}}~
        \subfloat[\label{subfig:2mu_to_bbvv_continuum}]{\includegraphics[width=0.21\columnwidth]{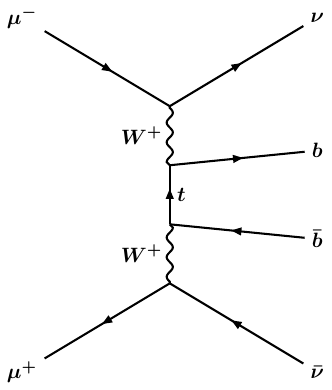}
        }
    \caption{ {Some representative Feynman diagrams leading to $b \bar{b} \nu \bar{\nu}$ final state in $\mu^+ \mu^-$ collisions within the SM.}}
    \label{fig:final_state_background}
\end{figure}

{The dominant SM backgrounds that contribute to the {\it $b\bar{b}$ plus missing energy} final state, namely, $\mu^+ \mu^- \to b \bar{b} \nu \bar{\nu}$ process comes from
all those sub-processes that lead to $b \bar{b} \nu \bar{\nu}$ final state at muon collider. The various Feynman diagrams representing these sub-processes can be sub-divided into $2 \to 2$, $2 \to 3$ and $2 \to 4$ categories.}
{Some representative Feynman diagrams corresponding to these categories are shown in Fig.~\ref{fig:final_state_background}.}

{We separately generate and analyze $Zh$, $ZZ$ , $h \nu \bar{\nu}$ and $Z \nu \bar{\nu}$ processes along with the total $bb\nu \bar{\nu}$ process for comparative study of the SM background contribution. In Table~\ref{tab:xsec_BGs}, we tabulate the cross sections for individual SM backgrounds.
The total cross-section for the $b \bar{b} \nu \bar{\nu}$ background is obtained 
 summing over all contributing Feynman diagrams leading to this final state. 
 We would like to emphasize that the cross-section quoted for $b \bar{b} \nu \bar{\nu}$ (total) is {\it not equal to} the algebraic sum of the cross-sections of the individual processes listed in the same table. }

% Table-4 : Cross section for SM background processes
\begin{table}[H]
\centering
\resizebox{0.5\columnwidth}{!}{
\begin{tabular}{ c c c }
  \toprule
  \toprule
  Background & $\sigma$ [fb]  & $\sigma$ [fb] \\
  processes \footnotemark & at $\sqrt{s}=3$ TeV & at $\sqrt{s}=10$ TeV\\
  \midrule
  % \hline
{$b \bar{b} \nu \bar{\nu}$ ($Zh$)} & $0.18$ & $0.02$ \\
  % \hline
{$b \bar{b} \nu \bar{\nu}$ ($ZZ$)} & $1.45$ & $0.18$ \\
    % \midrule
  {$b \bar{b} \nu \bar{\nu}$ ($h \nu \bar{\nu}$)} & $318$ & $540$ \\
  {$b \bar{b} \nu \bar{\nu}$ ($Z\nu \bar{\nu}$)} & $305$ & $526$ \\
  {$\bm{b \bar{b} \nu \bar{\nu}}$ \bf{(total)}}  & $\bm{641}$ & $\bm{1101}$ \\
  \bottomrule
  \bottomrule
\end{tabular}
}
\caption{{LO cross-section (in fb) at $\sqrt{s}=3$ TeV and $10$ TeV muon collider for various SM background processes, $Zh$, $ZZ$, $h \nu \bar{\nu}$, $Z \nu \bar{\nu}$ leading to $b\bar{b} \nu \bar{\nu}$ final state. The cross-sections quoted in the last row ({\bf boldface}) represent the total cross-sections for the $\mu^+ \mu^- \to b\bar{b} \nu \bar{\nu}$ process obtained by summing over all contributing Feynman diagrams (see Fig. \ref{fig:final_state_background}) that leads to this final state and {\it not equal to} the algebraic sum of the cross-sections of individual  $Zh$, $ZZ$, $h \nu \bar{\nu}$, $Z \nu \bar{\nu}$ processes mentioned above.
 }}
\label{tab:xsec_BGs}
\end{table}

\begin{figure}[H]
    \centering
    \resizebox{\columnwidth}{!}
    {
        \subfloat[\label{subfig:jet_multiplicity_distribution_for_Ecm_3tev}]{\includegraphics[width=0.5\columnwidth]{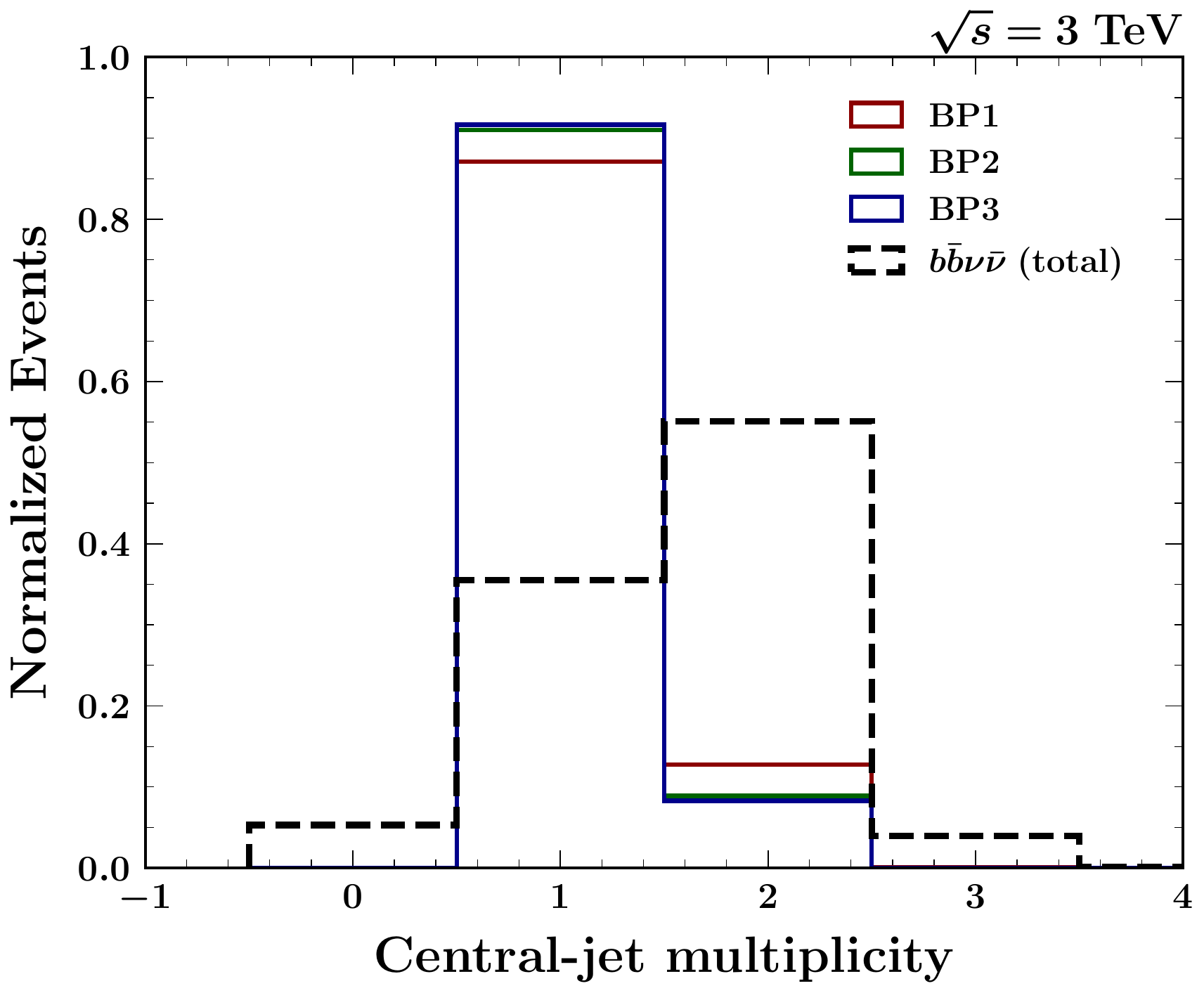}}~
        \hspace{0.4cm}        \subfloat[\label{subfig:jet_multiplicity_distribution_Ecm_10tev}]{\includegraphics[width=0.5\columnwidth]{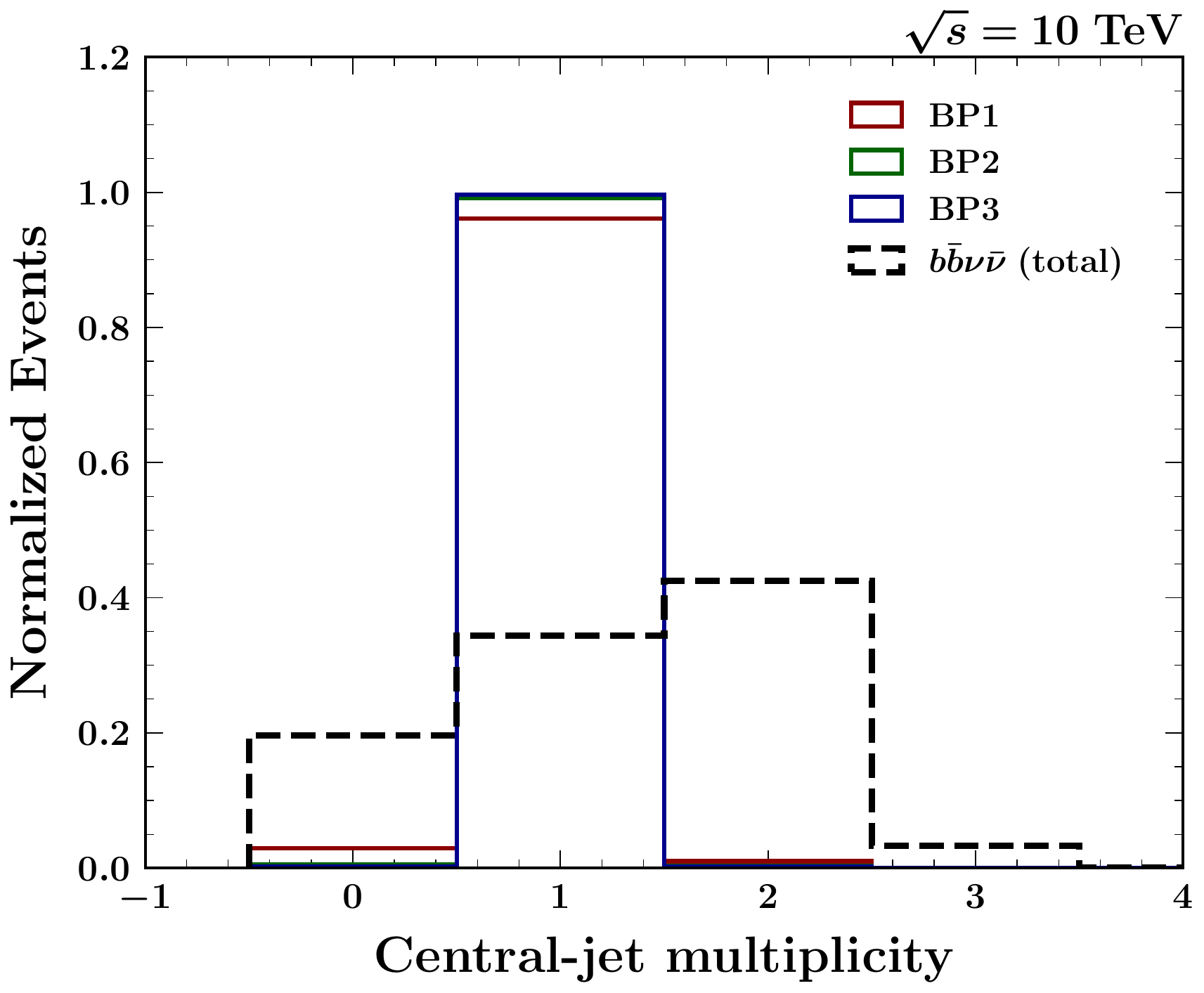}}
    }
    \caption{{Central-jet multiplicity distribution} at $\sqrt{s}=3$ TeV and $10$ TeV for three different signal benchmark points and the total SM background.}
    \label{fig:jet_multiplicity_distribution}
  \end{figure}

\footnotetext{In Table~\ref{tab:xsec_BGs}, the cross section quoted for $h \nu \bar\nu$ includes contributions from both
$\mu^+ \mu^- \to Zh \to b \bar{b} \nu \bar{\nu}~(2 \to 2)$ and
$\mu^+ \mu^- \to h \nu\bar{\nu} \to b \bar{b} \nu \bar{\nu}~(2 \to 3)$ sub-processes with an on-shell Higgs and on-shell/off-shell $Z$ boson. 
Similarly, for $Z \nu \bar{\nu}$ contributions from all relevant Feynman diagrams with one on-shell $Z$ are considered.
}
Events generated using the event generator \textsc{MadGraph5\_aMC@NLO} for
both the signal and the SM background processes are then interfaced with  \textsc{Pythia8} (version 8.3) \cite{Bierlich:2022pfr} for showering, hadronization, decay and further collider analysis. The detector response in the context of future muon collider is parametrized using a Gaussian smearing function as given in Eq.~\eqref{eq:smearing_function} following the Ref. \cite{park_2018_ww43z-k8316,CMS:2016lmd}. 
%
% smearing function
\begin{eqnarray}
\label{eq:smearing_function}
        \frac{\sigma_{_E}}{E} 
        &=& 
        \sqrt{
        \left( \frac{N}{E} \right)^2 
        + 
        \left( \frac{S}{\sqrt{E}} \right)^2 
        + 
        C^2
        }
\end{eqnarray}

where $E$ is the energy associated with the final state object and $\sigma_{_E}$ is the standard deviation in the measurement of $E$. The parameters $N,~S$ and $C$ represents respectively the noise, stochastic and the constant terms. Depending on the detector these parameters have different values for specific final state objects. In the context of future muon collider for jets in the central region ($|\eta| < 2.5$), we have used the following set of values for these parameters \cite{Pezzotti:2022ndj}
\begin{eqnarray*}
    N = 0,~~S = 0.3,~~C = 0.05  
\end{eqnarray*}
  We have used anti-kt jet clustering algorithm \cite{Cacciari:2008gp} with clustering radius R=0.4 using FastJet (version 3.5.0) \cite{Cacciari:2011ma}.
  {We also work with jets in the central region ({\it central-jet}), defined as 
  \begin{eqnarray*}
      p_{_T} > 20~{\rm GeV~} \quad {\rm and} ~ \quad |\eta| < 2.5
  \end{eqnarray*}
  \noindent
  In Fig.~\ref{fig:jet_multiplicity_distribution}, we plot the central-jet 
 multiplicity distributions for various signal benchmark points.}
 This has been done keeping in mind the fact that at the muon collider center-of-mass energies ($\sqrt{s}=3~\text{TeV}~\text{and}~10~\text{TeV})$, both the Higgs and the dark photon are always highly boosted, each sharing almost half of the available center-of-mass energies. Therefore, most of the signal events are expected to have one collimated $b\bar{b}$ pair originating from the Higgs decays resulting into a single narrow jet. Only a fraction of signal events  due to final state radiation, showering and hadronization effect result into events with more than one central-jet.

It turns out at 3 TeV center-of-mass energy a large fraction ($\sim 87-90$\%) of the signal events consists of {\it exactly one central-jet} and remaining $10-13$\% corresponds to {\it exactly two central-jet} events. This pattern changes at 10 TeV center-of-mass energy. In this case, almost all the signal events ($\gtrsim 96$\%) correspond to one central-jet, as expected.
The {central-jet multiplicity} distribution corresponding to the total SM background in the $b\bar{b}\nu\bar{\nu}$ final state is also displayed on the same figure.
One can see that the SM background events are distributed among $0,1,2~\text{or}~3$ {{\it central-jet} events categories}\footnote{Here, the {\it zero  central-jet} event category represents all events with no jets in the central region (i.e., $|\eta|<2.5$) having $p_{_{T}} > 20$ GeV and is not considered for any further analysis. }. At $\sqrt{s}=3$ TeV the {central-jet multiplicity} distribution has a peak around {two} containing almost $54$\% of the total background events. At 10 TeV center-of-mass energy, there is a significant decrease (increase) in number of events in exactly two (zero) central-jet event categories.

  \begin{figure}[H]
    \centering
    \resizebox{\columnwidth}{!}
    {
        \subfloat[\label{subfig:jet_multiplicity_distribution_Ecm_3tev_backgrounds}]{\includegraphics[width=0.5\columnwidth]{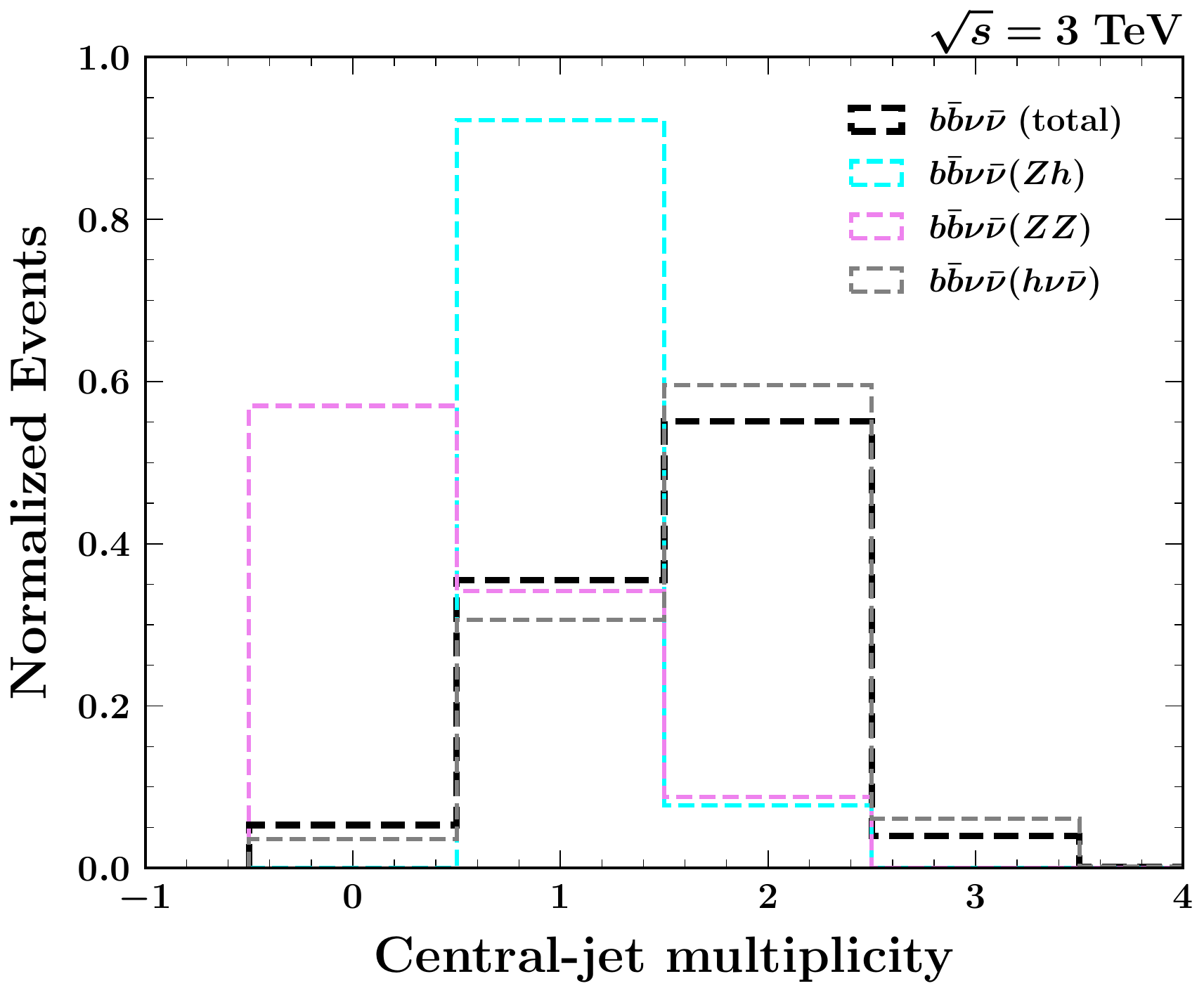}}~
        \hspace{0.4cm}
        \subfloat[\label{subfig:jet_multiplicity_distribution_Ecm_10tev_backgrounds}]{\includegraphics[width=0.5\columnwidth]{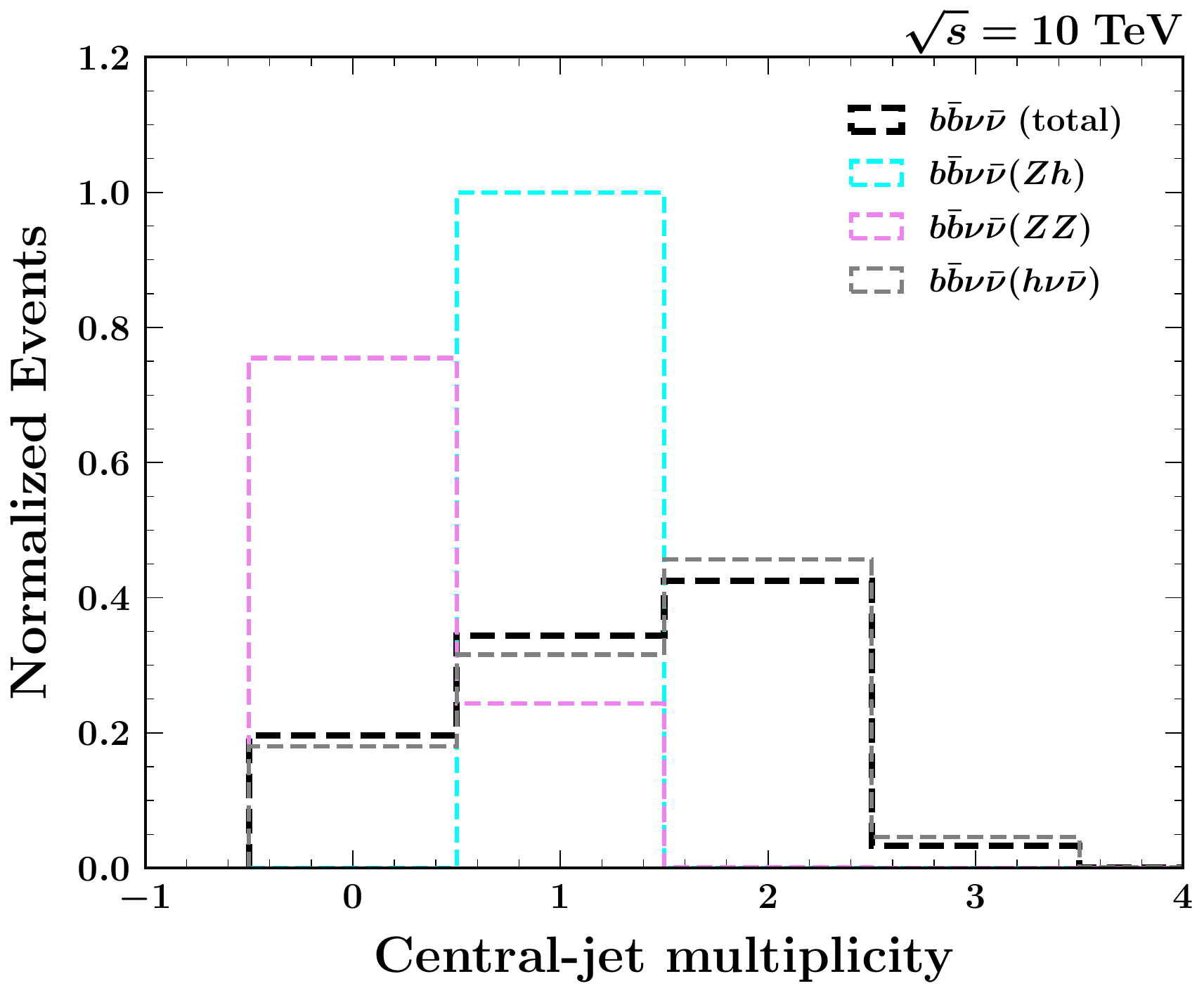}}
    }
    \caption{{Central-jet multiplicity distribution} at $\sqrt{s}=3$ TeV and $10$ TeV for various {SM backgrounds and the total $b \bar{b} \nu \bar{\nu}$ background}.}
    \label{fig:jet_multiplicity_distribution_backgrounds}
  \end{figure}
{
The behavior of the central-jet multiplicity distribution for the total background shown in Fig.~\ref{fig:jet_multiplicity_distribution_backgrounds} can be understood as follows: 
The Higgs or $Z$ boson produced in association with a non-resonant neutrino pair in a $2 \to 3$ process is generally not boosted, often resulting in multi-jet final states.
Similar arguments apply to the continuum $b\bar{b}\nu\bar{\nu}$ background.
In contrast, the SM $Zh$ background,
with both $Z$ and $h$ on-shell, 
being an $s$-channel $2 \to 2$ process, displays a different central-jet number distribution, mostly peaking around one. The SM $ZZ$ background with both $Z$ on-shell, despite being a $2 \to 2$ process, is expected to give more zero central-jet events than $Zh$ as it is produced via $t/u$-channel muon mediation. These features are also reflected in Fig.~\ref{fig:jet_multiplicity_distribution_backgrounds}. Since the cross sections for SM $Zh$ and $ZZ$ processes are very small compared to the total background cross section, the normalized central-jet multiplicity distribution mostly follows that of the $h \nu \bar\nu$ (and $Z\nu \bar{\nu}$).  }

% 
% 
% 
% 
%
% % 
% % 
% 
% 
% 
% 
% 
%
% 
%}

%

This motivates us to focus on two different final states, namely,  {\it leptonically quiet exactly one central-jet {\bf(LQ1CJ)} and exactly two central-jets {\bf(LQ2CJ)}, both in conjunction with missing energy} at two different $\mu^+ \mu^-$ collision energies ($\sqrt{s}=3$ TeV and $10$ TeV).

 \begin{figure}[H]
    \centering
    \resizebox{\columnwidth}{!}
    {
        \subfloat[\label{subfig:jet_pT_distribution_ncjet1_Ecm_3tev}]{\includegraphics[width=0.5\columnwidth]{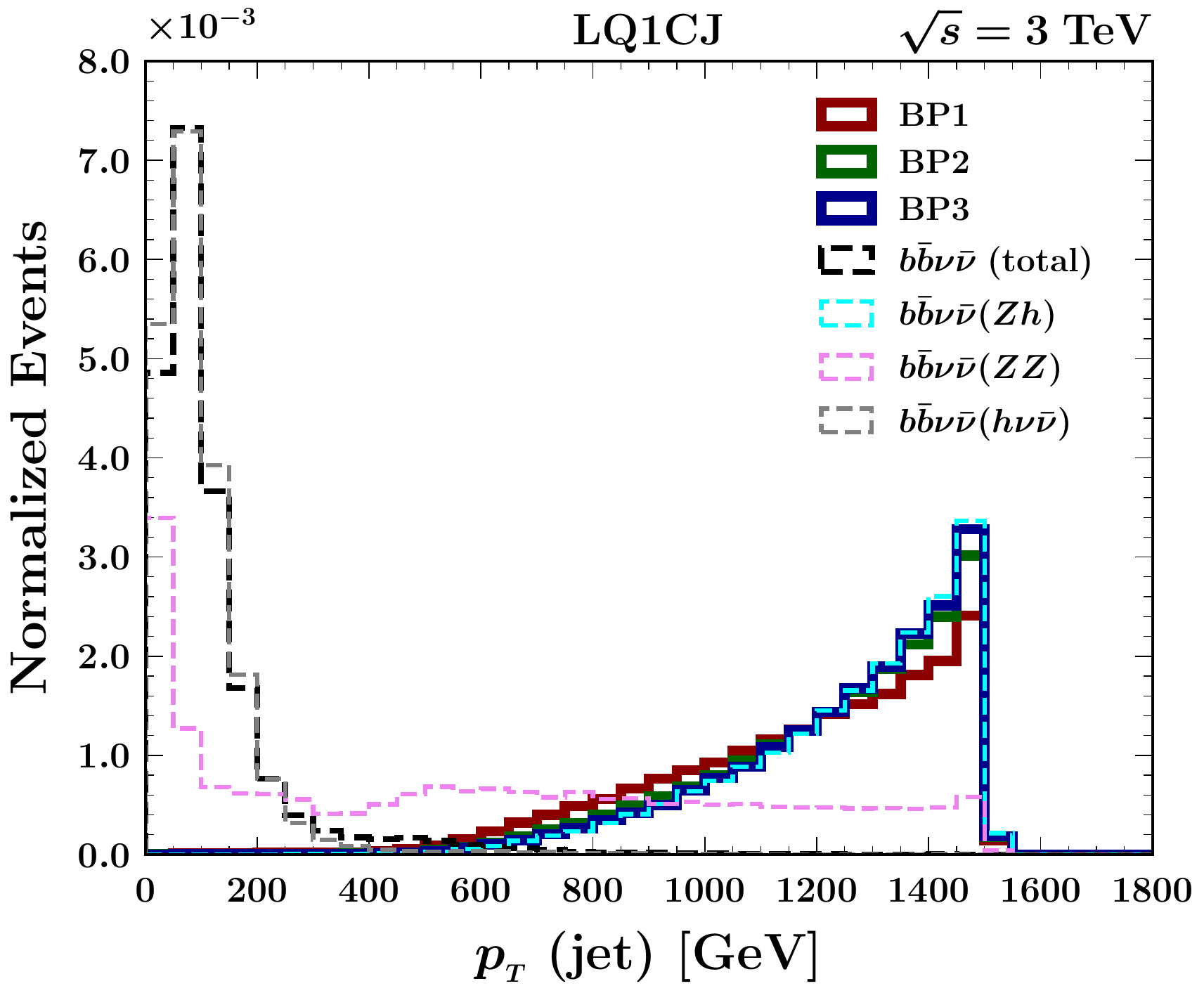}}~
        \hspace{0.3cm}
        \subfloat[\label{subfig:jet_pT_distribution_ncjet2_Ecm_3tev}]{\includegraphics[width=0.5\columnwidth]{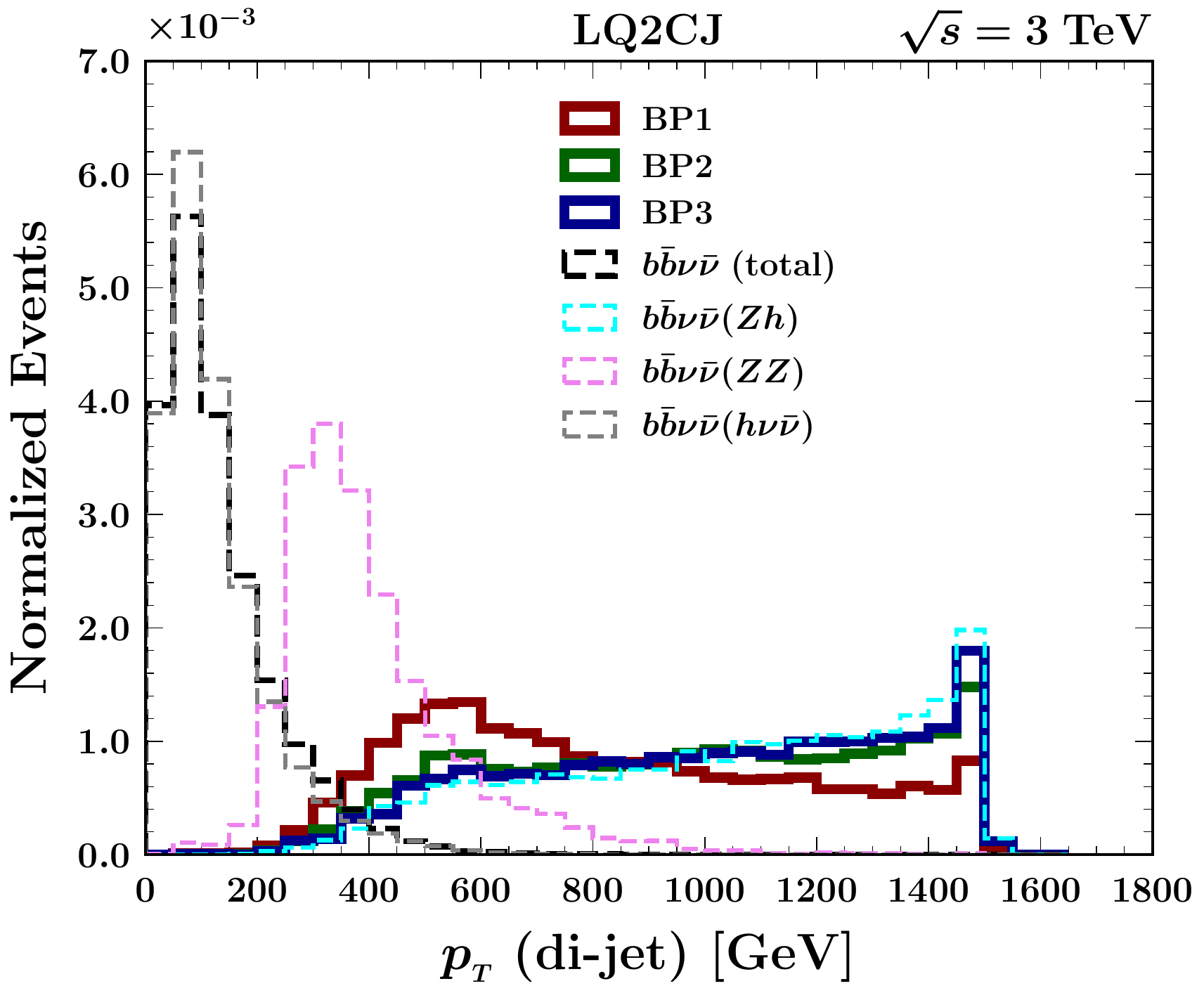}}
    }
    \caption{Transverse momentum ($p_{_T}$) distributions of the (a) jet and (b) the di-jet in the {LQ1CJ and LQ2CJ} final states, respectively, at $3$ TeV muon collider center-of-mass energy for both signal and the SM background events.}
    \label{fig:jet_pT_distribution_Ecm_3tev}
\end{figure}

\begin{figure}[H]
    \centering
    \resizebox{\columnwidth}{!}
    {
        \subfloat[\label{subfig:Jet_pT_distribution_ncjet1_Ecm_10tev}]{\includegraphics[width=0.5\columnwidth]{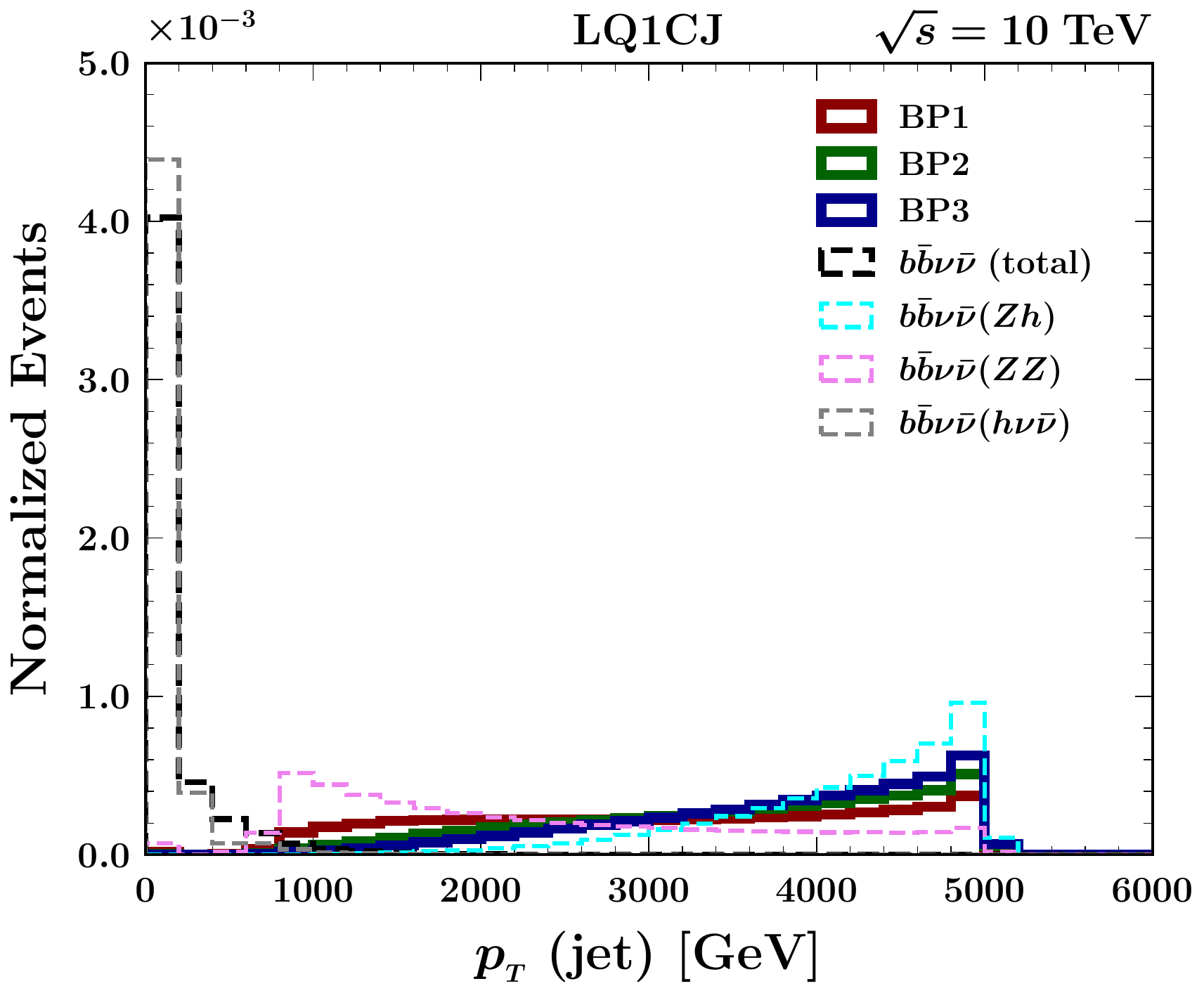}}~
        \hspace{0.3cm}
       \subfloat[\label{subfig:Jet_pT_distribution_ncjet2_Ecm_10tev}]{\includegraphics[width=0.5\columnwidth]{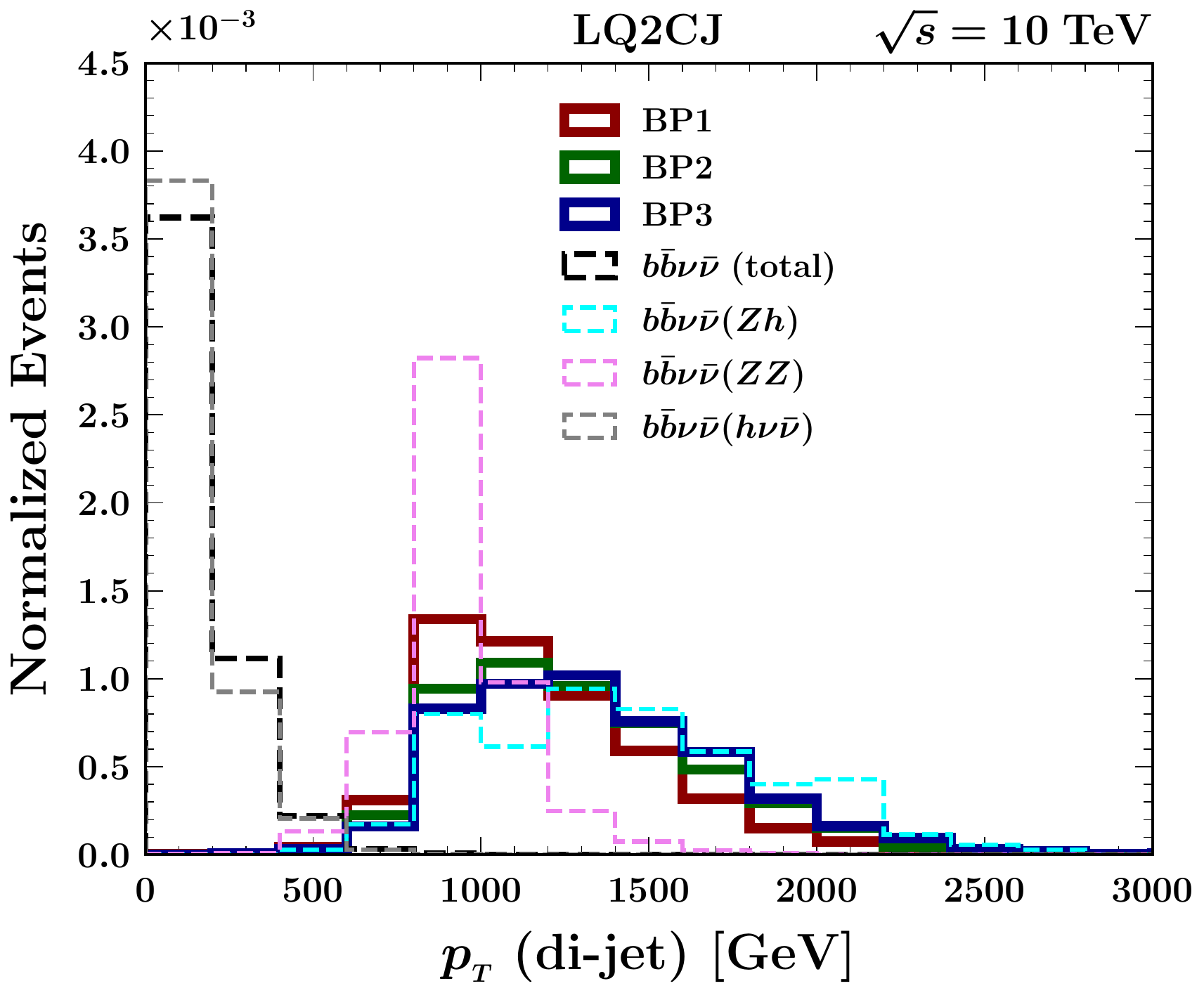}}
    }
    \caption{Transverse momentum ($p_{_T}$) distributions of the (a) jet and (b) the di-jet in the {LQ1CJ and LQ2CJ} final states, respectively, at $10$ TeV muon collider center-of-mass energy for both signal and the SM background events. }
    \label{fig:jet_pT_distribution_Ecm_10tev}
\end{figure}

 As already mentioned, Higgs boson produced at the future muon collider center-of-mass energies in a $2 \to 2$ process are typically highly boosted. The Higgs subsequently decays into a pair of bottom quarks which due to their small angular separation, manifest as a single collimated jet in the detector. In Fig.~\ref{fig:jet_pT_distribution_Ecm_3tev} and \ref{fig:jet_pT_distribution_Ecm_10tev}, we plot the transverse momentum distributions of a jet or di-jet system, as applicable, for two different muon collider center-of-mass energies.
 The purpose of these distributions are two-fold. They provide an explanation of Fig.~\ref{fig:jet_multiplicity_distribution} and \ref{fig:jet_multiplicity_distribution_backgrounds} and also give a motivation for the following discussion.

To identify and analyze the boosted Higgs jets, we employ jet substructure analysis and dedicated Higgs tagging techniques. 
For this purpose we have used {\it MassDropTagger} \cite{Butterworth:2008iy} to identify a boosted Higgs structure inside jets. These jets are obtained after re-clustering all the hadrons in a particular final state with  Cambridge-Aachen (CA) jet clustering algorithm \cite{Bentvelsen:1998ug} and energy-scheme recombination assuming jet-clustering radius\footnote{We call these jets as CA$_{1.2}$ jets. Also the above analysis is not too sensitive to the choice of the jet clustering radius values. The algorithm works fine in the range $0.8\leq R \leq 1.4$.} $R=1.2$. 
The CA algorithm is chosen because it clusters particles based solely on angular distance, making it well-suited for subsequent substructure analyses.  

Below we briefly summarize the algorithm behind the jet substructure analysis implemented in {\it MassDropTagger}:
\begin{enumerate}
    \item The leading-$p_{_T}$ jet (with transverse momentum $p_{_T} > 200~\mathrm{GeV}$) is first passed through the {\it MassDropTagger} procedure. The algorithm iteratively de-clusters the jet to find an imprint of a heavy-particle by applying the following two conditions characterized by the dimensionless parameters $\mu$ and $y_{\text{cut}}$:
        \begin{itemize}
          \item[i)] Mass-drop condition : $\displaystyle \frac{m_{j_1}}{m_j} < \mu$
          \item[ii)]  $\displaystyle \text{Symmetry condition:}
            ~y = 
            \frac{\min\!\left(p_{Tj_1}^2,\, p_{Tj_2}^2\right)}{m_j^2} 
            \, \Delta R_{j_1 j_2}^2 
            \;>\; y_{\text{cut}} \,.$
        \end{itemize}

    where $j_1$ and $j_2$ (with $m_{j_1} > m_{j_2}$) denote the two sub-jets obtained by undoing the final stage of clustering of a parent jet $j$. Here, $\Delta R_{j_1 j_2}$ (with $\Delta R = \sqrt{\Delta \eta^2 + \Delta \phi^2}$) is the angular separation between the two sub-jets in the $\eta - \phi$ plane. The values of the dimensionless parameters $\mu$ and the $y_{\rm cut}$ are $0.67$ and $0.09$, respectively \cite{Butterworth:2008iy}. 
    
    \item If both the above criteria are satisfied, the jet is tagged as a boosted heavy-particle candidate consisting of two sub-jets $j_1$ and $j_2$. Otherwise, the lighter sub-jet is discarded, and the procedure is repeated by treating $j_1$ as the parent jet until a valid substructure is identified. We select an event for further analysis if it contains at least one ${\rm CA}_{1.2}$ jet with a substructure.
\end{enumerate}

{The application of jet-substructure techniques or the use of the transverse momentum of the $b\bar{b}$ system, in combination with missing energy, prove highly effective in suppressing a large fraction of the total SM background and only the $2 \to 2$ processes with on-shell $Zh$ and $ZZ$ backgrounds surpass the substructure requirement. However, the cross sections for these processes are significantly small compared to the total SM background cross section (see Table~\ref{tab:xsec_BGs}).}

%%%% subsection-3 : Distribution of kinematic variables %%%%
\subsection{Kinematic variables}
\label{subsec:KM}
\begin{figure}[H]
    \centering
    \resizebox{\columnwidth}{!}
    {
        \subfloat[\label{subfig:inv_mass_distribution_ncjet1_Ecm_3tev}]{\includegraphics[width=0.5\columnwidth]{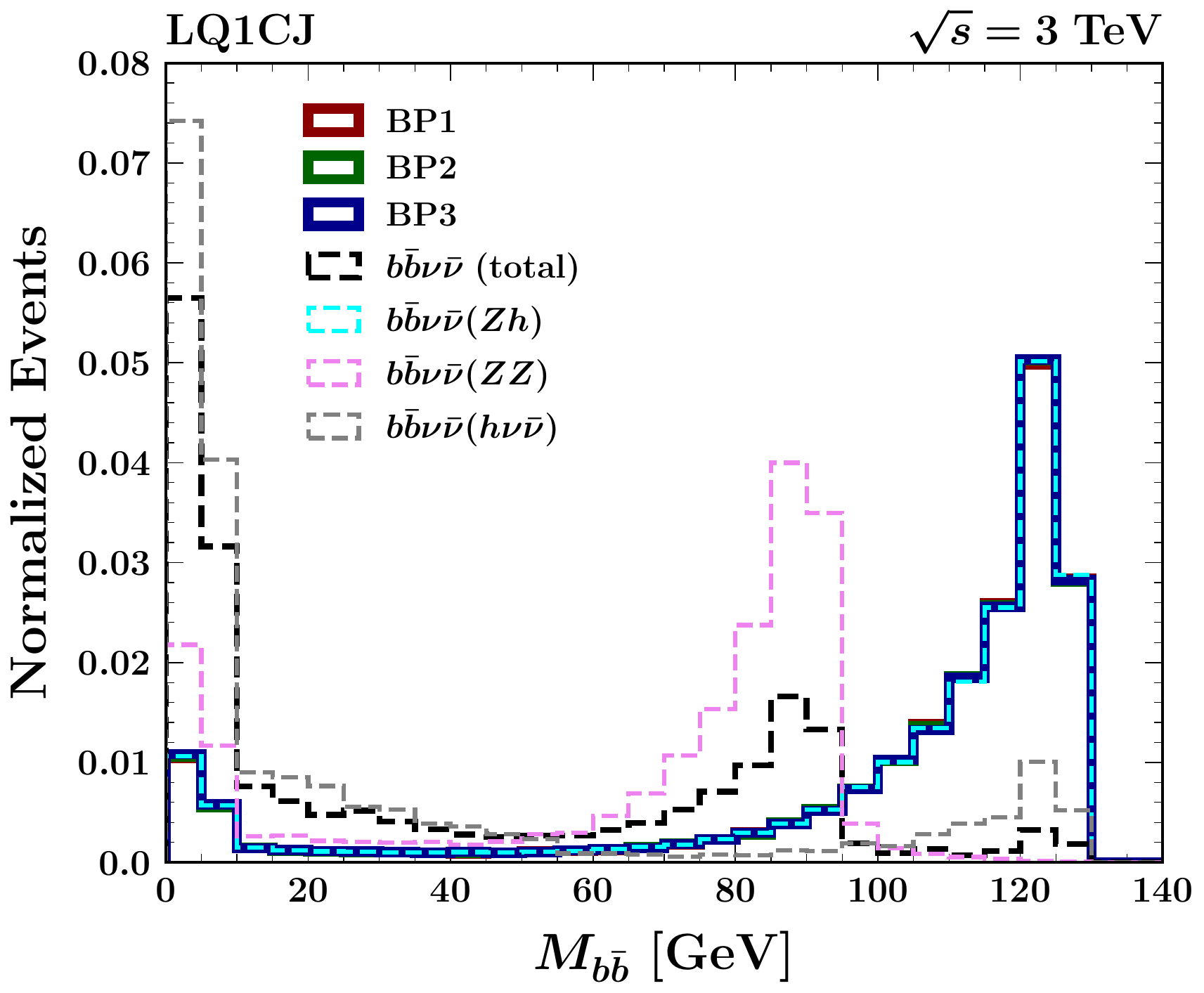}}~
        \hspace{0.3cm}
        \subfloat[\label{subfig:inv_mass_distribution_ncjet2_Ecm_3tev}]{\includegraphics[width=0.5\columnwidth]{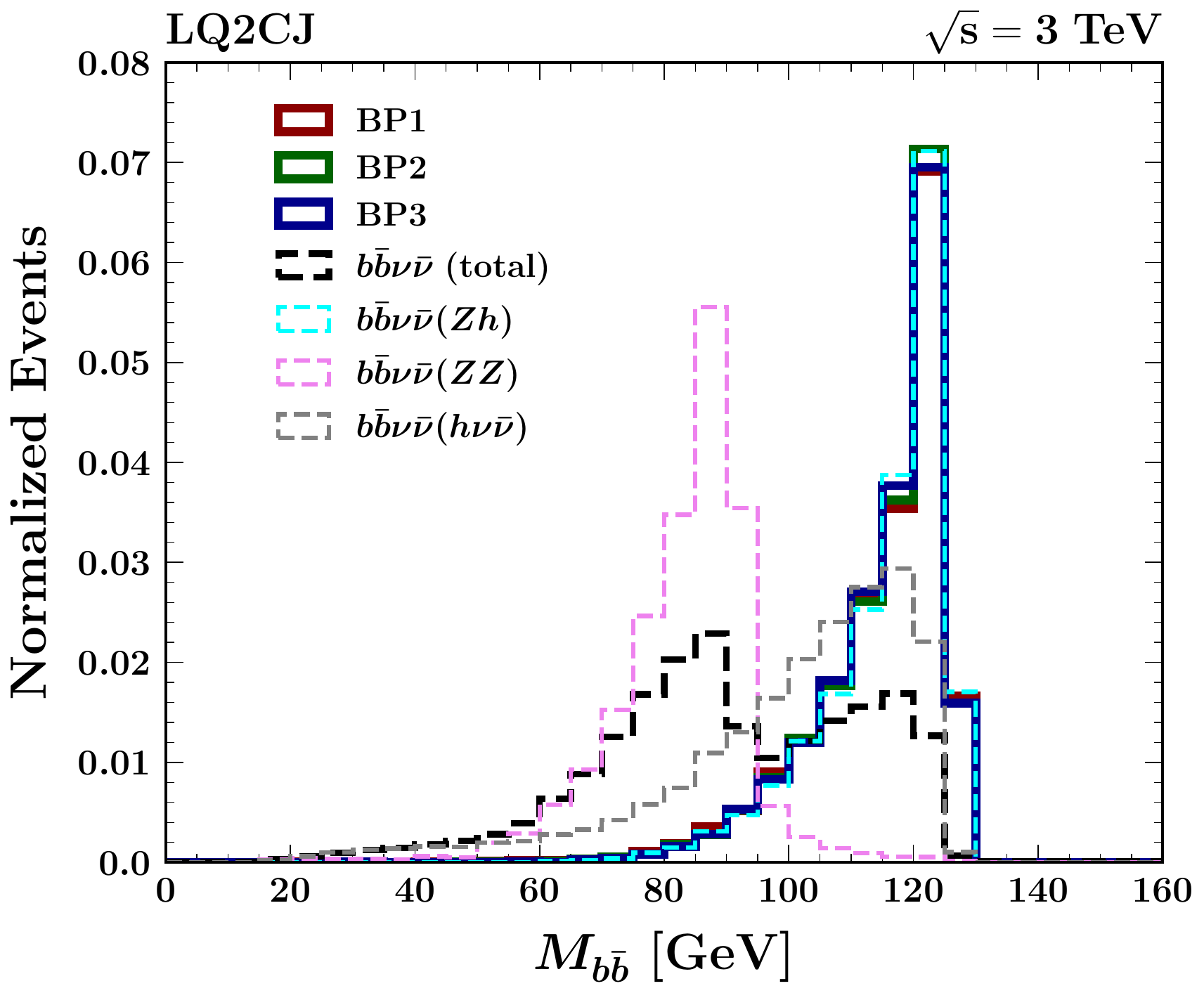}}
    }
    \caption{Invariant mass ($M_{b\bar{b}}$) distribution at $\sqrt{s}=3$ TeV in the {(a) LQ1CJ and (b) LQ2CJ} final state for both signal and background events.}
    \label{fig:inv_mass_Ecm_3tev}
\end{figure}

\begin{figure}[H]
    \centering
    \resizebox{\columnwidth}{!}
    {
        \subfloat[\label{subfig:inv_mass_distribution_ncjet1_Ecm_10tev}]{\includegraphics[width=0.5\columnwidth]{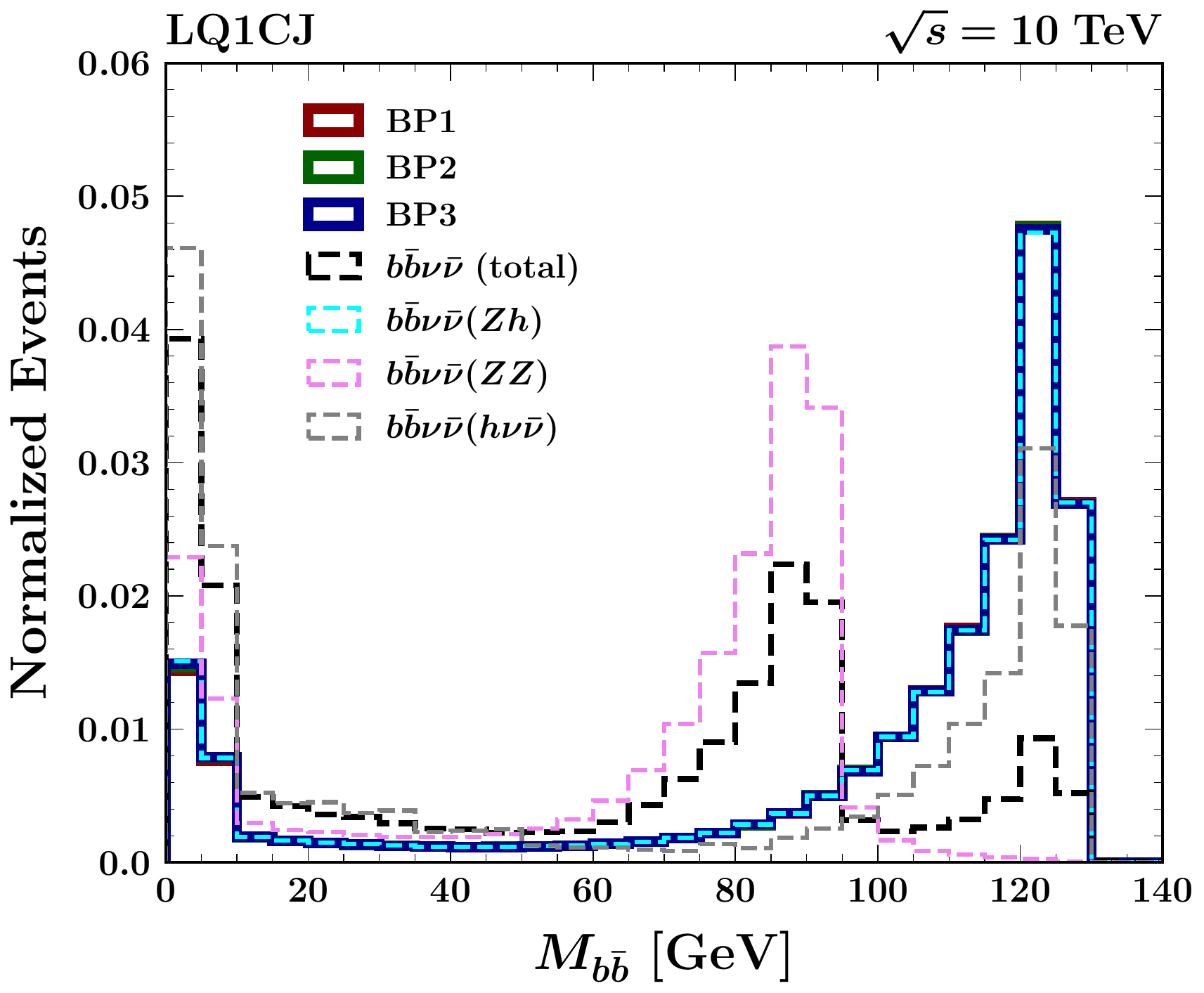}}~
        \hspace{0.3cm}
        \subfloat[\label{subfig:inv_mass_distribution_ncjet2_Ecm_10tev}]{\includegraphics[width=0.5\columnwidth]{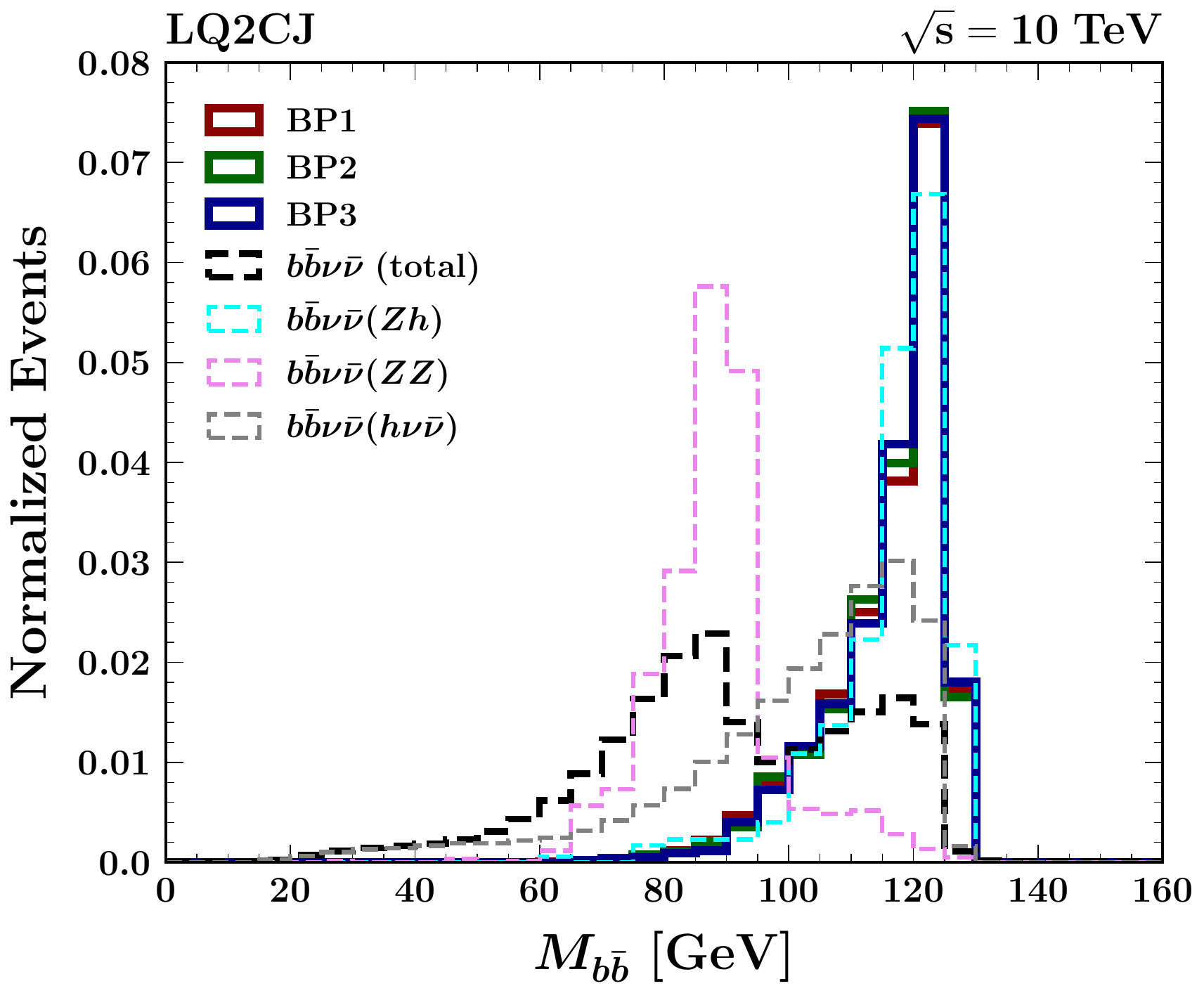}}
    }
    \caption{Invariant mass ($M_{b\bar{b}}$) distribution at $\sqrt{s}=10$ TeV in the (a) {LQ1CJ and (b) LQ2CJ} final state for both signal and background events.}
    \label{fig:inv_mass_Ecm_10tev}
\end{figure}

In this subsection we discuss few useful kinematic variables and choice of selection cuts. 
The next stage of our cut based analysis utilizes both the missing energy and invariant mass of a jet (or di-jet system) to reduce/eliminate the SM backgrounds. 
\begin{itemize}
\item \textbf{Missing energy ($\slashed{E}$):} In collider experiments, missing energy corresponds to the total energy carried by the invisible particles which remain unobserved at the detector level. In a particular final state consisting of both visible and invisible particles, using the energy conservation we can find the missing energy of the system as 
\begin{eqnarray}
  \slashed{E} = \sqrt{s} - E_{\text{vis}}
  \label{eq:Emiss_1}
\end{eqnarray}
where, $\sqrt{s}$ is center-of-mass energy of the collisions and $E_{\text{vis}}$ is the total energy of all the visible final state particles. 

In collider investigations, the visible particles in general consist of the hadrons clustered as jets, isolated photons and leptons, and soft unclustered components. Thus, the expression of missing energy at the detector level can be written as 
\begin{eqnarray}
  \slashed{E} = \sqrt{s} - \left( 
\sum_{i \in \text{iso leps}} E(i) +
\sum_{i \in \text{iso phos}} E(i) +
\sum_{i \in \text{jets}} E(i) +
\sum_{i \in \text{soft}} E(i)
\right)
\label{eq:Emiss_2}
\end{eqnarray}
In Fig.~\ref{fig:missing_energy_distribution_Ecm_3tev}, we plot the missing energy distributions for the signal events for various benchmark scenarios corresponding to two different final states at center-of-mass energy $\sqrt{s}=3~\text{TeV}$. The corresponding distributions at 10 TeV center-of-mass energy are shown in Fig.~\ref{fig:missing_energy_distribution_Ecm_10tev}. The contribution to the missing energy for the signal events is due to the invisible dark photon and the distribution is expected to peak $\sim {\sqrt{ \rm s}/2}$ (neglecting the finite mass effect) in a $2 \to 2$ process. However, due to the presence of additional neutrinos coming from $B$ meson decays the missing energy distribution extends beyond ${\sqrt{ \rm s}/2}$. {Since the SM $Zh$ and $ZZ$ form a very small fraction of the total background cross section, the normalized missing energy distribution for the total SM background follows that of the $h \nu \bar{\nu}/Z \nu \bar{\nu} $ process.
The pair of neutrinos in the $h\nu\bar{\nu}$ or $Z \nu \bar{\nu}$ processes carries maximum share of the available center-of-mass energy  leading to a peak around $\sqrt{s}$ in the missing energy distribution.  This is reflected in the missing energy distributions for individual SM backgrounds and the total SM background as depicted in the same figures.}
Therefore, missing energy variable is an extremely useful discriminator between the signal and the background.
\begin{figure}[H]
    \centering
    \resizebox{\columnwidth}{!}
    {
        \subfloat[\label{subfig:missing_energy_distribution_ncjet1_Ecm_3tev}]{\includegraphics[width=0.5\columnwidth]{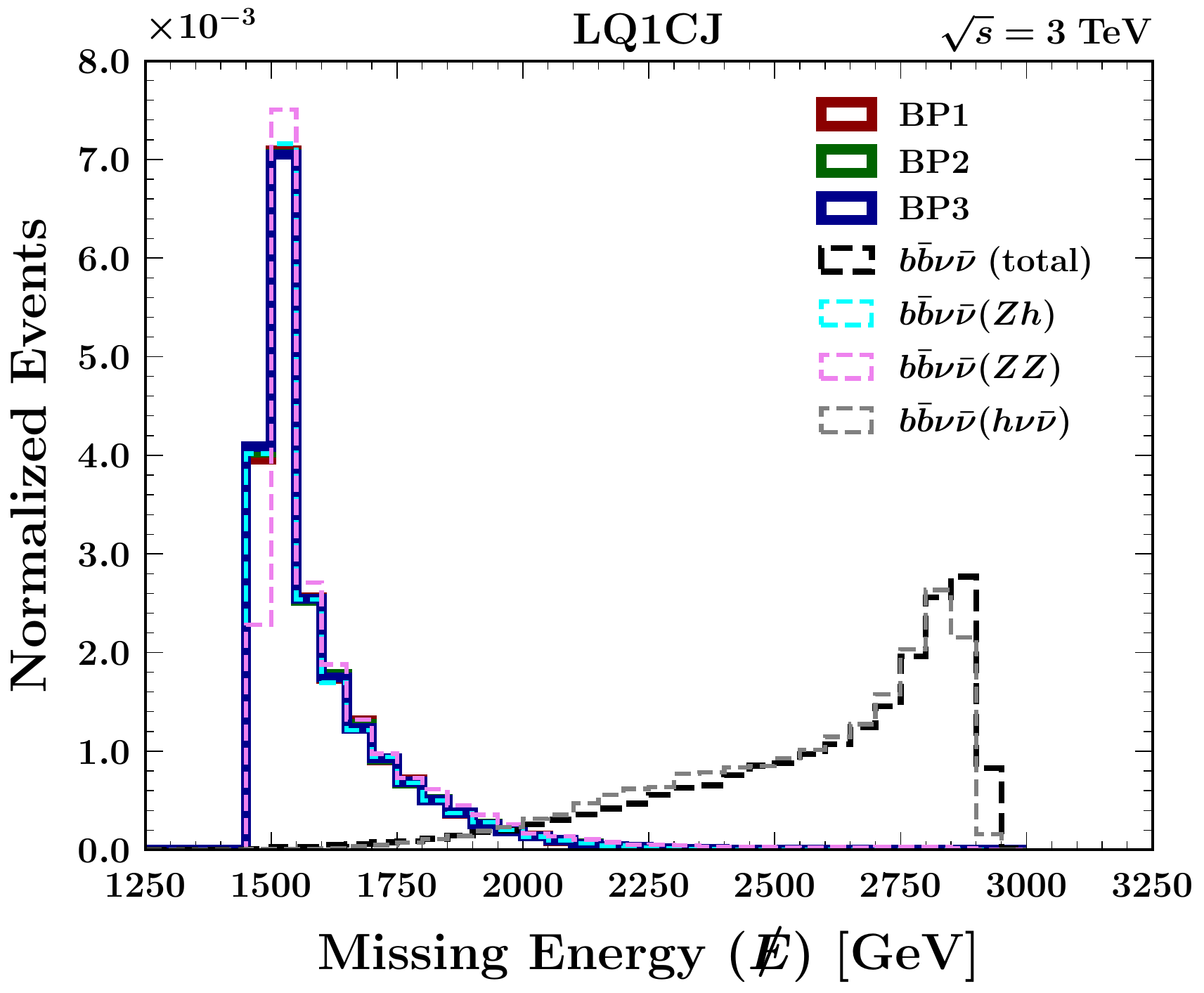}}~

        \hspace{0.3cm}
        \subfloat[\label{subfig:missing_energy_distribution_ncjet2_Ecm_3tev.pdf}]{\includegraphics[width=0.5\columnwidth]{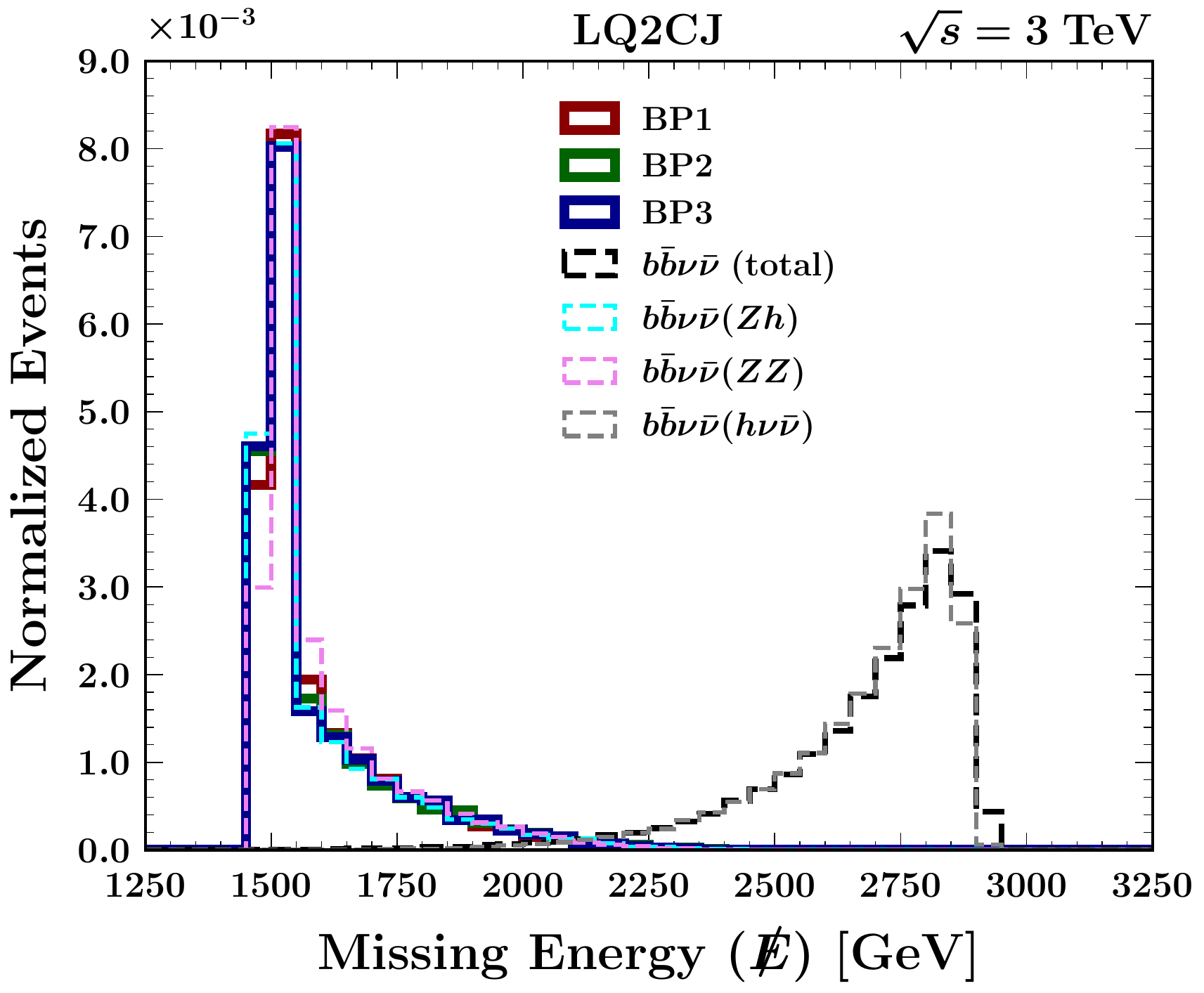}}~
       
    }
    \caption{Missing energy ($\slashed{E}$) distributions at $\sqrt{s}=3$ TeV in the {(a) LQ1CJ and (b) LQ2CJ} final states for both signal and SM background events.}
    \label{fig:missing_energy_distribution_Ecm_3tev}
\end{figure}
\begin{figure}[H]
    \centering
    \resizebox{\columnwidth}{!}
    {        \subfloat[\label{subfig:Ecm_10tev_set1_missing_energy_distribution_ncjet1}]{\includegraphics[width=0.5\columnwidth]{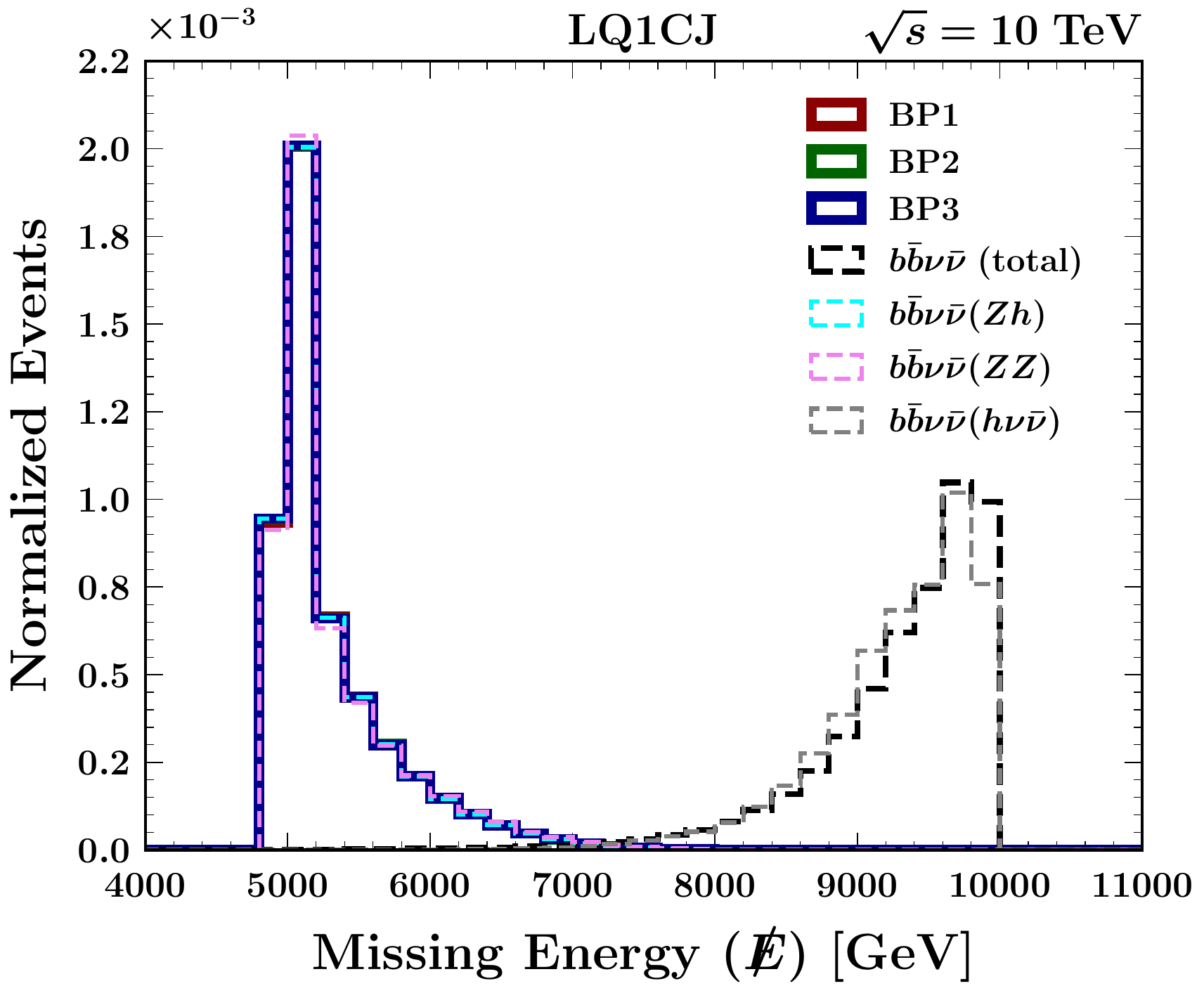}}
    \hspace{0.3cm}
    \subfloat[\label{subfig:Ecm_10tev_set1_missing_energy_ncjet2_distribution}]{\includegraphics[width=0.5\columnwidth]{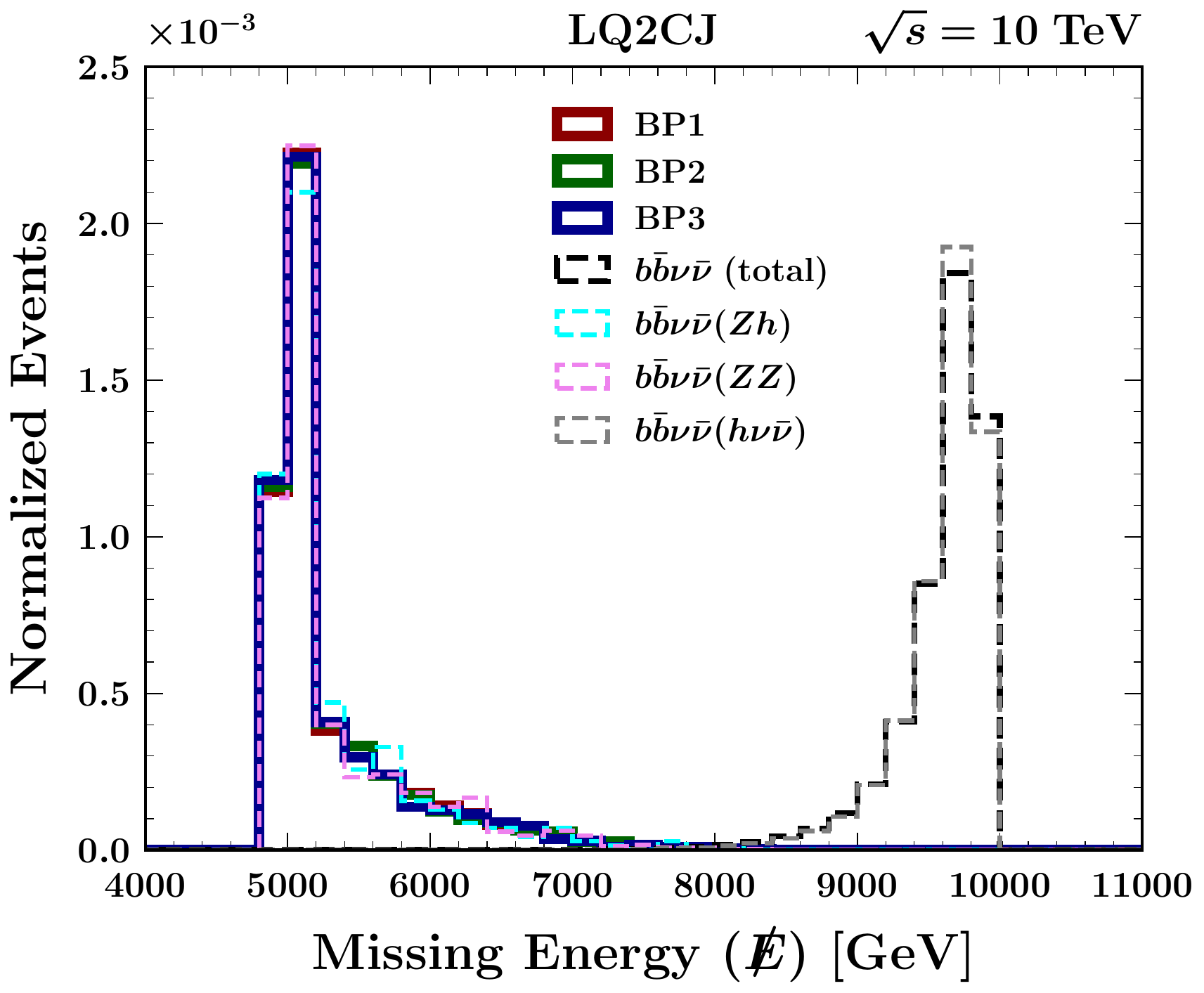}}
    }
    \caption{Missing energy ($\slashed{E}$) distributions at $\sqrt{s}=10$ TeV in the {(a) LQ1CJ and (b) LQ2CJ} final states for both signal and SM background events.}
    \label{fig:missing_energy_distribution_Ecm_10tev}
\end{figure}
\item  {\bf $\bm {b\bar{b}}$ invariant mass ($M_{b\bar{b}}$) }:
In our analysis the invariant mass associated with the $b\bar{b}$ system is extracted in two different ways using the following formula
  \begin{eqnarray}
    M_{12}^2 = (p_1 + p_2)^2 = (E_1 + E_2)^2 - |\vec{p_1} + \vec{p_2}|^2
  \end{eqnarray}
where $p_1 = (E_1, |\vec{p_1}|)$ and $p_2 = (E_2,|\vec{p_2}|)$ are the four momentum associated with the two sub-jets (jets) in the final state comprising one (two) central-jet. 
In events with {\it exactly one central-jet} we combine the four momentum of the two sub-jets inside a boosted $\rm{CA}_{1.2}$ jet to get the invariant mass distribution. Whereas in the {case of {\it exactly two central-jet events} ,} we combine the four momentum of two AK4 jets in order to get the invariant mass associated with the di-jet system. The corresponding distributions in these two final states are plotted in Fig.~\ref{fig:inv_mass_Ecm_3tev} and \ref{fig:inv_mass_Ecm_10tev}. One can see that for signal events this distribution has a peak near the Higgs mass. {However, the same for the SM backgrounds} peak at varying locations, namely, at the mass of the Higgs and $Z$ boson as expected. One also sees a peak at a very low invariant mass which has origin in the final state radiation and hadronization effect{s} from one of the bottom quarks only, and may give rise to jet substructure.
{The $b \bar{b}$ invariant mass distribution in the LQ2CJ final state has two clear peaks at the values of $Z$ and Higgs mass.  
% \textcolor{red}{This is due to the presence of $ZZ,~ Z \nu \bar{\nu}$ component as well as $hZ$ and $h\nu\bar{\nu}$ components of the total SM background. }
% 
This is due to the presence of an on-shell $Z$ and an on-shell Higgs, decaying to a bottom quark pair, in the total SM background.}
The tail in all these invariant mass distributions at values lower than the peak values is due to the missing momentum component associated with the invisible neutrinos produced in $B$ decays.
\end{itemize}

 % \textbf{Subjet invariant mass distribution}

 One can see that the {\it normalized} signal distributions are mostly insensitive to the intermediate heavy muon mass.
While presenting the kinematic distributions for the signal benchmark points (BPs) we have chosen (BP1, BP2 and BP3) as our representative BPs corresponding to three different heavy muon masses ($1,~2~\text{and}~3$ TeV).  Similar distributions exist for other signal benchmark points \{BP4--BP6\} corresponding to $m_{\gamma_d} = 50~\text{GeV}$ as well. However, the effect of the dark photon mass on the normalized distributions are not significant enough. Hence, we choose not to display them here as they closely resemble the distributions already presented.

{The study of SM $h\nu\bar{\nu}/Z\nu\bar{\nu}$ is useful to understand the behavior of the total background and its rates at various stages of our kinematic selections. However, to get an idea about the behavior of the resonant $\nu \bar\nu$ component of the total background, we separately analyze the SM $Zh/ZZ$ processes. For significance estimate, we only use the total $b \bar{b} \nu \bar{\nu}$ background which is obtained by summing over all possible Feynman diagrams leading to $b \bar b \nu \bar{\nu}$ final state and is not equal to the algebraic sum of individual SM background processes used to illustrate various kinematic distributions and rates.}

\subsection{Cuts}
\label{subsec:cuts}
To separate the signal from the corresponding SM background, we have used the following selection cuts on the observables associated with the individual final states.

\begin{itemize}

    \item $\slashed{E} < \slashed{E}_0$, where $\slashed{E}_0 = 2~(7)$ TeV for $\sqrt{s} = 3~(10)$ TeV.
    \item Jet substructure requirement
    \item $M_{b\bar{b}} > M_0$, where $M_0=95$ GeV.
 
\end{itemize}

%%% Sec-4: Results %%
\section{Results}
\label{sec:results}
In this section, we present the results of our analysis of the process Higgs production in association with a dark photon at the future muon collider. In the previous section,  we have discussed the production rate of $\mu^+ \mu^- \to h \gamma_d$ and its comparison with the SM $Zh$ production 
in presence of the portal vector-like lepton. It also includes the choice of various kinematic variables and selection criteria. In Table~\ref{tab:cutflow_table_BPs_Ecm_3tev} and \ref{tab:cutflow_table_BPs_Ecm_10tev}, we tabulate the production cross sections for the signal process as well as the effects of our selection criteria on it for various choices of our benchmark points (BP1-BP6) in the {{\it leptonically quiet exactly one and two central-jet(s)} final states} at $\sqrt{s} =3$ TeV and $10$ TeV, respectively.
 The signal cross sections mentioned in these tables assume $h\to b \bar{b}$ decay branching ratio (BR) to be $100$\% .
{The effect of cut-flow on the individual SM background processes ($ Zh,~ ZZ,~ h \nu\bar{\nu},~ Z \nu \bar{\nu}$) and the total SM background in the $b\bar{b} \nu \bar{\nu}$ final state are presented in Table~\ref{tab:cutflow_table_backgrounds_Ecm_3tev} and Table~\ref{tab:cutflow_table_backgrounds_Ecm_10tev} at muon collider center-of-mass energies $3$ and $10$ TeV,  respectively. A significant fraction of the $\mu^+\mu^- \to h \nu \bar{\nu}$ background can be effectively reduced by rejecting events beyond a certain missing energy value ($\slashed{E}_0$) as can be seen from Fig.~\ref{fig:missing_energy_distribution_Ecm_3tev} and \ref{fig:missing_energy_distribution_Ecm_10tev}.
These cross sections for the SM backgrounds are quoted taking into account appropriate decay BRs, as applicable. }

% 

% 

%%%%%%%%%%%%%%%%%%%%%
%%% Cutflow table BPs at Ecm = 3 TeV

\begin{table}[H]
\centering
\resizebox{\columnwidth}{!}{
\begin{tabular}{lc|cccc|cccc}
% \hline
% \hline
\toprule
\toprule
% \rowfont\normalfont
BPs & Production & 
\multicolumn{4}{c|}{Cross sections (fb): {LQ1CJ} final state} &
\multicolumn{4}{c}{Cross sections (fb): {LQ2CJ} final state} \\ \cline{3-10}
% \rowfont\normalfont
 & cross-section & Basic & $\slashed{E}$ cut & Jet substructure & $M_{b\bar{b}}$ cut
 & Basic & $\slashed{E}$ cut & Jet substructure & $M_{b\bar{b}}$ cut \\
 % \rowfont\normalfont
 & (fb) & Cuts & ($\slashed{E} < \slashed{E}_0$) & requirement  & ($M_{b\bar{b}} > M_0$)
 & Cuts & ($\slashed{E} < \slashed{E}_0$) & requirement & ($M_{b\bar{b}} > M_0$) \\
% \hline
% \hline
\midrule
\midrule

BP1 & $11.1$ & $9.70$ & $9.51$ & $9.39$ & $7.36$ & $1.42$ & $1.38$ & $1.30$ & $1.24$ \\
BP2 & $22.8$ & $20.7$ & $20.3$ & $20.1$ & $15.7$ & $2.03$ & $1.97$ & $1.84$ & $1.76$ \\
BP3 & $0.064$ & $0.059$ & $0.058$ & $0.057$ & $0.044$ & $0.005$ & $0.005$ & $0.005$ & $0.005$ \\
BP4 & $47.2$ & $41.1$ & $40.3$ & $39.8$ & $31.3$ & $5.97$ & $5.79$ & $5.49$ & $5.24$ \\
BP5 & $114$ & $104$ & $102$ & $101$ & $78.8$ & $9.95$ & $9.63$ & $8.97$ & $8.65$ \\
BP6 & $0.436$ & $0.401$ & $0.393$ & $0.388$ & $0.303$ & $0.035$ & $0.034$ & $0.031$ & $0.030$ \\
\hline
\hline
\end{tabular}
}
\caption{The effect of cut-flow on the signal cross section for various representative benchmark points in two different final states {( LQ1CJ and LQ2CJ)} at $\sqrt{s}=3$~TeV assuming BR($h \to b \bar{b}$) $=1$.  
}
\label{tab:cutflow_table_BPs_Ecm_3tev}
\end{table}

%%%%%%%%%%%%%%%%%%%%%%%%
%%% Cutflow table backgrounds at Ecm=3 TeV

\begin{table}[H]
\centering
\resizebox{\columnwidth}{!}{
\begin{tabular}{lc|cccc|cccc}
% \hline
% \hline
\toprule
\toprule
% \rowfont\normalfont
Background & Production & 
\multicolumn{4}{c|}{Cross sections (fb): {LQ1CJ} final state} &
\multicolumn{4}{c}{Cross sections (fb): {LQ2CJ} final state} \\ \cline{3-10}
% \rowfont\normalfont
processes & cross-section & Basic & $\slashed{E}$ cut & Jet substructure & $M_{b\bar{b}}$ cut
 & Basic & $\slashed{E}$ cut & Jet substructure & $M_{b\bar{b}}$ cut \\
 % \rowfont\normalfont
 & (fb) & Cuts & ($\slashed{E} < \slashed{E}_0$) & requirement  & ($M_{b\bar{b}} > M_0$)
 & Cuts & ($\slashed{E} < \slashed{E}_0$) & requirement & ($M_{b\bar{b}} > M_0$) \\
% \hline
% \hline
\midrule
\midrule

$b \bar{b} \nu \bar{\nu}~(Zh)$ & $0.178$ & $0.164$ & $0.161$ & $0.160$ & $0.124$ &
$0.014$ & $0.013$ & $0.012$ & $0.012$ \\
$b \bar{b} \nu \bar{\nu}~(ZZ)$ & $1.45$ & $0.495$ & $0.473$ & $0.368$ & $0.014$
& $0.127$ & $0.124$ & $0.120$ & $0.007$ \\
{$b \bar{b} \nu \bar{\nu}~(h\nu\bar{\nu})$} & {$318$} & {$96.2$} & {$4.44$} & {$0.745$} & {$0.420$} &
{$187$} & {$1.38$} & {$1.03$} & {$0.914$} \\
{$b \bar{b} \nu \bar{\nu}~(Z\nu\bar{\nu})$} & {$305$} & {$121$} & {$5.81$} & {$3.07$} & {$0.113$} &
{$152$} & {$2.73$} & {$2.47$} & {$0.168$} \\
$\bm{ b \bar{b} \nu \bar{\nu}}$ {\bf (tot)} & $\bm{641}$ & $\bm{222}$ & $\bm{10.9}$ & $\bm{4.00}$ & $\bm{0.538}$ & $\bm{351}$ & $\bm{4.75}$ & $\bm{3.97}$ & $\bm{1.44}$ \\
\hline
\hline
\end{tabular}
}
\caption{{ The effect of cut-flow on the cross-sections for various SM background processes and the {\bf total $b\bar{b} \nu \bar\nu$} background  in two different final states ( LQ1CJ and LQ2CJ) at $\sqrt{s}=3$~TeV. Here the total background cross-sections quoted after each selection cut are not equal to the algebraic sums of the corresponding cross-sections of the individual background processes mentioned above.}
}
\label{tab:cutflow_table_backgrounds_Ecm_3tev}
\end{table}

%%%%%%%%%%%%%%%%%%%%%%%%%%%%%%%%%%%%%%%%
%% BP cut-flow table for Ecm = 10 tev %%

\begin{table}[H]
\centering
\resizebox{\columnwidth}{!}{
\begin{tabular}{lc|cccc|cccc}
% \hline
% \hline
\toprule
\toprule
% \rowfont\normalfont
BPs & Production & 
\multicolumn{4}{c|}{Cross sections (fb): {LQ1CJ} final state} &
\multicolumn{4}{c}{Cross sections (fb): {LQ2CJ} final state} \\ \cline{3-10}
% \rowfont\normalfont
 & cross-section & Basic & $\slashed{E}$ cut & Jet substructure & $M_{b\bar{b}}$ cut
 & Basic & $\slashed{E}$ cut & Jet substructure & $M_{b\bar{b}}$ cut \\
 % \rowfont\normalfont
 & (fb) & Cuts & ($\slashed{E} < \slashed{E}_0$) & requirement  & ($M_{b\bar{b}} > M_0$)
 & Cuts & ($\slashed{E} < \slashed{E}_0$) & requirement & ($M_{b\bar{b}} > M_0$) \\
% \hline
% \hline
\midrule
\midrule

BP1 & $2.66$ & $2.55$ & $2.53$ & $2.49$ & $1.84$ &
$0.027$ & $0.026$ & $0.024$ & $0.023$ \\
BP2 & $10.9$ & $10.8$ & $10.7$ & $10.5$ & $7.72$ &
$0.040$ & $0.039$ & $0.036$ & $0.035$ \\
BP3 & $0.056$ & $0.056$ & $0.056$ & $0.055$ & $0.040$ &
$\sim 10^{-4}$ & $\sim 10^{-4}$ & $\sim 10^{-4}$ & $\sim 10^{-4}$ \\
BP4 & $11.4$ & $10.9$ & $10.8$ & $10.6$ & $7.85$ &
$0.114$ & $0.112$ & $0.102$ & $0.099$ \\
BP5 & $55.0$ & $54.5$ & $54.0$ & $53.1$ & $38.9$ &
$0.207$ & $0.203$ & $0.184$ & $0.180$ \\
BP6 & $0.388$ & $0.386$ & $0.383$ & $0.376$ & $0.276$ &
$\sim 6 \cdot 10^{-4}$ & $\sim 6 \cdot 10^{-4}$ & $\sim 6 \cdot 10^{-4}$ & $\sim 6 \cdot 10^{-4}$ \\
\hline
\hline
\end{tabular}
}
\caption{ The effect of cut-flow on the signal cross section for various representative benchmark points in two different final states {( LQ1CJ and LQ2CJ)} at $\sqrt{s}=10$~TeV assuming BR($h \to b \bar{b}$) $=1$.
}
\label{tab:cutflow_table_BPs_Ecm_10tev}
\end{table}

%%%%%%%%%%%%%%%%%%%%%%%%%%%%%%%%
%%% Cutflow table backgrounds at Ecm=10 TeV
% 

\begin{table}[H]
\centering
\resizebox{\columnwidth}{!}{
\begin{tabular}{lc|cccc|cccc}
% \hline
% \hline
\toprule
\toprule
% \rowfont\normalfont
Background & Production & 
\multicolumn{4}{c|}{Cross sections (fb): {LQ1CJ} final state} &
\multicolumn{4}{c}{Cross sections (fb): {LQ2CJ} final state} \\ \cline{3-10}
% \rowfont\normalfont
processes & cross-section & Basic & $\slashed{E}$ cut & Jet substructure & $M_{b\bar{b}}$ cut
 & Basic & $\slashed{E}$ cut & Jet substructure & $M_{b\bar{b}}$ cut \\
 % \rowfont\normalfont
 & (fb) & Cuts & ($\slashed{E} < \slashed{E}_0$) & requirement  & ($M_{b\bar{b}} > M_0$)
 & Cuts & ($\slashed{E} < \slashed{E}_0$) & requirement & ($M_{b\bar{b}} > M_0$) \\
% \hline
% \hline
\midrule
\midrule

$b \bar{b} \nu \bar{\nu}~(Zh) $ & $0.016$ & $0.016$ & $0.016$ & $0.015$ & $0.011$ &
$< 10^{-5}$ & $< 10^{-5}$ & $< 10^{-5}$ & $< 10^{-5}$ \\
$b \bar{b} \nu \bar{\nu}~(ZZ)$ & $0.183$ & $0.045$ & $0.044$ & $0.043$ & $0.002$ &
$\sim 2 \cdot 10^{-4}$ & $\sim 2 \cdot 10^{-4}$ & $\sim 2 \cdot 10^{-4}$ & $< 10^{-4}$ \\
{$b \bar{b} \nu \bar{\nu}~(h\nu\bar{\nu})$} & {$540$} & {$167$} & {$0.718$} & {$0.583$} & {$0.470$} &
{$245$} & {$0.086$} & {$0.086$} & {$0.081$} \\
{$b \bar{b} \nu \bar{\nu}~(Z\nu\bar{\nu})$} & {$526$} & {$197$} & {$2.80$} & {$2.74$} & {$0.158$} &
{$204$} & {$0.111$} & {$0.100$} & {$0.016$} \\
$\bm{b \bar{b} \nu \bar{\nu}}$ {\bf(tot)} & $\bm{1101}$ & $\bm{373}$ & $\bm{4.03}$ & $\bm{3.53}$ & $\bm{0.673}$ & $\bm{466}$ & $\bm{0.488}$ & $\bm{0.427}$ & $\bm{0.310}$ \\
\hline
\hline
\end{tabular}
}
\caption{{ The effect of cut-flow on the cross-sections for various SM background processes and the {\bf total $b \bar{b} \nu \bar{\nu}$ background}  in two different final states ( LQ1CJ and LQ2CJ) at $\sqrt{s}=10$~TeV. Here the total background cross-sections quoted after each selection cut are not equal to the algebraic sums of the corresponding cross-sections of the individual background processes mentioned above.}
}
\label{tab:cutflow_table_backgrounds_Ecm_10tev}
\end{table}

It is clear from the Tables~\ref{tab:cutflow_table_BPs_Ecm_3tev} and \ref{tab:cutflow_table_BPs_Ecm_10tev} that at muon collider center-of-mass energies the signal events dominantly contribute in the {LQ1CJ }final state than the other one because of the fact that it is a $2 \to 2 $ process. Moreover, the contribution of signal events in {this} final state increases with increasing mass of the {heavy muon partner} in the $t/u$ channel. This is evident if we compare the event rate after basic selection criteria for various benchmark points (\{BP1-BP3\} or \{BP4-BP6\}) which corresponds to {heavy muon mass} in the range \{1-3\} TeV. This feature is also visible in Fig.~\ref{fig:jet_multiplicity_distribution}.

% 

%cutflow effect
The effect of our selection criteria on the signal events is less severe in both the final states as can be seen from the cut-flow tables.
 Whereas one has a factor of $\sim \frac{1}{400}$ and $\frac{1}{550} $ suppression in the SM background rate in the {\it exactly one central-jet} final state after applying the combination of cuts mentioned in section~\ref{subsec:cuts} at $\sqrt{s} = 3 $ TeV and $10 $ TeV, respectively. This factor for the {\it exactly two central-jet} final state is $\sim \frac{1}{250}$ for $\sqrt{s}=3$ TeV and even larger at 10 TeV center-of-mass energy. The significant effect on the background comes from the missing energy cut as already explained in section~\ref{sec:collider_analysis}. 

 The cut-flow tables corresponding to the signal events also suggest that the effect of the dark photon mass on the {\it kinematics}, in particular, on the cut efficiencies is very small as long as $m_{\gamma_d} \ll \sqrt{s}$. However, it plays a significant role in the estimation of the cross section as it depends on values the of dark photon mass as well as the dark gauge coupling.

%Significance

In Table~\ref{tab:significance_table}, we quote the signal significances for $3$ TeV and $10$ TeV muon collider center-of-mass energies for various choices of our benchmark points for two different final states assuming an integrated luminosity of $100$ fb$^{-1}$. To estimate the signal significance ($\sigma$) and set a $2 \sigma$ exclusion limit we have used the following formula \cite{Cowan:2010js,Kumar:2015tna,Bhattiprolu:2020mwi}

\bea
\sigma \;=\; \sqrt{\,2 \left( S - B \ln\!\left(1 + \frac{S}{B}\right) \right)} .
\eea

{where $S$ and $B$ refer to the number of signal and background events obtained after applying the selection criteria mentioned earlier at a given integrated luminosity. }

While estimating the significance we have also taken into account the appropriate branching ratio ($57\%$) for $h \to b \bar{b}$ decay to get the actual number of signal events. { The number of background events $B$ is estimated using the total $b \bar{b} \nu \bar\nu $ event sample, and not adding the contributions of the individual background processes.}

%%%%%%%%%%%%%%%%%%%%%%%%%%%%%%%

\begin{table}[H]
\centering
\resizebox{0.7\columnwidth}{!}{
\begin{tabular}{lcc|cc}
\toprule
\toprule
 & \multicolumn{4}{c}{Signal significance ($\sigma$)} \\ 
\cmidrule(lr){2-5}
BPs & \multicolumn{2}{c|}{$\sqrt{s}=3$ TeV} & \multicolumn{2}{c}{$\sqrt{s}=10$ TeV} \\ 
\cmidrule(lr){2-5} 
 & {LQ1CJ} & {LQ2CJ} & {LQ1CJ} & {LQ2CJ} \\
\midrule
BP1 & $24.6$  & { $5.15$} & $9.12$ & $0.23$ \\
BP2 & $38.5$ & {$6.97$} & $24.6$ & $0.35$ \\
BP3 & $0.34$ & {$0.02$} & $0.27$ & $-$ \\
BP4 & $56.4$ & $16.6$ & $24.9$ & $0.96$ \\
BP5 & $92.2$ & $23.6$ & $62.9$ & $1.67$ \\
BP6 & $2.14$ & $0.14$ & $1.78$ & $-$ \\
\bottomrule
\bottomrule
\end{tabular}}
\caption{Estimated signal significance in the context of $\mu^+ \mu^- \to h \gamma_d$ process in two different final states  {( LQ1CJ and LQ2CJ)} under consideration at two different muon collider center-of-mass energies ($\sqrt{s}=3$ and $10$ TeV) assuming an integrated luminosity of 100  fb$^{-1}$.}
\label{tab:significance_table}
\end{table}

 The increase in significance for BP2 compared to BP1 is due to higher signal cross section at this benchmark point even though the mass of the pVLL in the $t/u$-channel propagator is higher. Also one has a better estimate of the signal significance for the set of BPs corresponding to dark photon mass 50 GeV compared to that for dark photon mass 90 GeV as the cross section for $h \gamma_d$ production is expected to increase with decreasing dark photon mass.

\begin{figure}[H]
    \centering
    \resizebox{\columnwidth}{!}
    {
        \subfloat[\label{subfig:2sig_3tev_set1}]{\includegraphics[width=0.5\columnwidth]{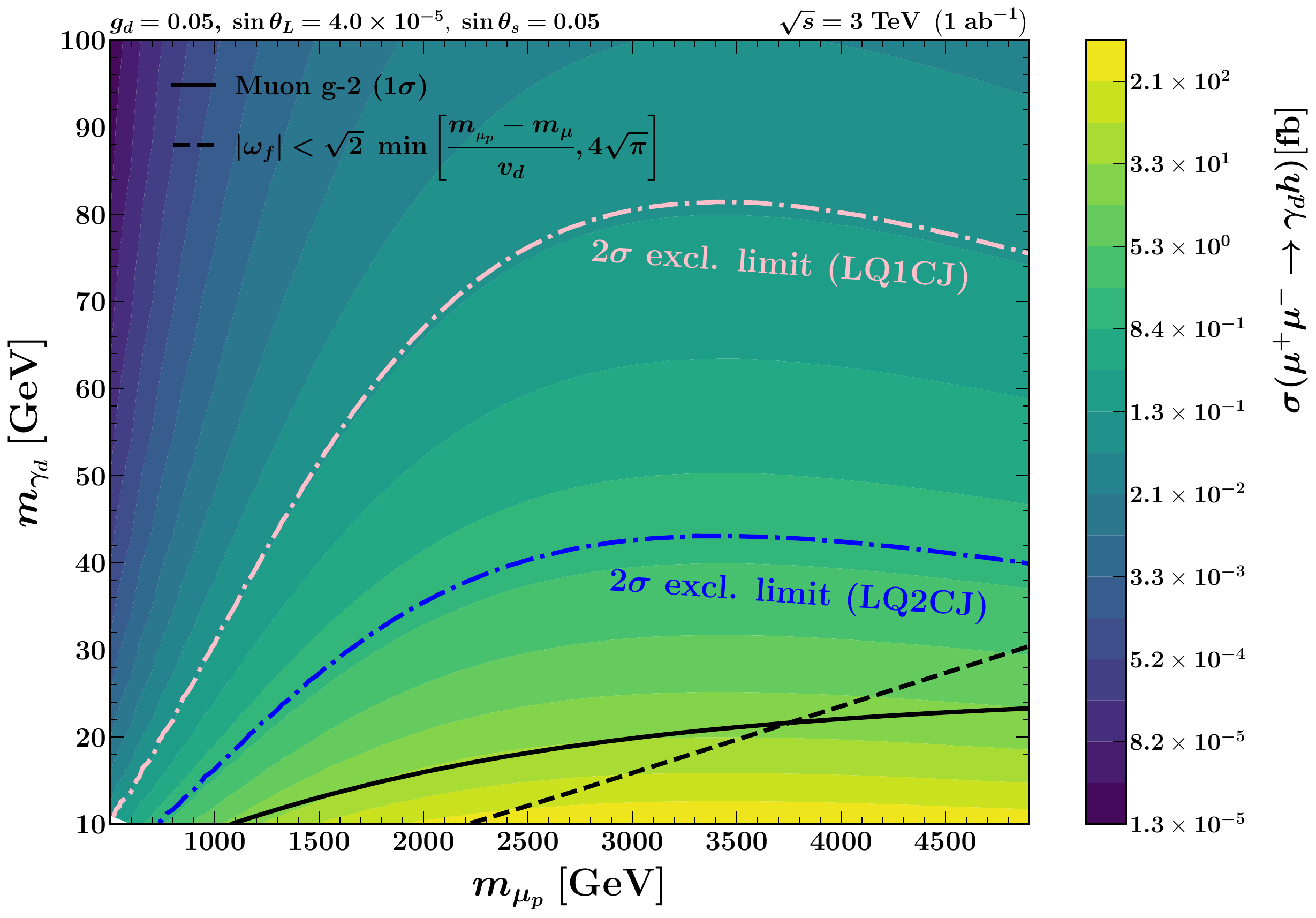}}~
        \hspace{0.3cm}\subfloat[\label{subfig:2sig_3tev_set2}]{\includegraphics[width=0.5\columnwidth]{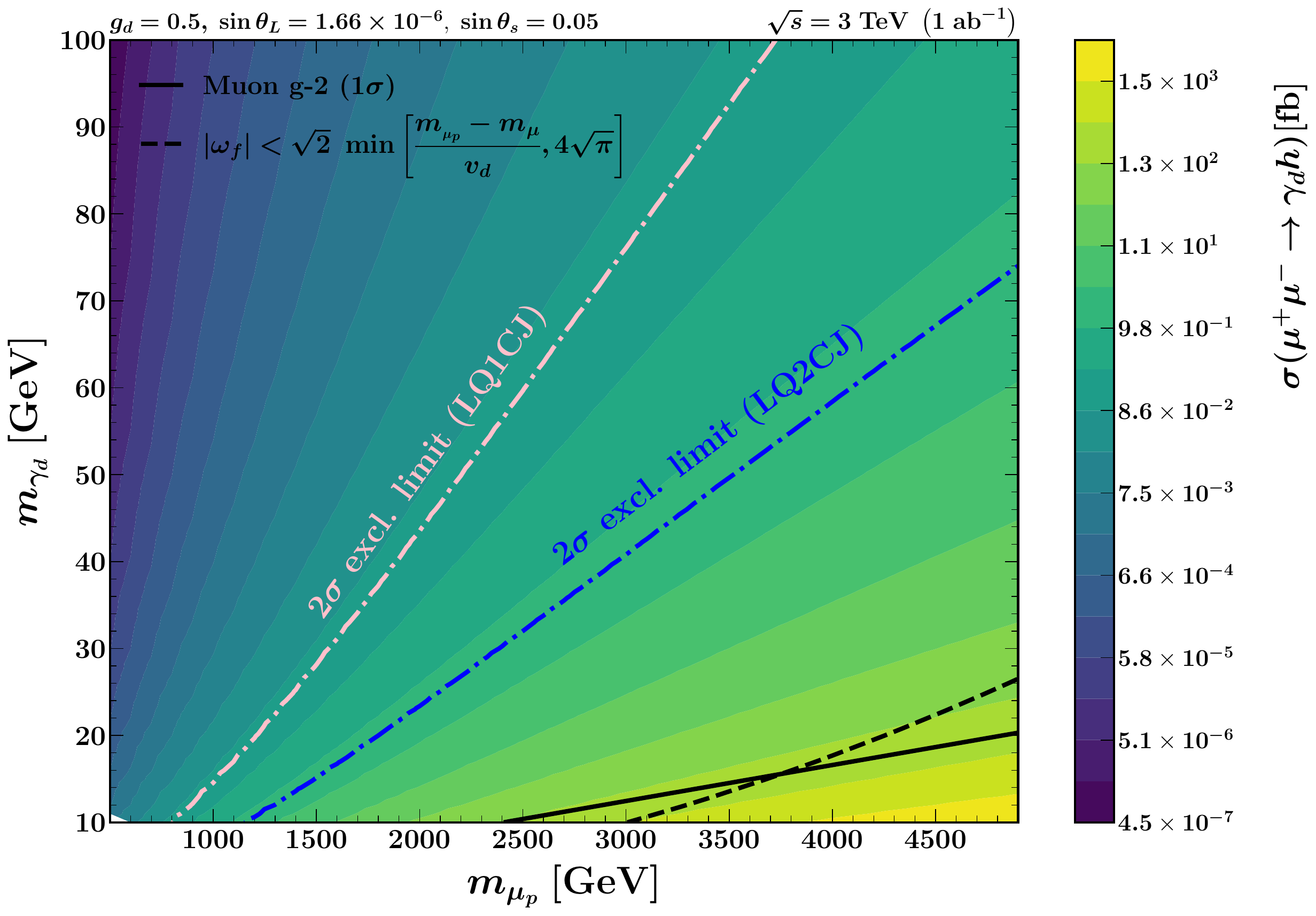}}
    }
    \caption{$2\sigma$ exclusion limit in the $m_{\gamma_d}-m_{\mu_p}$ plane for two different final states at $3$ TeV muon collider center-of-mass energy for (a) $g_d=0.05, ~\sin\theta_L = 4 \times 10^{-5}, ~\sin\theta_s = 0.05$ and (b)  $g_d=0.5,~ \sin\theta_L = 1.66 \times 10^{-6}, ~\sin\theta_s = 0.05$ assuming an integrated luminosity of 1 ab$^{-1}$.
    The exclusion limits in {LQ1CJ} (pink) and {LQ2CJ} (blue) final states are represented by the region below the dash-dot lines. The region above the dashed (black) line is allowed by the constraint represented in Eq.~\eqref{eq:perturbative_unitarity_bound} while the muon $g-2$ allowed region is represented by the one above solid (black) line.} 
    \label{fig:2sig_3tev}
\end{figure}

\begin{figure}[H]
    \centering
    \resizebox{\columnwidth}{!}
    {
        \subfloat[\label{subfig:2sig_10tev_set1}]{\includegraphics[width=0.5\columnwidth]{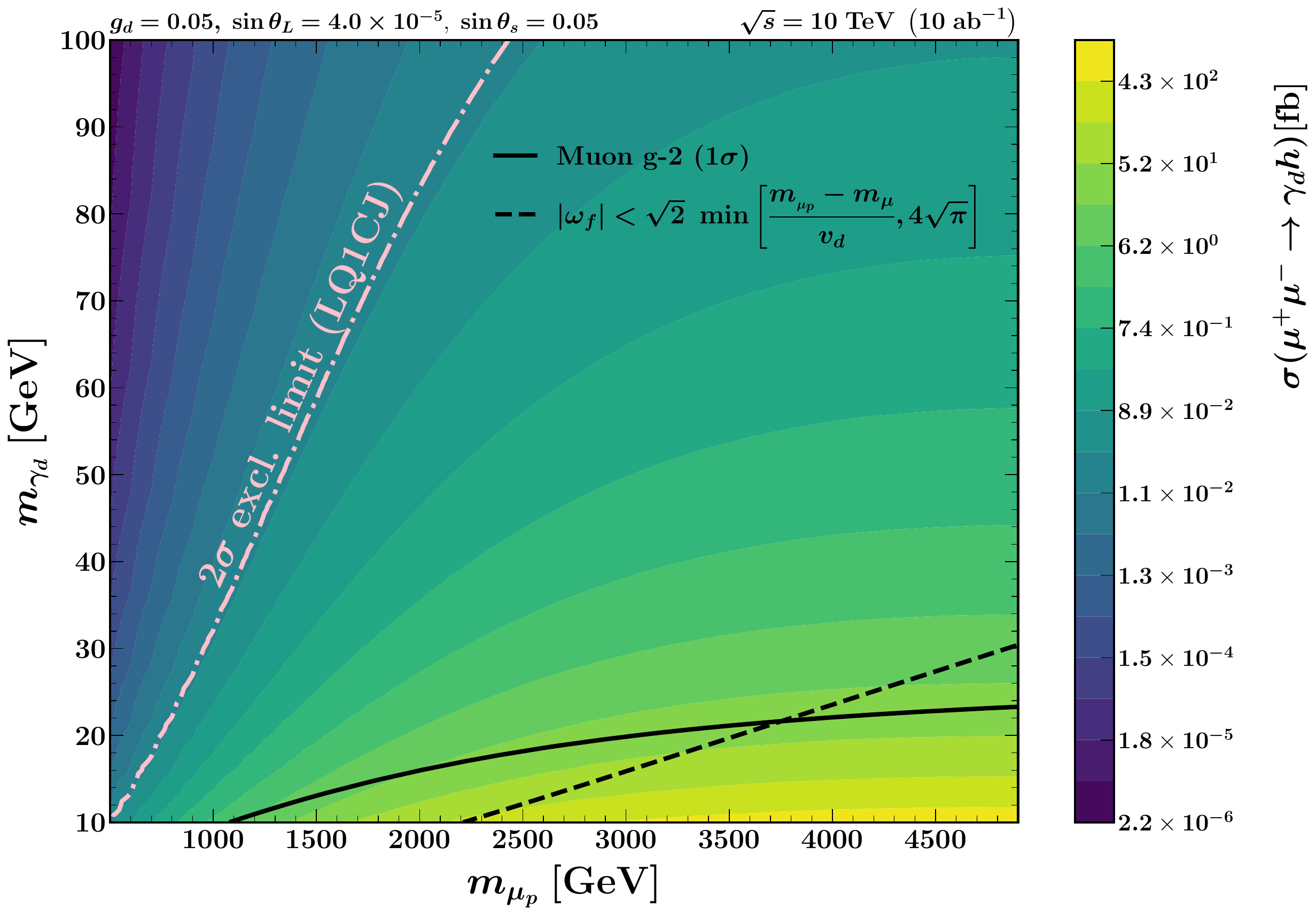}}~
        \hspace{0.3cm}\subfloat[\label{subfig:2sig_10tev_set2}]{\includegraphics[width=0.5\columnwidth]{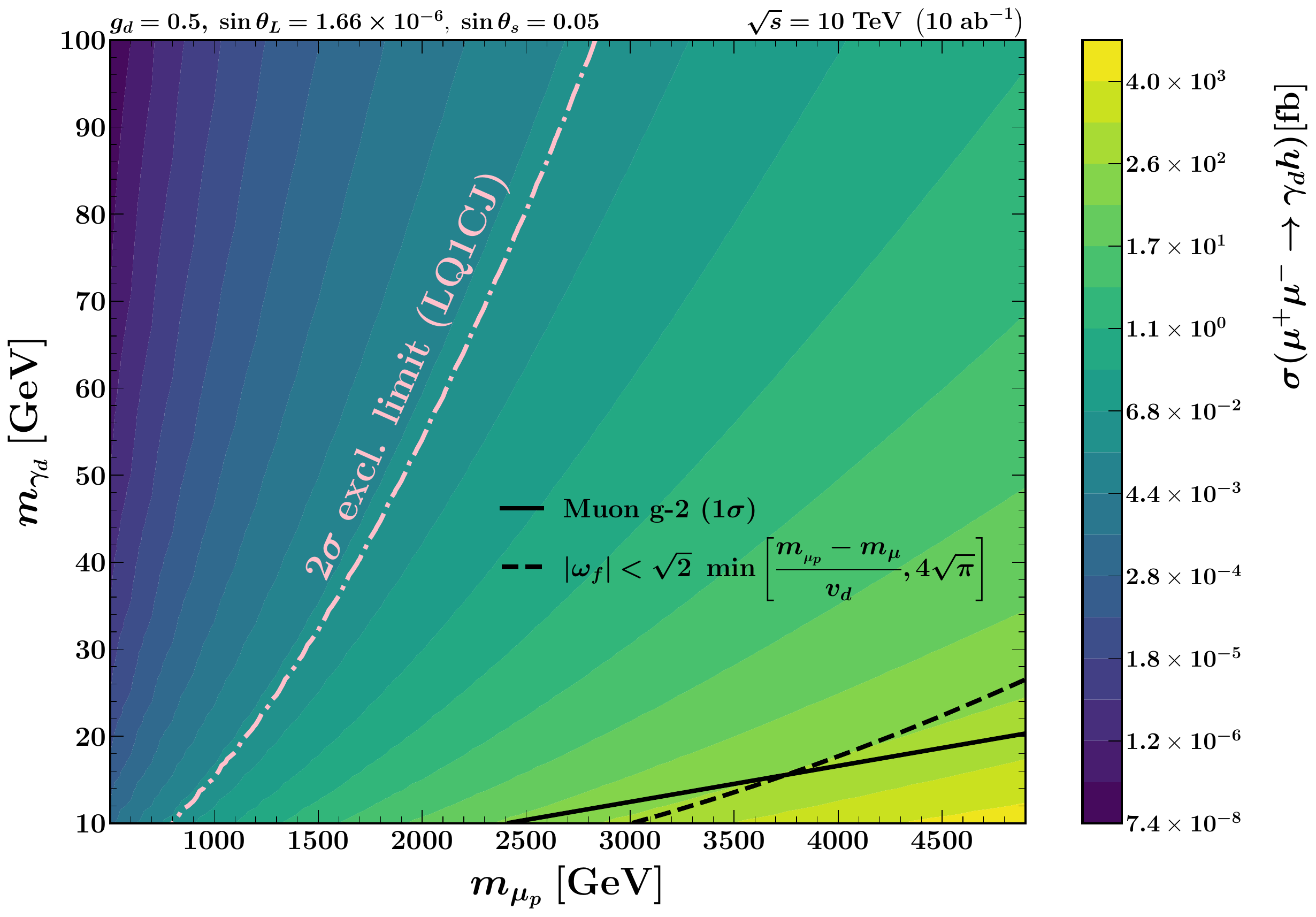}}
        }
    \caption{$2\sigma$ exclusion limit in the $m_{\gamma_d}-m_{\mu_p}$ plane in {LQ1CJ} final state at $10$ TeV muon collider center-of-mass energy for (a) $g_d=0.05, ~\sin\theta_L = 4 \times 10^{-5}, ~\sin\theta_s = 0.05$ and (b)  $g_d=0.5,~ \sin\theta_L = 1.66 \times 10^{-6}, ~\sin\theta_s = 0.05$ assuming an integrated luminosity of 10 ab$^{-1}$.
    The $2\sigma$ exclusion limit is  represented by dash-dot line (pink). The region above the dashed (black) line is allowed by the constraint represented in Eq.~\eqref{eq:perturbative_unitarity_bound} while the muon $g-2$ allowed region is represented by the one above solid (black) line.} 
    \label{fig:2sig_10tev}
\end{figure}

\begin{figure}[H]
    \centering
    \resizebox{\columnwidth}{!}
    {
        \subfloat[\label{subfig:mzp_vs_gd_Ecm_3tev_set1}]{\includegraphics[width=0.5\columnwidth]{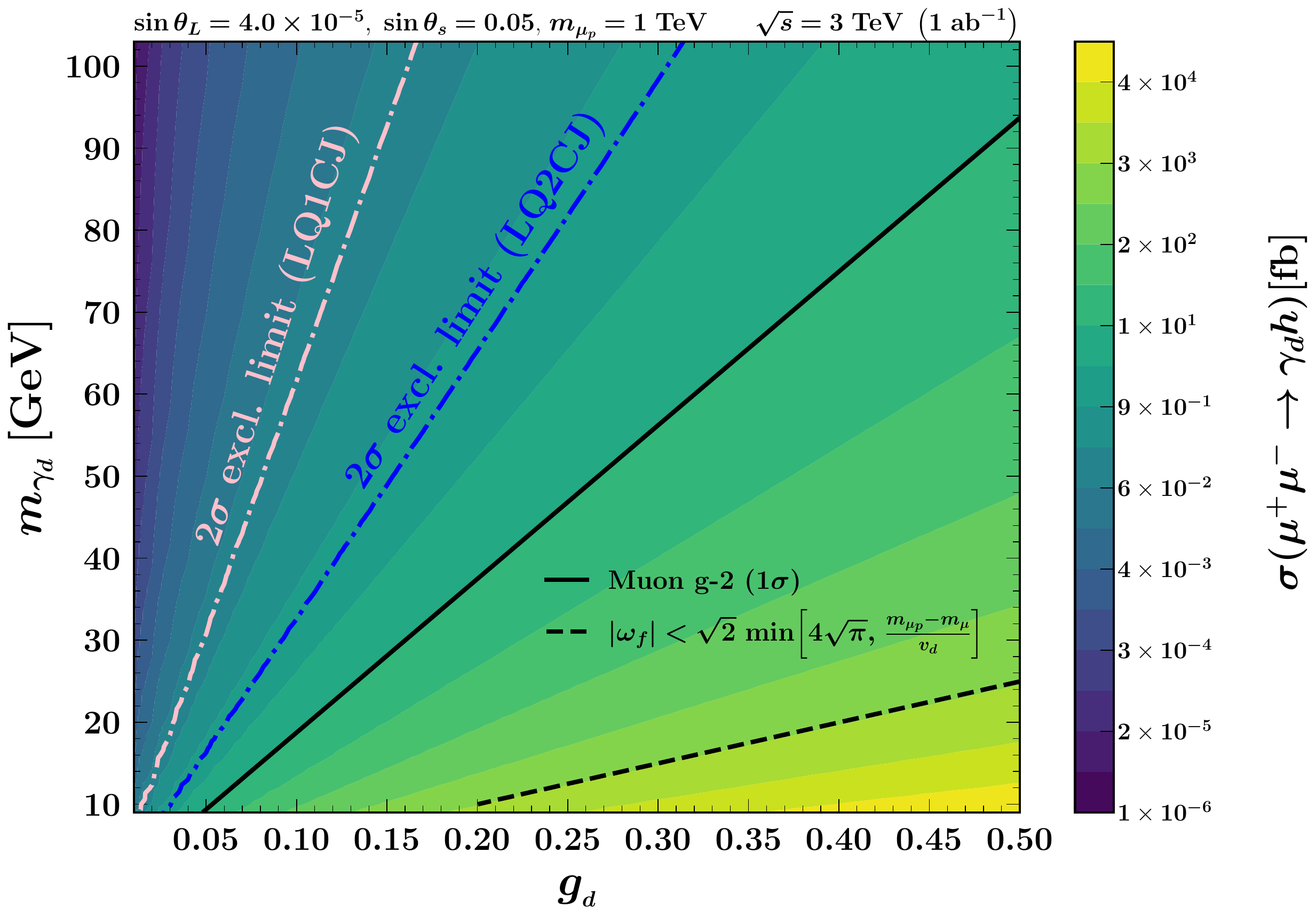}}~
        \hspace{0.3cm}\subfloat[\label{subfig:mzp_vs_gd_Ecm_3tev_set2}]{\includegraphics[width=0.5\columnwidth]{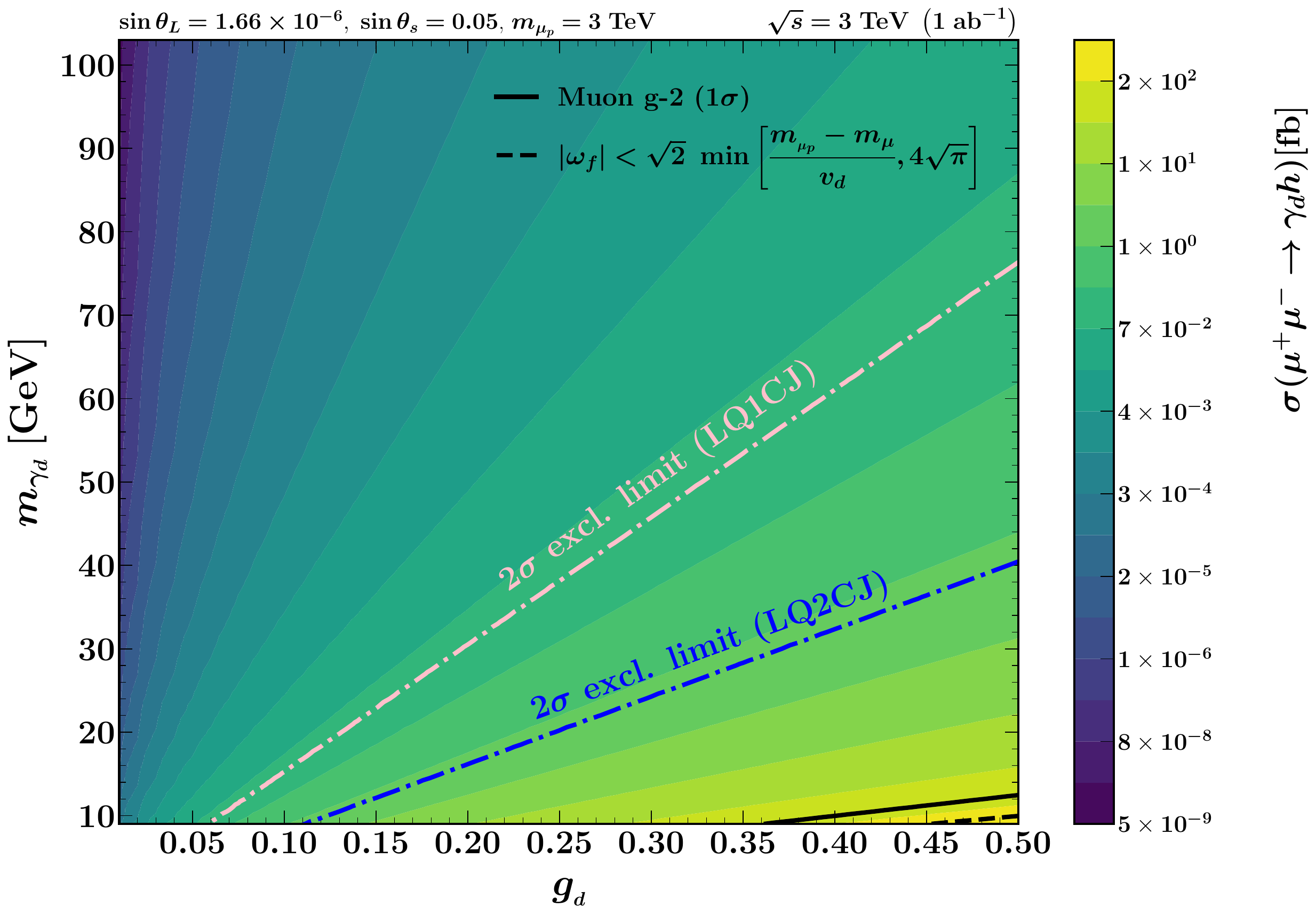}}
    }
    \caption{$2\sigma$ exclusion limit (represented by dashed-dot line) in the $g_d-m_{\gamma_d}$ plane using the data corresponding to $\sqrt{s} = 3$ TeV and an integrated luminosity of $1$ ab$^{-1}$ in the {LQ1CJ} (pink) and {LQ2CJ} (blue) final states for two different sets of $\sin\theta_L$ and $m_{\mu_p}:$ (a) $\sin\theta_L = 4 \times 10^{-5}$, $m_{\mu_p}=1$ TeV and (b) $\sin\theta_L = 1.66 \times 10^{-6}$, $m_{\mu_p}=3$ TeV. The region above the dashed (black) line is allowed by the constraint represented in Eq.~\eqref{eq:perturbative_unitarity_bound} while the muon $g-2$ allowed region is represented by the one above solid (black) line. } 
    \label{fig:mzp_vs_gd_results}
\end{figure}

In Fig.~\ref{fig:2sig_3tev}, we present the $2\sigma$ exclusion limit in the $m_{\gamma_d}-m_{\mu_p}$ plane at $\sqrt{s}=3$ TeV assuming an integrated luminosity of 1 ab$^{-1}$ in two different final states for two sets of $g_d$ and $\sin\theta_L$ keeping $\sin\theta_s$ fixed at $0.05$. The same at $\sqrt{s}= 10$ TeV and an integrated luminosity of $10$ ab$^{-1}$ is presented in Fig.~\ref{fig:2sig_10tev}. The Fig.~\ref{fig:mzp_vs_gd_results} represents the $2\sigma$ exclusion limit in the $g_d-m_{\gamma_d}$ plane corresponding to $\sqrt{s} = 3$ TeV and an integrated luminosity of $1$ ab$^{-1}$. The regions allowed by the constraint presented in Eq.~(\ref{eq:perturbative_unitarity_bound}) and the current muon $g-2$ measurements at $1\sigma$ level are represented by the one above the dashed (black) and the solid (black) lines, respectively, in all these figures.

\section{Conclusion}
\label{sec:conclusion}

Motivated by the prospects of Higgs physics at proposed future lepton colliders and precise measurements of the Higgs properties, we have investigated the implications of a {\it portal vector-like lepton} on the Higgs production in association with a neutral gauge boson ($Z/\gamma_d$) at the proposed future muon collider experiment and its consequence on dark photon. In particular, we focus on the Higgs and dark photon associated production via $t$ and $u$ channel pVLL exchange in addition to the dark photon mediated $s$-channel contribution. To avoid constraints coming from the low energy flavor changing observables we consider pVLL mixing only with one generation of lepton family, in particular, the second generation charged lepton.

{
We have investigated the role of the pVLL in the context of the muon $g-2$ anomaly. Compatibility with the current value of muon $g-2$ suggests that one can have a sizable value of $\sin\theta_R$ ($\sim 10^{-1}$), while $\sin\theta_L$ can remain small ($\sim 10^{-5}$). We have shown that there exists a wide region of parameter space in the $g_d\!-\!m_{\gamma_d}$ plane that is consistent with the observed value of muon $g-2$ as well as other existing constraints coming from searches for additional scalar and fermions. The presence of the pVLL also plays a crucial role in the context of DM relic abundance calculation. In the absence of the pVLL state, 
 one requires large kinetic mixing ($\varepsilon \sim 10^{-2}$) away from resonance to obtain an underabundant DM relic density for low mass dark photon ($m_\chi < m_{\gamma_d}$). Such a large kinetic mixing is disfavored by direct detection experiments. However, the presence of the pVLL state not only enhances dark matter annihilation into a pair of muons—if the pVLL mixes with the Standard Model second-generation charged lepton—even for small kinetic mixing, but also helps in evading direct detection bounds.}

{Next we investigate the implication of the pVLL on the associated Higgs production at muon collider and} present a detailed comparison of the $h\gamma_d$ production rate with that of the $hZ$ production at muon collider.
The later process is considered both in the context of SM and in presence of the portal VLL extension of the SM. 
The $\mu^+ \mu^- \to h \gamma_d$ process is highly sensitive to the following independent model parameters: the dark photon mass ($m_{\gamma_d}$), dark gauge coupling ($g_d$), and both fermion ($\sin\theta_L$) and scalar ($\sin\theta_s$) mixing angles.
We explore the {\it non-decoupling} effect owing to the presence of the muon-heavy muon-dark photon vertex and show that a significant region of parameter space consistent with the perturbative unitarity, electroweak precision measurements, lepton flavor universality, allowed ranges of scalar mixing angle and other relevant constraints can be found where the $h\gamma_d$ cross-section is $\sim$ few fb to $10^2$ fb. Consequently, one can achieve up to $10\%$ better precision in the measurement of Higgs properties w.r.t. that achievable in the $hZ$ channel.

A crucial outcome of our study is that this process not only provides an enhanced Higgs production rate at the muon collider but also helps to set a strong limit on the dark photon mass even for a fermion mixing angle as low as $10^{-6}$, independent of the gauge kinetic mixing parameter.

We have proposed two final states, i.e., {\it leptonically quiet exactly one and exactly two central-jet(s) along with missing energy}. These final states originate from the $h\gamma_d$ process when the Higgs boson decays to a pair of bottom quarks and the dark photon is treated as an invisible particle at the detector. In order to set a $2\sigma$ exclusion limit in the $m_{\gamma_d}- m_{\mu_p}$ plane using the $h\gamma_d$ process in the aforementioned final states, we have also considered an exhaustive study of the SM background processes which contribute to these final states. With the help of certain kinematic variables such as missing energy, requirement of the presence of substructure in the leading central CA$_{1.2}$ jet, and the invariant mass associated with the $b\bar{b}$ system to lie around the Higgs mass it is possible to substantially reduce the SM background contributions and efficiently isolate the signal of $h\gamma_d$ production at the future muon collider.

A $2\sigma$ exclusion limit upto $m_{\gamma_d}=80~(76)$ GeV can be obtained using our analysis in the {\it exactly one central-jet plus missing energy} final state at $3$ TeV muon collider center-of-mass energy for $g_d=0.05~(0.5),~ \sin\theta_L = 4 \times 10^{-5}~(1.66 \times 10^{-6}),~ \sin\theta_s = 0.05$ for a heavy muon mass $m_{\mu_p}\sim 3$ TeV.
The corresponding bounds in the {\it exactly two central-jets plus missing energy} final state are $m_{\gamma_d}=43~\text{GeV and}~46$ GeV for a heavy muon mass $m_{\mu_p}\sim 3.3$ TeV using the same set of choices of the other parameters as mentioned above. We also present the $2\sigma$ exclusion limit for the {\it exactly one central-jet plus missing energy} final state in the $m_{\mu_p}$--$m_{\gamma_d}$ plane for a higher-energy muon collider scenario with $\sqrt{s}=10$~TeV and corresponding integrated luminosity of $10\ \mathrm{ab}^{-1}$.
Assuming 
$~m_{\mu_p}=1~\text{TeV},~\sin\theta_L=4 \times 10^{-5}~\text{and}~\sin\theta_s=0.05$ one can exclude the value of the dark gauge coupling above {$0.15~(0.28)$} for $m_{\gamma_d}\leq 90~\text{GeV}$ at $\sqrt{s} = 3$ TeV with an integrated luminosity of $1$ ab$^{-1}$ in the {\it exactly one (two) central-jet(s) plus missing energy} final state. The corresponding limits on $g_d$ for $~m_{\mu_p}=3~\text{TeV},~\sin\theta_L=1.66 \times 10^{-6},~\sin\theta_s=0.05$ and $m_{\gamma_d}<40$ GeV are respectively 0.26 and 0.5 for these two final states assuming the same muon collider collision energy and integrated luminosity.  
{Thus, the Higgs production in association with a dark photon at future muon collider facilities provide a sensitive probe of a large region of parameter space in the $g_d$--$m_{\gamma_d}$ and $m_{\mu_p}$--$m_{\gamma_d}$ planes while addressing the muon $g-2$ and DM relic density.}

\section*{Acknowledgement}
We gratefully acknowledge Satyanarayan Mukhopadhyay for his valuable insights and useful comments. KT thanks the Department of Science and Technology (DST), Government of India, for financial support under the INSPIRE Fellowship Programme (IF220170). {The work of SV at Physical Research Laboratory was supported by Department of Space, Govt. of India.} KT is also grateful to Sujay Mondal for providing access to high-performance computational facilities which significantly aided the progress of this work. KT further extends her thanks to the organizers and instructors of the Iwate Collider School 2025 for their excellent lectures and tutorial sessions. 

%%%%%%%%%%%%

\appendix

\setcounter{equation}{0}
\numberwithin{equation}{section}

\section{Vertex factors}
\label{app:vert_fact}
In this appendix, we present Feynman rules for the relevant vertices assuming all the associated four momenta are coming out of the vertex {in the zero kinetic mixing limit ($\varepsilon \to 0$)}. Below the $L$ and $R$ represent the {\it left- and right- chiral} fermions, respectively.

\subsection{Vertices with Dark Photon:}
\label{app:vertices_with_zp}
\begin{enumerate}
    \item $\bm{\mu^+_{p_{_R}}- \mu^-_L - \gamma_d}~/~\bm{\mu^+_R - \mu^-_{p_{_L}}-\gamma_d}$

\begin{eqnarray}
    \text{VF}_{1} &=&  -i g_d \sin\theta_L \cos\theta_L \gamma^\mu 
\end{eqnarray}

    \item  $\bm{\mu^+_{p_{_L}}- \mu^-_R - \gamma_d}~/~\bm{\mu^+_L - \mu^-_{p_{_R}}-\gamma_d}$

\begin{eqnarray}
    \text{VF}_{2} &=&  -i g_d \sin\theta_R \cos\theta_R \gamma^\mu 
    \label{Appn_eq:vf2}
\end{eqnarray}

    \item  $\bm{\mu^+_{{L}}- \mu^-_R - \gamma_d}$

\begin{eqnarray}
    \text{VF}_{3} &=&  -i g_d \sin^2\theta_R  \gamma^\mu 
\end{eqnarray}

 \item  $\bm{\mu^+_{{R}}- \mu^-_L - \gamma_d}$

\begin{eqnarray}
    \text{VF}_{4} &=&  -i g_d \sin^2\theta_L  \gamma^\mu 
\end{eqnarray}

\end{enumerate}

\subsection{Vertices with \texorpdfstring{$Z$}{Z} boson:}
\label{app:vert_with_Z}
\begin{enumerate}
    \item $\bm{\mu^+_{p_{_R}}- \mu^-_L - Z}~/~\bm{\mu^+_R - \mu^-_{p_{_L}}-Z}$

\begin{eqnarray}
    \text{VF}_{5} &=&  -\dfrac{i}{2} g_z \sin\theta_L \cos\theta_L \gamma^\mu 
\end{eqnarray}

    \item  $\bm{\mu^+_{{L}}- \mu^-_R - Z}$

\begin{eqnarray}
    \text{VF}_{6} &=&  i g_z \sin^2\theta_W  \gamma^\mu 
\end{eqnarray}

 \item  $\bm{\mu^+_{{R}}- \mu^-_L - Z}$

\begin{eqnarray}
    \label{eq:mur_mul_z_ver}
    \text{VF}_{7} &=&  i~\dfrac{g_z}{2} (2 \sin^2\theta_W - \cos^2\theta_L) \gamma^\mu 
\end{eqnarray}

\end{enumerate}

\subsection{Vertices with Higgs:}
\label{app:vert_with_H}
\begin{enumerate}
    \item $\bm{\mu^+_{p_{_L}}- \mu^-_L - h}~/~\bm{\mu^+_R - \mu^-_{p_{_R}}-h}$

\begin{eqnarray}
    \text{VF}_{8} &=&  -\dfrac{i}{\sqrt{2}} \sin\theta_R ( \sin\theta_L \sin\theta_s \omega_f + \cos\theta_L \cos\theta_s y_m) 
\end{eqnarray}

 \item  $\bm{\mu^+_{p_{_R}}- \mu^-_R - h}~/~\bm{\mu^+_L - \mu^-_{p_{_L}}-h}$

\begin{eqnarray}
    \text{VF}_{9} &=&  \dfrac{i}{\sqrt{2}} \cos\theta_R ( \cos\theta_L \sin\theta_s \omega_f - \sin\theta_L \cos\theta_s y_m) 
\end{eqnarray}

 \item  $\bm{\gamma_d- \gamma_d - h}$

\begin{eqnarray}
    \text{VF}_{10} &=&  -2i g_d m_{\gamma_d} \sin\theta_s 
\end{eqnarray}

 \item  $\bm{Z- Z - h}$

\begin{eqnarray}
    \text{VF}_{11} &=& i g_z m_z \cos\theta_s  
\end{eqnarray}
\end{enumerate}

\section{Squared matrix element for 
\texorpdfstring{$\mu^+\mu^- \to h\,\gamma_d$}{mu+mu- -> h gd}}\label{app:matrix_element}

\begin{figure}[H]
    \centering
    \resizebox{0.7\columnwidth}{!}
    {
        \subfloat[\label{subfig:t_channel_mmup_mediated}]{\includegraphics[width=0.5\columnwidth]{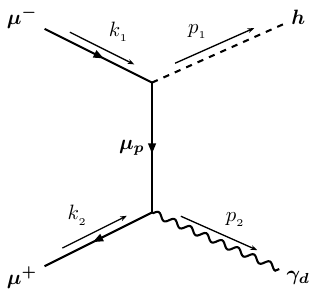}}~
        \hspace{2cm}\subfloat[\label{subfig:u_channel_mmup_mediated}]{\includegraphics[width=0.5\columnwidth]{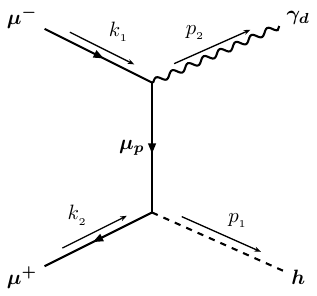}}
    }
 \caption{Feynman diagrams for the process $\mu^+\mu^- \to h\,\gamma_d$ mediated by the portal matter $m_{\mu_p}$ through the (a) $t$-channel and (b) $u$-channel.}

    \label{fig:t_u_diagrams_mmup_mediated}
\end{figure}

The $t$- and $u$- channel diagrams for the $\mu^+\mu^- \to h\,\gamma_d$ process via portal vector-like lepton mediation ($\mu_p$) are presented in Fig~\ref{fig:t_u_diagrams_mmup_mediated}.

The amplitude for the $t$-channel diagram is given by 
\bea
i\mathcal{M}_t &=& -i \left[
\bar{v}_{s_2}\left(k_2\right)
\gamma^\mu \left(c_{_{V}}^t +c_{_{A}}^t \gamma^5\right)
\epsilon^{*\lambda}_\mu\left(\vec{p}_2\right)
\frac{(\slashed{k}_1-\slashed{p}_1 + m_{\mu_p})}{\left(k_1 - p_1\right)^2 - m^2_{\mu_p}}
\left(c_s^t +c_p^t \gamma^5\right)
u_{s_1}\left(k_1\right) \right]
\label{eq:Mt_expression}
\eea

where $c_s^t,~c_p^t,~c_{_{V}}^t$ and $c_{_{A}}^t$ are given by
\begin{eqnarray*}
        c_s^t &=& \dfrac{m_{\mu_p}}{4} \sin{2\theta_L} \big[ \dfrac{1}{D^2} \dfrac{g_d}{m_{\gamma_d}} \sin{\theta_s} (1- \dfrac{m^2_{\mu}}{m^2_{\mu_p}}) (\sin^2{\theta_L} - \dfrac{m_{\mu}}{m_{\mu_p}} \cos^2{\theta_L})~ +~ \dfrac{1}{v_{_{\rm{EW}}}} \cos{\theta_s} ( 1 + \dfrac{m_{\mu}}{m_{\mu_p}}) \big ] \\ 
        c_p^t &=& -\dfrac{m_{\mu_p}}{4} \sin{2\theta_L} \big[ \dfrac{1}{D^2} \dfrac{g_d}{m_{\gamma_d}} \sin{\theta_s} (1- \dfrac{m^2_{\mu}}{m^2_{\mu_p}}) (\sin^2{\theta_L} + \dfrac{m_{\mu}}{m_{\mu_p}} \cos^2{\theta_L})~ +~ \dfrac{1}{v_{_{\rm{EW}}}} \cos{\theta_s} ( 1 - \dfrac{m_{\mu}}{m_{\mu_p}}) \big ]  \\ 
         c_{_{V}}^t &=& \dfrac{g_d}{2} \cos{\theta_L} \sin{\theta_L} \big( \dfrac{m_\mu}{m_{\mu_p}} \dfrac{1}{D^2} + 1 \big ) \\
      c_{_{A}}^t &=&  \dfrac{g_d}{2} \cos{\theta_L} \sin{\theta_L} \big( \dfrac{m_\mu}{m_{\mu_p}} \dfrac{1}{D^2} - 1 \big ) 
\end{eqnarray*}

The amplitude of the $u$-channel diagram is given by
\bea
i\mathcal{M}_u &=& 
-i\left[\bar{v}_{s_2}\left(k_2\right) \left(c_s^u + c_p^u \gamma^5 \right)
\frac{\left(\slashed{p}_1 -\slashed{k}_2 + m_{\mu_p}\right)}{\left(p_1 - k_2\right)^2 -m^2_{\mu_p}}
\gamma^\mu \left(c_{_{V}}^u +c_{_{A}}^u \gamma^5\right)\epsilon^{*\lambda}_\mu\left(\vec{p}_2\right)u_{s_1}\left(k_1\right) \right]
\label{eq:Mu_expression}
\eea

where $c_s^u,~c_p^u,~c_{_{V}}^u$ and $c_{_{A}}^u$ are given by
\begin{eqnarray*}
        c_s^u &=& \dfrac{m_{\mu_p}}{4} \sin{2\theta_L} \big[ \dfrac{1}{D^2} \dfrac{g_d}{m_{\gamma_d}} \sin{\theta_s} (1- \dfrac{m^2_{\mu}}{m^2_{\mu_p}}) (\sin^2{\theta_L} - \dfrac{m_{\mu}}{m_{\mu_P}} \cos^2{\theta_L})~ +~ \dfrac{1}{v_{_{\rm{EW}}}} \cos{\theta_s} ( 1 + \dfrac{m_{\mu}}{m_{\mu_p}}) \big ]\\
        c_p^u &=& \dfrac{m_{\mu_p}}{4} \sin{2\theta_L} \big[ \dfrac{1}{D^2} \dfrac{g_d}{m_{\gamma_d}} \sin{\theta_s} (1- \dfrac{m^2_{\mu}}{m^2_{\mu_p}}) (\sin^2{\theta_L} + \dfrac{m_{\mu}}{m_{\mu_P}} \cos^2{\theta_L})~ +~ \dfrac{1}{v_{_{\rm{EW}}}} \cos{\theta_s} ( 1 - \dfrac{m_{\mu}}{m_{\mu_p}}) \big ] \\
          c_{_{V}}^u &=& \dfrac{g_d}{2} \cos{\theta_L} \sin{\theta_L} \big( \dfrac{m_\mu}{m_{\mu_P}} \dfrac{1}{D^2} + 1 \big ) \\
      c_{_{A}}^u &=&  \dfrac{g_d}{2} \cos{\theta_L} \sin{\theta_L} \big( \dfrac{m_\mu}{m_{\mu_P}} \dfrac{1}{D^2} - 1 \big ) \\
    \end{eqnarray*}

The spin-averaged squared amplitude, neglecting the s channel dark photon mediated and t/u channel muon mediated contribution, is

\begin{eqnarray}
\label{mod_msquared_general}
\overline{|\mathcal{M}|^2} &=& 
\overline{|\mathcal{M}_t|^2} 
+ \overline{|\mathcal{M}_u|^2} 
+ 2\,\operatorname{Re}\!\left( \overline{\mathcal{M}_t \mathcal{M}_u^{*}} \right)
\end{eqnarray}
where the $t$-channel contribution in the limit $m_{\mu} \to 0$ is 
\bea
\overline{|\mathcal{M}_t|^2} &=& 
\frac{2}{\left[ \left(k_1 - p_1\right)^2 - m^2_{\mu_p}  \right]^2}\Bigg[A_1 \left\{ 2 \left(k_1 \cdot p_1\right) \left(k_2 \cdot p_1\right) - p_1^2 \left(k_1 \cdot k_2\right) \right\}+ A_2~ m_{\mu_p}^2 \left(k_1 \cdot k_2\right)\nonumber\\ 
&& + \frac{1}{m_{\gamma_d}^2}\Bigg( A_1 \Big\{2 \left(k_1 \cdot p_1\right) \left(k_2 \cdot  p_2\right) \left(p_1 \cdot p_2\right) - p_1^2 \left(k_1 \cdot p_2\right)\left(k_2 \cdot p_2\right) - p_2^2 \left(k_1 \cdot p_1\right)\left(k_2 \cdot p_1\right) \nonumber\\ 
&&  + \frac{p_1^2 ~p_2^2}{2} \left(k_1 \cdot k_2\right) \Big\} + A_2 ~m_{\mu_p}^2\left\{\left(k_1 \cdot p_2\right) \left(k_2 \cdot p_2\right) -\frac{p_2^2}{2} \left(k_1 \cdot k_2\right) \right\}
\Bigg)\Bigg]
\eea
Similarly, the $u$- channel contribution in the same limit reads 
\bea
\overline{|\mathcal{M}_u|^2} &=& 
\frac{2}{\left[ \left(p_1 - k_2\right)^2 - m^2_{\mu_p} \right]^2}
\Bigg[
A_1 \left\{ 2 \left(k_1 \cdot p_1\right)\left(k_2 \cdot p_1\right) - p_1^2 \left(k_1 \cdot k_2\right) \right\}
+ A_2 ~m_{\mu_p}^2 \left(k_1 \cdot k_2\right) \nonumber\\ 
&& 
+ \frac{1}{m_{\gamma_d}^2}
\Bigg(
A_1 \Big\{
2 \left(k_1 \cdot p_2\right)\left(k_2 \cdot p_1\right)\left(p_1 \cdot p_2\right)
- p_1^2 \left(k_1 \cdot p_2\right)\left(k_2 \cdot p_2\right)  - p_2^2 \left(k_1 \cdot p_1\right)\left(k_2 \cdot p_1\right)
 \nonumber\\ 
&& 
+ \frac{p_1^2~ p_2^2}{2} \left(k_1 \cdot k_2\right)
\Big\} + A_2 ~m_{\mu_p}^2 
\left\{
\left(k_1 \cdot p_2\right)\left(k_2 \cdot p_2\right)
- \frac{p_2^2}{2} \left(k_1 \cdot k_2\right)
\right\}
\Bigg)
\Bigg]
\eea
Here, the coefficients $A_1$ and $A_2$ are as follows

\begin{eqnarray}
    A_1 &=& \left(c_{_{V}}^2 +c_{_{A}}^2\right)\left(c_s^2 +c_p^2\right) - 4c_s ~c_p ~c_{_{V}}c_{_{A}}\\ \nonumber
    A_2 &=& \left(c_{_{V}}^2 +c_{_{A}}^2\right)\left(c_s^2 +c_p^2\right) + 4c_s ~c_p ~c_{_{V}}c_{_{A}}
\end{eqnarray}
with the coupling identifications
\[
c_s = c_s^t = c_s^u,\qquad
c_p = c_p^t = -\,c_p^u,\qquad
c_V = c_V^t = c_V^u,\qquad
c_A = c_A^t = c_A^u.
\]
The corresponding interference term is 
\bea
\overline{\mathcal{M}_t \mathcal{M}_u^{*}} &=&
\frac{2}{\left[ \left(k_1 - p_1\right)^2 - m_{\mu_p}^2 \right] 
           \left[ \left(p_1 - k_2\right)^2 - m_{\mu_p}^2 \right]}
\Bigg[
2 A_3 \left(k_1 \cdot p_1\right)\left(k_2 \cdot p_1\right) 
+ A_2~  m_{\mu_p}^2 \left(k_1 \cdot k_2\right)  \nonumber\\
&& 
- \frac{1}{m_{\gamma_d}^2}
\Bigg(
A_3 \Big\{
p_1^2 \left(k_1 \cdot p_2\right)\left(k_2 \cdot p_2\right)
+ p_2^2 \left(k_1 \cdot p_1\right)\left(k_2 \cdot p_1\right)
- \left(p_1 \cdot p_2\right)\left(k_1 \cdot p_2\right)\left(k_2 \cdot p_1\right)
\nonumber\\
&& - \left(p_1 \cdot p_2\right)\left(k_1 \cdot p_1\right)\left(k_2 \cdot p_2\right)
+ \left(p_1 \cdot p_2\right)^2 \left(k_1 \cdot k_2\right)
- \frac{p_1^2 ~p_2^2}{2}\left(k_1 \cdot k_2\right)
\Big\} \nonumber\\
&& 
+ \frac{A_2}{2} m_{\mu_p}^2
\left\{
p_2^2 \left(k_1 \cdot k_2\right)
- \left(k_1 \cdot p_2\right)\left(k_2 \cdot p_2\right)
\right\}
\Bigg)
\Bigg]
\eea
where the coefficient $A_3$ takes the form
\begin{eqnarray}
    A_3 &=& (c^2_s - c^2_p) (c_{_{V}}^2 - c_{_{A}}^2)
\end{eqnarray}
In the center-of-mass frame and in the limit {$m_h,~m_{\gamma_d}\ll E_{_{\rm{CM}}}$}, the $\overline{|\mathcal{M}|^2}$ for the pure $t$- and $u$- channels and the corresponding interference contribution take the form
\bea
% \begin{eqnarray}
\overline{|\mathcal{M}_t|^2} &=&
\frac{E_{_{\rm CM}}^2}{\left(a-b\cos\theta\right)^2}
\Bigg[
    A_1\left(
        \frac{E_{_{\rm CM}}^2}{4}\sin^2\theta
        - m_h^2
    \right)
    + A_2~ m_{\mu_p}^2  +\frac{1}{2  m_{\gamma_d}^2}
    \Bigg( {A_1}~ \Bigg\{
       \frac{E_{_{\rm CM}}^4}{4} ~(1-\cos\theta)^2
\nonumber \\  &&  - \frac{ \left( m_h^2 + m_{\gamma_d}^2 \right) E_{_{\rm CM}}^2  }{4} \sin^2\theta
            + m_h^2~ m_{\gamma_d}^2 
        \Bigg\}  + A_2~{m_{\mu_p}^2}
        \left\{
            \frac{E_{_{\rm CM}}^2}{4}\sin^2\theta
            - m_{\gamma_d}^2
        \right\}
    \Bigg)
\Bigg]
\eea

\bea
\overline{|\mathcal{M}_u|^2} &=&
\frac{E_{_{\rm CM}}^2}{(a+b\cos\theta)^2}
\Bigg[
    A_1\left( \frac{E_{_{\rm CM}}^2}{4}\sin^2\theta- m_h^2 \right)
    + A_2~ m_{\mu_p}^2 +\frac{1}{2 m_{\gamma_d}^2}
    \Bigg( {A_1} \Bigg\{  \frac{E_{_{\rm CM}}^4}{4} (1+\cos\theta)^2 \nonumber\\ &&
    - \frac{ \left( m_h^2 + m_{\gamma_d}^2 \right) E_{_{\rm CM}}^2  }{4} \sin^2\theta
            + m_h^2 m_{\gamma_d}^2 
        \Bigg\}  +\, A_2\,{m_{\mu_p}^2}
        \left\{
            \frac{E_{_{\rm CM}}^2}{4}\sin^2\theta
            - m_{\gamma_d}^2
        \right\}
    \Bigg)
\Bigg]
\eea

\bea
  \overline{\mathcal{M}_t \mathcal{M}_u^{*}}  &=& 
 \frac{{E_{_{\rm CM}}^2}}{(a^2 - b^2\cos^2\theta)}
\Bigg[ \left(A_3
        \frac{E_{_{\rm CM}}^2}{4}\sin^2\theta
+ A_2~ m_{\mu_p}^2 \right)  +\frac{1}{2  m_{\gamma_d}^2}
    \Bigg(  {A_3} \Bigg\{m^2_h m^2_{\gamma_d}\nonumber 
\\&& - \frac{ \left( E_{_{\rm CM}}^2 + m_h^2 +  m_{\gamma_d}^2 \right) {E_{_{\rm CM}}^2}}{4} \sin^2\theta  \Bigg\} + A_2~m_{\mu_p}^2 \left\{\frac{E_{_{\rm CM}}^2}{8}\sin^2\theta- m_{\gamma_d}^2 \right\}
\Bigg)
\Bigg]
% \end{eqnarray}
% 
\eea

Here,
\[
a = \frac{E_{\rm CM}^2}{2} + m_{\mu_p}^2,\qquad
b = \frac{E_{\rm CM}^2}{2}.
\]

\bibliographystyle{JHEP}
\bibliography{References}

\providecommand{\href}[2]{#2}\begingroup\raggedright\begin{thebibliography}{100}

\bibitem{ATLAS:2012yve}
{\scshape ATLAS} collaboration, G.~Aad et~al., \emph{{Observation of a new
  particle in the search for the Standard Model Higgs boson with the ATLAS
  detector at the LHC}},
  \href{http://dx.doi.org/10.1016/j.physletb.2012.08.020}{\emph{Phys. Lett. B}
  {\bf 716} (2012) 1--29}, [\href{https://arxiv.org/abs/1207.7214}{{\tt
  1207.7214}}].

\bibitem{CMS:2012qbp}
{\scshape CMS} collaboration, S.~Chatrchyan et~al., \emph{{Observation of a New
  Boson at a Mass of 125 GeV with the CMS Experiment at the LHC}},
  \href{http://dx.doi.org/10.1016/j.physletb.2012.08.021}{\emph{Phys. Lett. B}
  {\bf 716} (2012) 30--61}, [\href{https://arxiv.org/abs/1207.7235}{{\tt
  1207.7235}}].

\bibitem{Kilian:1995tr}
W.~Kilian, M.~Kramer and P.~M. Zerwas, \emph{{Higgsstrahlung and W W fusion in
  e+ e- collisions}},
  \href{http://dx.doi.org/10.1016/0370-2693(96)00100-1}{\emph{Phys. Lett. B}
  {\bf 373} (1996) 135--140}, [\href{https://arxiv.org/abs/hep-ph/9512355}{{\tt
  hep-ph/9512355}}].

\bibitem{Liu:2016zki}
Z.~Liu, L.-T. Wang and H.~Zhang, \emph{{Exotic decays of the 125 GeV Higgs
  boson at future $e^+e^-$ lepton colliders}},
  \href{http://dx.doi.org/10.1088/1674-1137/41/6/063102}{\emph{Chin. Phys. C}
  {\bf 41} (2017) 063102}, [\href{https://arxiv.org/abs/1612.09284}{{\tt
  1612.09284}}].

\bibitem{Akeroyd:1999gu}
A.~G. Akeroyd, A.~Arhrib and M.~Capdequi~Peyranere, \emph{{CP odd Higgs boson
  production in association with neutral gauge boson in high-energy e+ e-
  collisions}}, \href{http://dx.doi.org/10.1142/S0217732399002157}{\emph{Mod.
  Phys. Lett. A} {\bf 14} (1999) 2093--2108},
  [\href{https://arxiv.org/abs/hep-ph/9907542}{{\tt hep-ph/9907542}}].

\bibitem{Kanemura:1999tg}
S.~Kanemura, \emph{{Possible enhancement of the e+ e- ---{\ensuremath{>}} H+-
  W-+ cross-section in the two Higgs doublet model}},
  \href{http://dx.doi.org/10.1007/s100520000480}{\emph{Eur. Phys. J. C} {\bf
  17} (2000) 473--486}, [\href{https://arxiv.org/abs/hep-ph/9911541}{{\tt
  hep-ph/9911541}}].

\bibitem{Ouazghour:2024twx}
B.~A. Ouazghour, A.~Arhrib, K.~Cheung, E.-s. Ghourmin and L.~Rahili,
  \emph{{Associated charged Higgs boson production within the 2HDM: e-e+ versus
  {\ensuremath{\mu}}-{\ensuremath{\mu}}+ colliders}},
  \href{http://dx.doi.org/10.1103/PhysRevD.110.095026}{\emph{Phys. Rev. D} {\bf
  110} (2024) 095026}, [\href{https://arxiv.org/abs/2408.13952}{{\tt
  2408.13952}}].

\bibitem{LEPWorkingGroupforHiggsbosonsearches:2003ing}
{\scshape LEP Working Group for Higgs boson searches, ALEPH, DELPHI, L3, OPAL}
  collaboration, R.~Barate et~al., \emph{{Search for the standard model Higgs
  boson at LEP}},
  \href{http://dx.doi.org/10.1016/S0370-2693(03)00614-2}{\emph{Phys. Lett. B}
  {\bf 565} (2003) 61--75}, [\href{https://arxiv.org/abs/hep-ex/0306033}{{\tt
  hep-ex/0306033}}].

\bibitem{OPAL:2002sbl}
{\scshape OPAL} collaboration, G.~Abbiendi et~al., \emph{{Search for the
  standard model Higgs boson with the OPAL detector at LEP}},
  \href{http://dx.doi.org/10.1140/epjc/s2002-01092-3}{\emph{Eur. Phys. J. C}
  {\bf 26} (2003) 479--503}, [\href{https://arxiv.org/abs/hep-ex/0209078}{{\tt
  hep-ex/0209078}}].

\bibitem{FCC:2018evy}
{\scshape FCC} collaboration, A.~Abada et~al., \emph{{FCC-ee: The Lepton
  Collider}: {Future Circular Collider Conceptual Design Report Volume 2}},
  \href{http://dx.doi.org/10.1140/epjst/e2019-900045-4}{\emph{Eur. Phys. J. ST}
  {\bf 228} (2019) 261--623}.

\bibitem{Abramowicz:2016zbo}
H.~Abramowicz et~al., \emph{{Higgs physics at the CLIC
  electron{\textendash}positron linear collider}},
  \href{http://dx.doi.org/10.1140/epjc/s10052-017-4968-5}{\emph{Eur. Phys. J.
  C} {\bf 77} (2017) 475}, [\href{https://arxiv.org/abs/1608.07538}{{\tt
  1608.07538}}].

\bibitem{Franceschini:2021aqd}
R.~Franceschini and M.~Greco, \emph{{Higgs and BSM Physics at the Future Muon
  Collider}}, \href{http://dx.doi.org/10.3390/sym13050851}{\emph{Symmetry} {\bf
  13} (2021) 851}, [\href{https://arxiv.org/abs/2104.05770}{{\tt 2104.05770}}].

\bibitem{Hamada:2024ojj}
Y.~Hamada, R.~Kitano, R.~Matsudo, S.~Okawa, R.~Takai, H.~Takaura et~al.,
  \emph{{Higgs boson production at {\ensuremath{\mu}}+{\ensuremath{\mu}}+
  colliders}}, \href{http://dx.doi.org/10.1103/PhysRevD.110.113011}{\emph{Phys.
  Rev. D} {\bf 110} (2024) 113011},
  [\href{https://arxiv.org/abs/2408.01068}{{\tt 2408.01068}}].

\bibitem{Chun:2020uzw}
E.~J. Chun and T.~Mondal, \emph{{Explaining $g-2$ anomalies in two Higgs
  doublet model with vector-like leptons}},
  \href{http://dx.doi.org/10.1007/JHEP11(2020)077}{\emph{JHEP} {\bf 11} (2020)
  077}, [\href{https://arxiv.org/abs/2009.08314}{{\tt 2009.08314}}].

\bibitem{Asadi:2024jiy}
P.~Asadi, S.~Homiller, A.~Radick and T.-T. Yu, \emph{{Fermion-portal dark
  matter at a high-energy muon collider}},
  \href{http://dx.doi.org/10.1103/p7p8-wqqb}{\emph{Phys. Rev. D} {\bf 112}
  (2025) 055038}, [\href{https://arxiv.org/abs/2412.14235}{{\tt 2412.14235}}].

\bibitem{Crivellin:2020ebi}
A.~Crivellin, F.~Kirk, C.~A. Manzari and M.~Montull, \emph{{Global Electroweak
  Fit and Vector-Like Leptons in Light of the Cabibbo Angle Anomaly}},
  \href{http://dx.doi.org/10.1007/JHEP12(2020)166}{\emph{JHEP} {\bf 12} (2020)
  166}, [\href{https://arxiv.org/abs/2008.01113}{{\tt 2008.01113}}].

\bibitem{Kirk:2020wdk}
M.~Kirk, \emph{{Cabibbo anomaly versus electroweak precision tests: An
  exploration of extensions of the Standard Model}},
  \href{http://dx.doi.org/10.1103/PhysRevD.103.035004}{\emph{Phys. Rev. D} {\bf
  103} (2021) 035004}, [\href{https://arxiv.org/abs/2008.03261}{{\tt
  2008.03261}}].

\bibitem{Wojcik:2023ggt}
G.~N. Wojcik, L.~L. Everett, S.~T. Eu and R.~Ximenes, \emph{{Lepton flavor
  portal matter}},
  \href{http://dx.doi.org/10.1103/PhysRevD.108.055033}{\emph{Phys. Rev. D} {\bf
  108} (2023) 055033}, [\href{https://arxiv.org/abs/2303.12983}{{\tt
  2303.12983}}].

\bibitem{Poh:2017tfo}
Z.~Poh and S.~Raby, \emph{{Vectorlike leptons: Muon g-2 anomaly, lepton flavor
  violation, Higgs boson decays, and lepton nonuniversality}},
  \href{http://dx.doi.org/10.1103/PhysRevD.96.015032}{\emph{Phys. Rev. D} {\bf
  96} (2017) 015032}, [\href{https://arxiv.org/abs/1705.07007}{{\tt
  1705.07007}}].

\bibitem{Dermisek:2021ajd}
R.~Dermisek, K.~Hermanek and N.~McGinnis, \emph{{Muon g-2 in two-Higgs-doublet
  models with vectorlike leptons}},
  \href{http://dx.doi.org/10.1103/PhysRevD.104.055033}{\emph{Phys. Rev. D} {\bf
  104} (2021) 055033}, [\href{https://arxiv.org/abs/2103.05645}{{\tt
  2103.05645}}].

\bibitem{Lee:2021gnw}
H.~M. Lee, J.~Song and K.~Yamashita, \emph{{Seesaw lepton masses and muon $g-2$
  from heavy vector-like leptons}},
  \href{http://dx.doi.org/10.1007/s40042-021-00339-0}{\emph{J. Korean Phys.
  Soc.} {\bf 79} (2021) 1121--1134},
  [\href{https://arxiv.org/abs/2110.09942}{{\tt 2110.09942}}].

\bibitem{Kawamura:2019hxp}
J.~Kawamura, S.~Raby and A.~Trautner, \emph{{Complete vectorlike fourth family
  with U(1)' : A global analysis}},
  \href{http://dx.doi.org/10.1103/PhysRevD.101.035026}{\emph{Phys. Rev. D} {\bf
  101} (2020) 035026}, [\href{https://arxiv.org/abs/1911.11075}{{\tt
  1911.11075}}].

\bibitem{Endo:2020tkb}
M.~Endo and S.~Mishima, \emph{{Muon $g-2$ and CKM unitarity in extra lepton
  models}}, \href{http://dx.doi.org/10.1007/JHEP08(2020)004}{\emph{JHEP} {\bf
  08} (2020) 004}, [\href{https://arxiv.org/abs/2005.03933}{{\tt 2005.03933}}].

\bibitem{Manzari:2021prf}
C.~A. Manzari, \emph{{Vector-Like Leptons in Light of the Cabibbo-Angle
  Anomaly}},  in \emph{{Beyond Standard Model: From Theory to Experiment}}, 5,
  2021.
\newblock \href{https://arxiv.org/abs/2105.03399}{{\tt 2105.03399}}.

\bibitem{Capdevila:2020rrl}
B.~Capdevila, A.~Crivellin, C.~A. Manzari and M.~Montull, \emph{{Explaining
  $b\to s\ell^+\ell^-$ and the Cabibbo angle anomaly with a vector triplet}},
  \href{http://dx.doi.org/10.1103/PhysRevD.103.015032}{\emph{Phys. Rev. D} {\bf
  103} (2021) 015032}, [\href{https://arxiv.org/abs/2005.13542}{{\tt
  2005.13542}}].

\bibitem{Abdallah:2023pbl}
W.~Abdallah, M.~Ashry, J.~Kawamura and A.~Moursy, \emph{{Semivisible dark
  photon in a model with vectorlike leptons for the
  (g-2)e,{\,}{\ensuremath{\mu}} and W-boson mass anomalies}},
  \href{http://dx.doi.org/10.1103/PhysRevD.109.015031}{\emph{Phys. Rev. D} {\bf
  109} (2024) 015031}, [\href{https://arxiv.org/abs/2308.05691}{{\tt
  2308.05691}}].

\bibitem{Lee:2022nqz}
H.~M. Lee and K.~Yamashita, \emph{{A model of vector-like leptons for the muon
  $g-2$ and the W boson mass}},
  \href{http://dx.doi.org/10.1140/epjc/s10052-022-10635-z}{\emph{Eur. Phys. J.
  C} {\bf 82} (2022) 661}, [\href{https://arxiv.org/abs/2204.05024}{{\tt
  2204.05024}}].

\bibitem{Lee:2016wiy}
C.-H. Lee and R.~N. Mohapatra, \emph{{Vector-Like Quarks and Leptons, SU(5)
  $\otimes$ SU(5) Grand Unification, and Proton Decay}},
  \href{http://dx.doi.org/10.1007/JHEP02(2017)080}{\emph{JHEP} {\bf 02} (2017)
  080}, [\href{https://arxiv.org/abs/1611.05478}{{\tt 1611.05478}}].

\bibitem{Gursey:1975ki}
F.~Gursey, P.~Ramond and P.~Sikivie, \emph{{A Universal Gauge Theory Model
  Based on E6}},
  \href{http://dx.doi.org/10.1016/0370-2693(76)90417-2}{\emph{Phys. Lett. B}
  {\bf 60} (1976) 177--180}.

\bibitem{Kyae:2013hda}
B.~Kyae and C.~S. Shin, \emph{{Vector-like leptons and extra gauge symmetry for
  the natural Higgs boson}},
  \href{http://dx.doi.org/10.1007/JHEP06(2013)102}{\emph{JHEP} {\bf 06} (2013)
  102}, [\href{https://arxiv.org/abs/1303.6703}{{\tt 1303.6703}}].

\bibitem{Cvetic:2001tj}
M.~Cvetic, G.~Shiu and A.~M. Uranga, \emph{{Three family supersymmetric
  standard - like models from intersecting brane worlds}},
  \href{http://dx.doi.org/10.1103/PhysRevLett.87.201801}{\emph{Phys. Rev.
  Lett.} {\bf 87} (2001) 201801},
  [\href{https://arxiv.org/abs/hep-th/0107143}{{\tt hep-th/0107143}}].

\bibitem{Cleaver:1999mw}
G.~B. Cleaver, A.~E. Faraggi, D.~V. Nanopoulos and J.~W. Walker,
  \emph{{Phenomenological study of a minimal superstring standard model}},
  \href{http://dx.doi.org/10.1016/S0550-3213(00)00543-5}{\emph{Nucl. Phys. B}
  {\bf 593} (2001) 471--504}, [\href{https://arxiv.org/abs/hep-ph/9910230}{{\tt
  hep-ph/9910230}}].

\bibitem{Marchesano:2013ega}
F.~Marchesano, D.~Regalado and L.~Vazquez-Mercado, \emph{{Discrete flavor
  symmetries in D-brane models}},
  \href{http://dx.doi.org/10.1007/JHEP09(2013)028}{\emph{JHEP} {\bf 09} (2013)
  028}, [\href{https://arxiv.org/abs/1306.1284}{{\tt 1306.1284}}].

\bibitem{ATLAS:2024mrr}
{\scshape ATLAS} collaboration, G.~Aad et~al., \emph{{Search for vector-like
  leptons coupling to first- and second-generation Standard Model leptons in pp
  collisions at $ \sqrt{s} $ = 13 TeV with the ATLAS detector}},
  \href{http://dx.doi.org/10.1007/JHEP05(2025)075}{\emph{JHEP} {\bf 05} (2025)
  075}, [\href{https://arxiv.org/abs/2411.07143}{{\tt 2411.07143}}].

\bibitem{CMS:2022cpe}
{\scshape CMS} collaboration, A.~Tumasyan et~al., \emph{{Search for
  pair-produced vector-like leptons in final states with third-generation
  leptons and at least three b quark jets in proton-proton collisions at
  s=13TeV}},
  \href{http://dx.doi.org/10.1016/j.physletb.2023.137713}{\emph{Phys. Lett. B}
  {\bf 846} (2023) 137713}, [\href{https://arxiv.org/abs/2208.09700}{{\tt
  2208.09700}}].

\bibitem{CMS:2019hsm}
{\scshape CMS} collaboration, A.~M. Sirunyan et~al., \emph{{Search for
  vector-like leptons in multilepton final states in proton-proton collisions
  at $\sqrt{s}$ = 13 TeV}},
  \href{http://dx.doi.org/10.1103/PhysRevD.100.052003}{\emph{Phys. Rev. D} {\bf
  100} (2019) 052003}, [\href{https://arxiv.org/abs/1905.10853}{{\tt
  1905.10853}}].

\bibitem{Yue:2024sds}
C.-X. Yue, Y.-Q. Wang, H.~Wang, Y.-H. Wang and S.~Li, \emph{{Searching for
  singlet vector-like leptons via pair production at ILC}},
  \href{http://dx.doi.org/10.1016/j.nuclphysb.2024.116482}{\emph{Nucl. Phys. B}
  {\bf 1000} (2024) 116482}, [\href{https://arxiv.org/abs/2402.02072}{{\tt
  2402.02072}}].

\bibitem{Shang:2021mgn}
L.~Shang, M.~Wang, Z.~Heng and B.~Yang, \emph{{Search for the singlet
  vector-like lepton at future $e^+ e^-$ colliders}},
  \href{http://dx.doi.org/10.1140/epjc/s10052-021-09152-2}{\emph{Eur. Phys. J.
  C} {\bf 81} (2021) 415}.

\bibitem{Guo:2023jkz}
Q.~Guo, L.~Gao, Y.~Mao and Q.~Li, \emph{{Vector-like lepton searches at a muon
  collider in the context of the 4321 model}},
  \href{http://dx.doi.org/10.1088/1674-1137/ace5a7}{\emph{Chin. Phys. C} {\bf
  47} (2023) 103106}, [\href{https://arxiv.org/abs/2304.01885}{{\tt
  2304.01885}}].

\bibitem{Ghosh:2023xbj}
N.~Ghosh, S.~K. Rai and T.~Samui, \emph{{Search for a leptoquark and
  vector-like lepton in a muon collider}},
  \href{http://dx.doi.org/10.1016/j.nuclphysb.2024.116564}{\emph{Nucl. Phys. B}
  {\bf 1004} (2024) 116564}, [\href{https://arxiv.org/abs/2309.07583}{{\tt
  2309.07583}}].

\bibitem{Bhattiprolu:2019vdu}
P.~N. Bhattiprolu and S.~P. Martin, \emph{{Prospects for vectorlike leptons at
  future proton-proton colliders}},
  \href{http://dx.doi.org/10.1103/PhysRevD.100.015033}{\emph{Phys. Rev. D} {\bf
  100} (2019) 015033}, [\href{https://arxiv.org/abs/1905.00498}{{\tt
  1905.00498}}].

\bibitem{Rizzo:2022qan}
T.~G. Rizzo, \emph{{Portal Matter and Dark Sector Phenomenology at Colliders}},
   in \emph{{Snowmass 2021}}, 2, 2022.
\newblock \href{https://arxiv.org/abs/2202.02222}{{\tt 2202.02222}}.

\bibitem{Rueter:2019wdf}
T.~D. Rueter and T.~G. Rizzo, \emph{{Towards A UV-Model of Kinetic Mixing and
  Portal Matter}},
  \href{http://dx.doi.org/10.1103/PhysRevD.101.015014}{\emph{Phys. Rev. D} {\bf
  101} (2020) 015014}, [\href{https://arxiv.org/abs/1909.09160}{{\tt
  1909.09160}}].

\bibitem{Rizzo:2018vlb}
T.~G. Rizzo, \emph{{Kinetic Mixing and Portal Matter Phenomenology}},
  \href{http://dx.doi.org/10.1103/PhysRevD.99.115024}{\emph{Phys. Rev. D} {\bf
  99} (2019) 115024}, [\href{https://arxiv.org/abs/1810.07531}{{\tt
  1810.07531}}].

\bibitem{Belyaev:2025cgf}
A.~Belyaev, L.~Panizzi, N.~Thongyoi and F.~Wilhelm, \emph{{The Muonic Portal to
  Vector Dark Matter:connecting precision muon physics, cosmology, and
  colliders}},  \href{https://arxiv.org/abs/2510.18564}{{\tt 2510.18564}}.

\bibitem{Lahiri:2024rxc}
J.~Lahiri, D.~Pradhan and A.~Sarkar, \emph{{The influence of lepton portal on
  the WIMP-pFIMP framework}},
  \href{http://dx.doi.org/10.1007/JHEP08(2025)019}{\emph{JHEP} {\bf 08} (2025)
  019}, [\href{https://arxiv.org/abs/2410.19734}{{\tt 2410.19734}}].

\bibitem{Delahaye:2019omf}
J.~P. Delahaye, M.~Diemoz, K.~Long, B.~Mansouli{\'e}, N.~Pastrone, L.~Rivkin
  et~al., \emph{{Muon Colliders}},
  \href{https://arxiv.org/abs/1901.06150}{{\tt 1901.06150}}.

\bibitem{Schulte:2022brl}
{\scshape International Muon Collider} collaboration, D.~Schulte, \emph{{The
  Muon Collider}},
  \href{http://dx.doi.org/10.18429/JACoW-IPAC2022-TUIZSP2}{\emph{JACoW} {\bf
  IPAC2022} (2022) 821--826}.

\bibitem{Black:2022cth}
K.~M. Black et~al., \emph{{Muon Collider Forum report}},
  \href{http://dx.doi.org/10.1088/1748-0221/19/02/T02015}{\emph{JINST} {\bf 19}
  (2024) T02015}, [\href{https://arxiv.org/abs/2209.01318}{{\tt 2209.01318}}].

\bibitem{InternationalMuonCollider:2024jyv}
{\scshape International Muon Collider} collaboration, C.~Accettura et~al.,
  \emph{{Interim report for the International Muon Collider Collaboration
  (IMCC)}}, \href{http://dx.doi.org/10.23731/CYRM-2024-002}{\emph{CERN Yellow
  Rep. Monogr.} {\bf 2/2024} (2024) 176},
  [\href{https://arxiv.org/abs/2407.12450}{{\tt 2407.12450}}].

\bibitem{Holdom:1986eq}
B.~Holdom, \emph{{Searching for $\epsilon$ Charges and a New U(1)}},
  \href{http://dx.doi.org/10.1016/0370-2693(86)90470-3}{\emph{Phys. Lett. B}
  {\bf 178} (1986) 65--70}.

\bibitem{Fayet:1990wx}
P.~Fayet, \emph{{Extra U(1)'s and New Forces}},
  \href{http://dx.doi.org/10.1016/0550-3213(90)90381-M}{\emph{Nucl. Phys. B}
  {\bf 347} (1990) 743--768}.

\bibitem{Fabbrichesi:2020wbt}
M.~Fabbrichesi, E.~Gabrielli and G.~Lanfranchi, \emph{{The Dark Photon}},
  \href{https://arxiv.org/abs/2005.01515}{{\tt 2005.01515}}.

\bibitem{Alexander:2016aln}
J.~Alexander et~al., \emph{{Dark Sectors 2016 Workshop: Community Report}},
  tech. rep., {Fermi National Accelerator Laboratory (FNAL), Batavia, IL
  (United States)}, 8, 2016.

\bibitem{Babu:1997st}
K.~S. Babu, C.~F. Kolda and J.~March-Russell, \emph{{Implications of
  generalized Z - Z-prime mixing}},
  \href{http://dx.doi.org/10.1103/PhysRevD.57.6788}{\emph{Phys. Rev. D} {\bf
  57} (1998) 6788--6792}, [\href{https://arxiv.org/abs/hep-ph/9710441}{{\tt
  hep-ph/9710441}}].

\bibitem{Kim:2019oyh}
J.~H. Kim, S.~D. Lane, H.-S. Lee, I.~M. Lewis and M.~Sullivan, \emph{{Searching
  for Dark Photons with Maverick Top Partners}},
  \href{http://dx.doi.org/10.1103/PhysRevD.101.035041}{\emph{Phys. Rev. D} {\bf
  101} (2020) 035041}, [\href{https://arxiv.org/abs/1904.05893}{{\tt
  1904.05893}}].

\bibitem{Bauer:2018onh}
M.~Bauer, P.~Foldenauer and J.~Jaeckel, \emph{{Hunting All the Hidden
  Photons}}, \href{http://dx.doi.org/10.1007/JHEP07(2018)094}{\emph{JHEP} {\bf
  07} (2018) 094}, [\href{https://arxiv.org/abs/1803.05466}{{\tt 1803.05466}}].

\bibitem{APEX:2024jxw}
{\scshape APEX} collaboration, D.~He et~al., \emph{{Dark photon constraints
  from a 7.139~GHz cavity haloscope experiment}},
  \href{http://dx.doi.org/10.1103/PhysRevD.110.L021101}{\emph{Phys. Rev. D}
  {\bf 110} (2024) L021101}, [\href{https://arxiv.org/abs/2404.00908}{{\tt
  2404.00908}}].

\bibitem{Robens:2015gla}
T.~Robens and T.~Stefaniak, \emph{{Status of the Higgs Singlet Extension of the
  Standard Model after LHC Run 1}},
  \href{http://dx.doi.org/10.1140/epjc/s10052-015-3323-y}{\emph{Eur. Phys. J.
  C} {\bf 75} (2015) 104}, [\href{https://arxiv.org/abs/1501.02234}{{\tt
  1501.02234}}].

\bibitem{Robens:2019ynf}
T.~Robens, \emph{{Investigating extended scalar sectors at current and future
  colliders}}, \href{http://dx.doi.org/10.22323/1.350.0138}{\emph{PoS} {\bf
  LHCP2019} (2019) 138}, [\href{https://arxiv.org/abs/1908.10809}{{\tt
  1908.10809}}].

\bibitem{Robens:2022oue}
T.~Robens, \emph{{More Doublets and Singlets}},  in \emph{{56th Rencontres de
  Moriond on Electroweak Interactions and Unified Theories}}, 5, 2022.
\newblock \href{https://arxiv.org/abs/2205.06295}{{\tt 2205.06295}}.

\bibitem{Adhikari:2022yaa}
S.~Adhikari, S.~D. Lane, I.~M. Lewis and M.~Sullivan, \emph{{Complex Scalar
  Singlet Model Benchmarks for Snowmass}},  in \emph{{Snowmass 2021}}, 3, 2022.
\newblock \href{https://arxiv.org/abs/2203.07455}{{\tt 2203.07455}}.

\bibitem{Lane:2024vur}
S.~D. Lane, I.~M. Lewis and M.~Sullivan, \emph{{Resonant multiscalar production
  in the generic complex singlet model in the multi-TeV region}},
  \href{http://dx.doi.org/10.1103/PhysRevD.110.055017}{\emph{Phys. Rev. D} {\bf
  110} (2024) 055017}, [\href{https://arxiv.org/abs/2403.18003}{{\tt
  2403.18003}}].

\bibitem{Robens:2022cun}
T.~Robens, \emph{{Constraining extended scalar sectors at current and future
  colliders - an update}},  in \emph{{8th Workshop on Theory, Phenomenology and
  Experiments in Flavour Physics}: {Neutrinos, Flavor Physics and Beyond}}, 9,
  2022.
\newblock \href{https://arxiv.org/abs/2209.15544}{{\tt 2209.15544}}.

\bibitem{Robens:2025nev}
T.~Robens and R.~Santos, \emph{{BSM: Extended Scalar Sectors}},  tech. rep.,
  {Boskovic Inst., Zagreb, CERN}, 7, 2025.

\bibitem{ATLAS:2023tkt}
{\scshape ATLAS} collaboration, G.~Aad et~al., \emph{{Combination of searches
  for invisible decays of the Higgs boson using 139 fb{\ensuremath{-}}1 of
  proton-proton collision data at s=13 TeV collected with the ATLAS
  experiment}},
  \href{http://dx.doi.org/10.1016/j.physletb.2023.137963}{\emph{Phys. Lett. B}
  {\bf 842} (2023) 137963}, [\href{https://arxiv.org/abs/2301.10731}{{\tt
  2301.10731}}].

\bibitem{CMS:2023sdw}
{\scshape CMS} collaboration, A.~Tumasyan et~al., \emph{{A search for decays of
  the Higgs boson to invisible particles in events with a top-antitop quark
  pair or a vector boson in proton-proton collisions at $\sqrt{s} = 13\,\text
  {Te}\hspace{-.08em}\text {V} $}},
  \href{http://dx.doi.org/10.1140/epjc/s10052-023-11952-7}{\emph{Eur. Phys. J.
  C} {\bf 83} (2023) 933}, [\href{https://arxiv.org/abs/2303.01214}{{\tt
  2303.01214}}].

\bibitem{Papaefstathiou:2022oyi}
A.~Papaefstathiou, T.~Robens and G.~White, \emph{{Signal strength and W-boson
  mass measurements as a probe of the electro-weak phase transition at
  colliders - Snowmass White Paper}},  in \emph{{Snowmass 2021}}, 5, 2022.
\newblock \href{https://arxiv.org/abs/2205.14379}{{\tt 2205.14379}}.

\bibitem{CMS:2025ngq}
{\scshape CMS} collaboration, A.~Hayrapetyan et~al., \emph{{Combination of
  searches for nonresonant Higgs boson pair production in proton-proton
  collisions at $\sqrt{s}$= 13 TeV}},
  \href{https://arxiv.org/abs/2510.07527}{{\tt 2510.07527}}.

\bibitem{Cheng:2025aev}
{\scshape ATLAS} collaboration, A.~Cheng, \emph{{Combination of searches for
  resonant Higgs boson pair production using $pp$ collisions at $\sqrt s$ = 13
  TeV with the ATLAS detector}},
  \href{http://dx.doi.org/10.22323/1.478.0248}{\emph{PoS} {\bf LHCP2024} (2025)
  248}.

\bibitem{ATLAS:2023vdy}
{\scshape ATLAS} collaboration, G.~Aad et~al., \emph{{Combination of Searches
  for Resonant Higgs Boson Pair Production Using pp Collisions at
  s=13{\,}{\,}TeV with the ATLAS Detector}},
  \href{http://dx.doi.org/10.1103/PhysRevLett.132.231801}{\emph{Phys. Rev.
  Lett.} {\bf 132} (2024) 231801},
  [\href{https://arxiv.org/abs/2311.15956}{{\tt 2311.15956}}].

\bibitem{Buttazzo:2015bka}
D.~Buttazzo, F.~Sala and A.~Tesi, \emph{{Singlet-like Higgs bosons at present
  and future colliders}},
  \href{http://dx.doi.org/10.1007/JHEP11(2015)158}{\emph{JHEP} {\bf 11} (2015)
  158}, [\href{https://arxiv.org/abs/1505.05488}{{\tt 1505.05488}}].

\bibitem{Sopczak:1999ua}
A.~Sopczak, \emph{{Search for new particles at LEP}}, {\emph{Acta Phys. Polon.
  B} {\bf 30} (1999) 3229--3245},
  [\href{https://arxiv.org/abs/hep-ph/9911445}{{\tt hep-ph/9911445}}].

\bibitem{CMS:2022nty}
{\scshape CMS} collaboration, A.~Tumasyan et~al., \emph{{Inclusive nonresonant
  multilepton probes of new phenomena at $\sqrt s$=13{\,}{\,}TeV}},
  \href{http://dx.doi.org/10.1103/PhysRevD.105.112007}{\emph{Phys. Rev. D} {\bf
  105} (2022) 112007}, [\href{https://arxiv.org/abs/2202.08676}{{\tt
  2202.08676}}].

\bibitem{ATLAS:2023sbu}
{\scshape ATLAS} collaboration, G.~Aad et~al., \emph{{Search for
  third-generation vector-like leptons in $pp$ collisions at $\sqrt{s} =
  13\,\text{TeV}$ with the ATLAS detector}},
  \href{http://dx.doi.org/10.1007/JHEP07(2023)118}{\emph{JHEP} {\bf 07} (2023)
  118}, [\href{https://arxiv.org/abs/2303.05441}{{\tt 2303.05441}}].

\bibitem{CMS:2025urb}
{\scshape CMS} collaboration, V.~Chekhovsky et~al., \emph{{Search for
  vector-like leptons with long-lived particle decays in the CMS muon system in
  proton-proton collisions at $\sqrt{\text{s}}$ = 13 TeV}},
  \href{http://dx.doi.org/10.1007/JHEP08(2025)156}{\emph{JHEP} {\bf 08} (2025)
  156}, [\href{https://arxiv.org/abs/2503.16699}{{\tt 2503.16699}}].

\bibitem{ALEPH:2005ab}
{\scshape ALEPH, DELPHI, L3, OPAL, SLD, LEP Electroweak Working Group, SLD
  Electroweak Group, SLD Heavy Flavour Group} collaboration, S.~Schael et~al.,
  \emph{{Precision electroweak measurements on the $Z$ resonance}},
  \href{http://dx.doi.org/10.1016/j.physrep.2005.12.006}{\emph{Phys. Rept.}
  {\bf 427} (2006) 257--454}, [\href{https://arxiv.org/abs/hep-ex/0509008}{{\tt
  hep-ex/0509008}}].

\bibitem{Chen:2017hak}
C.-Y. Chen, S.~Dawson and E.~Furlan, \emph{{Vectorlike fermions and Higgs
  effective field theory revisited}},
  \href{http://dx.doi.org/10.1103/PhysRevD.96.015006}{\emph{Phys. Rev. D} {\bf
  96} (2017) 015006}, [\href{https://arxiv.org/abs/1703.06134}{{\tt
  1703.06134}}].

\bibitem{ParticleDataGroup:2024cfk}
{\scshape Particle Data Group} collaboration, S.~Navas et~al., \emph{{Review of
  particle physics}},
  \href{http://dx.doi.org/10.1103/PhysRevD.110.030001}{\emph{Phys. Rev. D} {\bf
  110} (2024) 030001}.

\bibitem{Muong-2:2025xyk}
{\scshape Muon g-2} collaboration, D.~P. Aguillard et~al., \emph{{Measurement
  of the Positive Muon Anomalous Magnetic Moment to 127~ppb}},
  \href{http://dx.doi.org/10.1103/7clf-sm2v}{\emph{Phys. Rev. Lett.} {\bf 135}
  (2025) 101802}, [\href{https://arxiv.org/abs/2506.03069}{{\tt 2506.03069}}].

\bibitem{Muong-2:2006rrc}
{\scshape Muon g-2} collaboration, G.~W. Bennett et~al., \emph{{Final Report of
  the Muon E821 Anomalous Magnetic Moment Measurement at BNL}},
  \href{http://dx.doi.org/10.1103/PhysRevD.73.072003}{\emph{Phys. Rev. D} {\bf
  73} (2006) 072003}, [\href{https://arxiv.org/abs/hep-ex/0602035}{{\tt
  hep-ex/0602035}}].

\bibitem{Aliberti:2025beg}
R.~Aliberti et~al., \emph{{The anomalous magnetic moment of the muon in the
  Standard Model: an update}},
  \href{http://dx.doi.org/10.1016/j.physrep.2025.08.002}{\emph{Phys. Rept.}
  {\bf 1143} (2025) 1--158}, [\href{https://arxiv.org/abs/2505.21476}{{\tt
  2505.21476}}].

\bibitem{Planck:2018vyg}
{\scshape Planck} collaboration, N.~Aghanim et~al., \emph{{Planck 2018 results.
  VI. Cosmological parameters}},
  \href{http://dx.doi.org/10.1051/0004-6361/201833910}{\emph{Astron.
  Astrophys.} {\bf 641} (2020) A6},
  [\href{https://arxiv.org/abs/1807.06209}{{\tt 1807.06209}}].

\bibitem{Mahapatra:2026fyv}
S.~Mahapatra, P.~K. Paul and N.~Sahu, \emph{{Forbidden dark matter assisted by
  first-order phase transition and associated gravitational waves}},
  \href{https://arxiv.org/abs/2601.12319}{{\tt 2601.12319}}.

\bibitem{Huang:2026qdc}
P.~Huang, A.~D. Medina and C.~E.~M. Wagner, \emph{{Rescuing Overabundant Dark
  Matter with a Strongly First Order Phase Transition in the Dark Sector}},
  \href{https://arxiv.org/abs/2602.16822}{{\tt 2602.16822}}.

\bibitem{Alloul_2014}
A.~Alloul, N.~D. Christensen, C.~Degrande, C.~Duhr and B.~Fuks, \emph{Feynrules
  2.0— a complete toolbox for tree-level phenomenology},
  \href{http://dx.doi.org/10.1016/j.cpc.2014.04.012}{\emph{Computer Physics
  Communications} {\bf 185} (Aug., 2014) 2250–2300}.

\bibitem{Alguero:2023zol}
G.~Alguero, G.~Belanger, F.~Boudjema, S.~Chakraborti, A.~Goudelis, S.~Kraml
  et~al., \emph{{micrOMEGAs 6.0: N-component dark matter}},
  \href{http://dx.doi.org/10.1016/j.cpc.2024.109133}{\emph{Comput. Phys.
  Commun.} {\bf 299} (2024) 109133},
  [\href{https://arxiv.org/abs/2312.14894}{{\tt 2312.14894}}].

\bibitem{XENON:2018voc}
{\scshape XENON} collaboration, E.~Aprile et~al., \emph{{Dark Matter Search
  Results from a One Ton-Year Exposure of XENON1T}},
  \href{http://dx.doi.org/10.1103/PhysRevLett.121.111302}{\emph{Phys. Rev.
  Lett.} {\bf 121} (2018) 111302},
  [\href{https://arxiv.org/abs/1805.12562}{{\tt 1805.12562}}].

\bibitem{XENON:2025vwd}
{\scshape XENON} collaboration, E.~Aprile et~al., \emph{{WIMP Dark Matter
  Search Using a 3.1 Tonne-Year Exposure of the XENONnT Experiment}},
  \href{http://dx.doi.org/10.1103/msw4-t342}{\emph{Phys. Rev. Lett.} {\bf 135}
  (2025) 221003}, [\href{https://arxiv.org/abs/2502.18005}{{\tt 2502.18005}}].

\bibitem{LZ:2024zvo}
{\scshape LZ} collaboration, J.~Aalbers et~al., \emph{{Dark Matter Search
  Results from 4.2{\,}{\,}Tonne-Years of Exposure of the LUX-ZEPLIN (LZ)
  Experiment}}, \href{http://dx.doi.org/10.1103/4dyc-z8zf}{\emph{Phys. Rev.
  Lett.} {\bf 135} (2025) 011802},
  [\href{https://arxiv.org/abs/2410.17036}{{\tt 2410.17036}}].

\bibitem{Degrande:2011ua}
C.~Degrande, C.~Duhr, B.~Fuks, D.~Grellscheid, O.~Mattelaer and T.~Reiter,
  \emph{{UFO - The Universal FeynRules Output}},
  \href{http://dx.doi.org/10.1016/j.cpc.2012.01.022}{\emph{Comput. Phys.
  Commun.} {\bf 183} (2012) 1201--1214},
  [\href{https://arxiv.org/abs/1108.2040}{{\tt 1108.2040}}].

\bibitem{Alwall_2014}
J.~Alwall, R.~Frederix, S.~Frixione, V.~Hirschi, F.~Maltoni, O.~Mattelaer
  et~al., \emph{The automated computation of tree-level and next-to-leading
  order differential cross sections, and their matching to parton shower
  simulations}, \href{http://dx.doi.org/10.1007/jhep07(2014)079}{\emph{Journal
  of High Energy Physics} {\bf 2014} (July, 2014) }.

\bibitem{Bierlich:2022pfr}
C.~Bierlich et~al., \emph{{A comprehensive guide to the physics and usage of
  PYTHIA 8.3}},
  \href{http://dx.doi.org/10.21468/SciPostPhysCodeb.8}{\emph{SciPost Phys.
  Codeb.} {\bf 2022} (2022) 8}, [\href{https://arxiv.org/abs/2203.11601}{{\tt
  2203.11601}}].

\bibitem{park_2018_ww43z-k8316}
T.~H. Park, \emph{Jet energy resolution measurement of the atlas detector using
  momentum balance},  Oct., 2018.

\bibitem{CMS:2016lmd}
{\scshape CMS} collaboration, V.~Khachatryan et~al., \emph{{Jet energy scale
  and resolution in the CMS experiment in pp collisions at 8 TeV}},
  \href{http://dx.doi.org/10.1088/1748-0221/12/02/P02014}{\emph{JINST} {\bf 12}
  (2017) P02014}, [\href{https://arxiv.org/abs/1607.03663}{{\tt 1607.03663}}].

\bibitem{Pezzotti:2022ndj}
I.~Pezzotti et~al., \emph{{Dual-Readout Calorimetry for Future Experiments
  Probing Fundamental Physics}},  \href{https://arxiv.org/abs/2203.04312}{{\tt
  2203.04312}}.

\bibitem{Cacciari:2008gp}
M.~Cacciari, G.~P. Salam and G.~Soyez, \emph{{The anti-$k_t$ jet clustering
  algorithm}},
  \href{http://dx.doi.org/10.1088/1126-6708/2008/04/063}{\emph{JHEP} {\bf 04}
  (2008) 063}, [\href{https://arxiv.org/abs/0802.1189}{{\tt 0802.1189}}].

\bibitem{Cacciari:2011ma}
M.~Cacciari, G.~P. Salam and G.~Soyez, \emph{{FastJet User Manual}},
  \href{http://dx.doi.org/10.1140/epjc/s10052-012-1896-2}{\emph{Eur. Phys. J.
  C} {\bf 72} (2012) 1896}, [\href{https://arxiv.org/abs/1111.6097}{{\tt
  1111.6097}}].

\bibitem{Butterworth:2008iy}
J.~M. Butterworth, A.~R. Davison, M.~Rubin and G.~P. Salam, \emph{{Jet
  substructure as a new Higgs search channel at the LHC}},
  \href{http://dx.doi.org/10.1103/PhysRevLett.100.242001}{\emph{Phys. Rev.
  Lett.} {\bf 100} (2008) 242001}, [\href{https://arxiv.org/abs/0802.2470}{{\tt
  0802.2470}}].

\bibitem{Bentvelsen:1998ug}
S.~Bentvelsen and I.~Meyer, \emph{{The Cambridge jet algorithm: Features and
  applications}}, \href{http://dx.doi.org/10.1007/s100520050232}{\emph{Eur.
  Phys. J. C} {\bf 4} (1998) 623--629},
  [\href{https://arxiv.org/abs/hep-ph/9803322}{{\tt hep-ph/9803322}}].

\bibitem{Cowan:2010js}
G.~Cowan, K.~Cranmer, E.~Gross and O.~Vitells, \emph{{Asymptotic formulae for
  likelihood-based tests of new physics}},
  \href{http://dx.doi.org/10.1140/epjc/s10052-011-1554-0}{\emph{Eur. Phys. J.
  C} {\bf 71} (2011) 1554}, [\href{https://arxiv.org/abs/1007.1727}{{\tt
  1007.1727}}].

\bibitem{Kumar:2015tna}
N.~Kumar and S.~P. Martin, \emph{{Vectorlike Leptons at the Large Hadron
  Collider}}, \href{http://dx.doi.org/10.1103/PhysRevD.92.115018}{\emph{Phys.
  Rev. D} {\bf 92} (2015) 115018},
  [\href{https://arxiv.org/abs/1510.03456}{{\tt 1510.03456}}].

\bibitem{Bhattiprolu:2020mwi}
P.~N. Bhattiprolu, S.~P. Martin and J.~D. Wells, \emph{{Criteria for projected
  discovery and exclusion sensitivities of counting experiments}},
  \href{http://dx.doi.org/10.1140/epjc/s10052-020-08819-6}{\emph{Eur. Phys. J.
  C} {\bf 81} (2021) 123}, [\href{https://arxiv.org/abs/2009.07249}{{\tt
  2009.07249}}].

\end{thebibliography}\endgroup

\end{document}